\documentclass[12pt,twoside]{mitthesis}
\usepackage{lgrind}
\usepackage{cmap}
\usepackage[T1]{fontenc}

\usepackage{amsmath,graphicx, amssymb, amsfonts, bbm}
\usepackage{bm}
\usepackage{color}
\usepackage{algorithmic}
\usepackage{graphicx}
\usepackage{textcomp}

\usepackage{soul}
\usepackage{url}
\usepackage[utf8]{inputenc}
\usepackage[small]{caption}
\usepackage{booktabs}
\usepackage{algorithm}
\usepackage{algorithmic}

\usepackage{calc}

\usepackage{multirow}
\usepackage{rotating}
\usepackage{balance}

\DeclareMathOperator{\E}{\mathbb{E}}

\def\L{{\cal L}}

\makeatletter
\def\Hline{%
\noalign{\ifnum0=`}\fi\hrule \@height 1pt \futurelet
\reserved@a\@xhline}
\makeatother

\def\BibTeX{{\rm B\kern-.05em{\sc i\kern-.025em b}\kern-.08emx
    T\kern-.1667em\lower.7ex\hbox{E}\kern-.125emX}}
    
\newcommand{\blockb}[1]{\multirow{3}{*}{ResBlock\(\left[\begin{array}{c}\text{1$\times$1}\\[-.1em] \text{3$\times$3}\\[-.1em] \text{1$\times$1}\end{array}\right]\)$\times$#1}
}

\newcommand{\blockc}[1]{\multirow{2}{*}{ResBlock\(\left[\begin{array}{c}\text{3$\times$3}\\[-.2em] \text{3$\times$3}\\[-.2em]\end{array}	\right]\)$\times$#1}
}

\def\all{all}
\ifx\files\all  \else 

\begin{document}

\title{The University of Tokyo}

\department{Department of Creative Informatics\\Graduate School of Information Science and Technology}

\author{Changhee Han}


\degreemonth{June}
\degreeyear{1990}
\thesisdate{May 18, 1990}


\supervisor{William J. Supervisor}{Associate Professor}

\chairman{Arthur C. Chairman}{Chairman, Department Committee on Graduate Theses}

\maketitle





\begin{center}
\ \\
\ \\
\ \\
\ \\
\ \\
\ \\
\ \\
\ \\
\ \\
\ \\

Dedicated with love to my mother, for always believing and supporting;\\and to my sister, for her unbounded trust and inspiration.
\end{center}

\newpage

\hrule height 0.05mm depth 1mm width 152mm
\vspace{-5mm}
\section*{Acknowledgments}
Most of all, I want to thank my mother and sister for raising me to value education and science---I would not be writing this thesis without their life-long dedicated support and disarming warmth. My second deepest appreciation goes to Prof. Hideki Nakayama for taking me under his wing and supporting me to explore creative and informative ideas, just like our department name. He kindly allowed me to challenge many different ideas despite my shortcomings. I am grateful to Research Center for Medical Big Data staff Prof. Fuyuko Kido, Dr. Youichirou Ninomiya, Fumika Tamura, and Prof. Shin'ichi Satoh, for offering me amazing research opportunities and helping me apply for Japanese permanent residency. In particular, I always enjoyed discussions and consequent publications together with you, Prof. Murao Kohei.

Many thanks to all people committed to our GAN projects. To physicians, including Dr. Yusuke Kawata and Dr. Fumiya Uchiyama (National Center for Global Health and Medicine), for their brilliant clinical insights and contributions. Especially, I owe entirely to Dr. Tomoyuki Noguchi (Kyushu Medical Center) for always sparing no effort to protect health and save lives with state-of-the-art technology---he introduced me many competent physicians for evaluating our GAN applications' clinical relevance. Dr. Yujiro Furukawa (Jikei University School of Medicine) has always supported our GAN projects from the very first project, trusting its potential despite repeated urgent requests. Dr. Kazuki Umemoto (Juntendo University School of Medicine) is AI-enthusiastic and very passionate about advancing healthcare.

This thesis would not have been achievable without Ryosuke Araki (Chubu University) and Prof. Hideaki Hayashi  (Kyushu University) because we launched and carried out these tough but fruitful projects together for so long. During my internship at FujiFilm with Dr. Yoshiro Kitamura, Akira Kudo, Akimichi Ichinose, and Dr. Yuanzhong Li, my best memory was drinking endless free beer.

I appreciate those participated in our questionnaire survey and GCL workshop: Dr. Yoshitaka Shida, Dr. Ryotaro Kamei  (National Center for Global Health and Medicine), Dr. Toshifumi Masuda (Kyushu Medical Center), Dr. Ryoko Egashira, Dr. Yoshiaki Egashira, Dr. Tetsuya Kondo (Saga University), Dr. Takafumi Nemoto (Keio University School of Medicine), Dr. Yuki Kaji, Miwa Aoki Uwadaira, Hajime Takeno (The University of Tokyo), and Kohei Yamamoto (Corpy\&Co., Inc.).

My sincere thanks go to Prof. Takeo Igarashi, Prof. Tatsuya Harada, Prof. Kazuhiko Ohe, Prof. Manabu Tsukada, and Prof. Murao Kohei for refereeing my Ph.D. thesis. Including Prof. Tsukada's mentoring, I was privileged to have continuous financial and human support from GCL program of my University by MEXT.

Clinically valuable research requires international/interdisciplinary collaboration. Backed up by the GCL, I achieved this goal thanks to Dr. Leonardo Rundo, real friend who invited me to Università degli Studi di Milano-Bicocca twice, University of Cambridge once, and his hometown in Sicily twice. After working hard together everyday from everywhere, he understands this thesis better than anyone in the world. Special thanks also to Prof. Giancarlo Mauri on his 70th birthday in Milan and Prof. Evis Sala in Cambridge, along with the other (mostly Italian) guys in their labs, including Prof. Daniela Besozzi, Prof. Paolo Cazzaniga, Prof. Marco Nobile, Dr. Andrea Tangherloni, and Simone Spolaor. After visiting 19 Italian cities, \textit{sono quasi Italiano!} Nicolas Y. Kröner was my geek tutor at Technische Universität München.

I wish to thank all my pleasant friends for making my Ph.D. days enjoyable and memorable in these intense years. Cheers to our long friendship, UTDS members, including Marishi Mochida and Kazuki Taniyoshi. Again, congratulations on your happy wedding, Yuki/Mari Inoue and Akito/Hitomi Misawa. Moreover, I shared far too many hilarious memories with KEK members Zoltán Ádám Milacski and Florian Gesser. Kurumi Nagasako drew this thesis' very first figure. Prof. Yumiko Furuichi and Miwako Hayasaka are always cheerful and smiling like Mother Teresa.

\textit{``My true title of glory is that I will live forever''} Last, but not least, I would like to thank Napoleon Bonaparte who taught me that \textit{nothing is impossible}. With courage and hope, I could conduct this research for both human prosperity and individual happiness. Listening to Beethoven's Symphony No.3 Eroica in the morning, I always try to be an everyday hero with great ambition like you. Your reign will never cease.

\cleardoublepage




\chapter*{Abstract}
Convolutional Neural Networks (CNNs) can play a key role in Medical Image Analysis under large-scale annotated datasets. However, preparing such massive dataset is demanding. In this context, Generative Adversarial Networks (GANs) can generate realistic but novel samples, and thus effectively cover the real image distribution. In terms of interpolation, the GAN-based medical image augmentation is reliable because medical modalities can display the human body's strong anatomical consistency at fixed position while clearly reflecting inter-subject variability; thus, we propose to use noise-to-image GANs  (e.g., random noise samples to diverse pathological images) for (\textit{i}) medical Data Augmentation (DA) and (\textit{ii}) physician training. Regarding the DA, the GAN-generated images can improve Computer-Aided Diagnosis based on supervised learning. For the physician training, the GANs can display novel desired pathological images and help train medical trainees despite infrastructural/legal constraints. This thesis contains four GAN projects aiming to present such novel applications' clinical relevance in collaboration with physicians. Whereas the methods are more generally applicable, this thesis only explores a few oncological applications.


In the first project, after proposing the two applications, we demonstrate that GANs can generate realistic/diverse 128 $\times$ 128 whole brain Magnetic Resonance (MR) images from noise samples---despite difficult training, such noise-to-image GAN can increase image diversity for further performance boost. Even an expert fails to distinguish the synthetic images from the real ones in Visual Turing Test.

The second project tackles image augmentation for 2D classification. Most CNN architectures adopt around 256 $\times$ 256 input sizes; thus, we use the noise-to-noise GAN, Progressive Growing of GANs (PGGANs), to generate realistic/diverse 256 $\times$ 256 whole brain MR images with/without tumors separately. Multimodal UNsupervised Image-to-image Translation further refines the synthetic images' texture and shape. Our two-step GAN-based DA boosts sensitivity 93.7\% to 97.5\% in 2D tumor/non-tumor classification. An expert classifies a few synthetic images as real.




The third project augments images for 2D detection. Further DA applications require pathology localization for detection and advanced physician training needs atypical image generation, respectively. To meet both clinical demands, we propose Conditional PGGANs (CPGGANs) that incorporates highly-rough bounding box conditions incrementally into the noise-to-image GAN (i.e., the PGGANs) to place realistic/diverse brain metastases at desired positions/sizes on $256 \times 256$ MR images; the bounding box-based detection requires much less physicians' annotation effort than segmentation. Our CPGGAN-based DA boosts sensitivity 83\% to 91\% in tumor detection with clinically acceptable additional False Positives (FPs). In terms of extrapolation, such pathology-aware GANs are promising because common and/or desired medical priors can play a key role in the conditioning---theoretically, infinite conditioning instances, external to the training data, exist and enforcing such constraints have an extrapolation effect \textit{via} model reduction.

Finally, we solve image augmentation for 3D detection. Because lesions vary in 3D position/appearance, 3D multiple pathology-aware conditioning is important. Therefore, we propose 3D Multi-Conditional GAN (MCGAN) that translates noise boxes into realistic/diverse $32 \times 32 \times 32$ lung nodules placed naturally at desired position/size/attenuation on Computed Tomography scans. Our 3D MCGAN-based DA boosts sensitivity in 3D nodule detection under any nodule size/attenuation at fixed FP rates. Considering the realism confirmed by physicians, it could perform as a physician training tool to display realistic medical images with desired abnormalities.

We confirm our pathology-aware GANs' clinical relevance for diagnosis \textit{via} two discussions: (\textit{i}) Conducting a questionnaire survey about our GAN projects for 9 physicians; (\textit{ii}) Holding a workshop about how to develop medical Artificial Intelligence (AI) fitting into a clinical environment in five years for 7 professionals with various AI and/or Healthcare background.





\pagestyle{plain}
\tableofcontents
\newpage
\listoffigures
\newpage
\listoftables
\newpage

\chapter*{List of Abbreviations}

\begin{table}[h]
\begingroup
\renewcommand{\arraystretch}{1.3}
\begin{tabular}{rl}
AI & Artificial  Intelligence \\
BRATS & Brain Tumor Image Segmentation Benchmark \\
CAD & Computer-Aided Diagnosis \\
CNN & Convolutional Neural Network \\
CPGGANs & Conditional Progressive Growing of Generative Adversarial Networks \\
CPM & Competition Performance Metric \\
CT & Computed Tomography \\
DA & Data Augmentation \\
DCGAN & Deep Convolutional Generative Adversarial Network \\
FLAIR & FLuid Attenuation Inversion Recovery \\
FP & False Positive \\
FROC & Free Receiver Operation Characteristic \\
GAN & Generative Adversarial Network \\
HGG & High-Grade Glioma \\
IoU & Intersection over Union \\
JS & Jensen-Shannon \\
LGG & Low-Grade Glioma \\
LSGANs & Least Squares Generative Adversarial Networks \\
MCGAN & Multi-Conditional Generative Adversarial Network \\
MRI & Magnetic Resonance Imaging \\
MUNIT & Multimodal UNsupervised Image-to-image Translation \\
PGGANs & Progressive Growing of Generative Adversarial Networks \\
ReLU & Rectified Linear Unit \\
\end{tabular}
\endgroup
\end{table}

\newpage

\begin{table}[h]
\begingroup
\renewcommand{\arraystretch}{1.3}
\begin{tabular}{rl}
ROI & Region Of Interest \\
SGD & Stochastic Gradient Descent \\
SimGAN & Simulated and unsupervised learning Generative Adversarial Network \\
t-SNE & t-distributed Stochastic Neighbor Embedding \\
T1 & T1-weighted \\
T1c & contrast enhanced T1-weighted \\
T2 & T2-weighted \\
UNIT & UNsupervised Image-to-image Translation \\
VAE & Variational AutoEncoder \\
VOI & Voxel Of Interest \\
WGAN & Wasserstein Generative Adversarial Network\\
WGAN-GP & Wasserstein Generative Adversarial Network with Gradient Penalty\\
\,\,\, & 
\end{tabular}
\endgroup
\end{table}
\setcounter{savepage}{\thepage}
\chapter{\LARGE Introduction}
\textit{``Life is short, and the Art long; the occasion fleeting; experience fallacious, and judgment difficult. The physician must not only be prepared to do what is right himself, but also to make the patient, the attendants, and externals cooperate.''}
\begin{flushright}
\textit{Hippocrates [460-375 BC]}
\end{flushright}

\section{Aims and Motivations}
Convolutional Neural Networks (CNNs) have revolutionized Medical Image Analysis by extracting valuable insights for better clinical examination and medical intervention; the CNNs occasionally outperformed even expert physicians in diagnostic accuracy when large-scale annotated datasets were available~\cite{hwang2018development,rajpurkar2017chexnet}. However, obtaining such massive datasets often involves the following intrinsic challenges~\cite{litjens2017survey,greenspan2016guest}: (\textit{i}) it is costly and laborious to collect medical images, such as Magnetic Resonance (MR) and Computed Tomography (CT) images, especially for rare disease; (\textit{ii}) it is time-consuming and observer-dependent, even for expert physicians, to annotate them due to the low pathological-to-healthy ratio. To tackle these issues, researchers have mainly focused on extracting as much information as possible from the available limited data~\cite{chapelle2009semi,vinyals2016matching}. Instead, Generative Adversarial Networks (GANs)~\cite{Goodfellow} can generate realistic but completely new samples \textit{via} many-to-many mappings, and thus effectively cover the real image distribution; they showed great promise in Data Augmentation (DA) using natural images, such as  21\% performance improvement in eye-gaze estimation~\cite{Shrivastava}.

Interpolation refers to new data point construction within a discretely-sampled data distribution. In terms of the interpolation, the GAN-based image augmentation is reliable on the medical images because medical modalites (e.g., X-ray, CT, MRI) can display the human body's strong anatomical consistency at fixed position while clearly reflecting inter-subject variability~\cite{hsieh2009computed,brown2014magnetic}---this is different from the natural images, where various objects can appear at any position; accordingly, to tackle large inter-subject, inter-pathology, and cross-modality variability~\cite{litjens2017survey,greenspan2016guest}, we propose to use noise-to-image GANs  (e.g., random noise samples to diverse pathological images) for (\textit{i}) medical DA and (\textit{ii}) physician training~\cite{Han1}. The noise-to-image GAN training is much more difficult than training image-to-image GANs  (e.g., a benign image to a malignant one); but, it can perform more global regularization (i.e., adding constraints when fitting a loss function on a training set to prevent overfitting) and increase image diversity for further performance boost.

Regarding the DA, the GAN-generated images can improve Computer-Aided Diagnosis (CAD) based on supervised learning~\cite{frid2018gan}. For the physician training, the GANs can display novel desired pathological images and help train medical trainees despite infrastructural and legal constraints~\cite{finlayson2018towards}. However, we cannot directly use conventional GANs for realistic/diverse high-resolution medical image augmentation. Moreover, we have to find effective loss functions and training schemes for each of those applications~\cite{han2019learning2}; the diversity matters more for the DA to sufficiently fill the real image distribution whereas the realism matters more for the physician training not to confuse the medical students and radiology trainees.


So, how can we perform clinically relevant GAN-based DA/physician training using only limited annotated training images? Always in collaboration with physicians, for improving 2D classification, we combine the noise-to-image~\cite{Han2,han2019combining}  (i.e., Progressive Growing of GANs, PGGANs~\cite{Karras}) and image-to-image GANs (i.e., Multimodal UNsupervised Image-to-image Translation, MUNIT~\cite{Huang}); the two-step GAN can generate and refine realistic/diverse original-sized $256 \times 256$ brain MR images with/without tumors separately. Nevertheless, further DA applications require pathology localization for detection (i.e., identifying target pathology positions in medical images) and advanced physician training needs atypical image generation, respectively. To meet both clinical demands, we propose novel 2D/3D bounding box-based GANs conditioned on pathology position/size/appearance; the bounding box-based detection requires much less physicians' annotation effort than rigorous segmentation.

Extrapolation refers to new data point estimation beyond a discretely-sampled data distribution. While it is not mutually-exclusive with the interpolation and both rely on a model's restoring force, it is more subject to uncertainty and thus a risk of meaningless data generation. In terms of the extrapolation, the pathology-aware GANs (i.e., the conditional GANs controlling pathology, such as tumors and nodules, based on position/size/appearance) are promising because common and/or desired medical priors can play a key role in the conditioning---theoretically, infinite conditioning instances, external to the training data, exist and enforcing such constraints have an extrapolation effect \textit{via} model reduction~\cite{stinis2019enforcing}; inevitable errors, not limited between two data points, caused by the model reduction forces a generator to synthesize images that the generator has never synthesized before.

For improving 2D detection, we propose Conditional PGGANs (CPGGANs) that incorporates highly-rough bounding box conditions incrementally into the noise-to-image GAN (i.e., the PGGANs) to place realistic/diverse brain metastases at desired positions/sizes on $256 \times 256$ MR images~\cite{han2019learning}. As its pathology-aware conditioning, we use 2D tumor position/size on MR images. Since lesions vary in 3D position/appearance, for improving 3D detection, we propose 3D Multi-Conditional GAN (MCGAN) that translates noise boxes into realistic/diverse $32 \times 32 \times 32$ lung nodules placed naturally at desired position/size/attenuation on CT scans~\cite{han2019synthesizing}; inputting the noise box with the surrounding tissues has the effect of combining the noise-to-image and image-to-image GANs. As its pathology-aware conditioning, we use 3D nodule position/size/attenuation on CT scans.

Lastly, we confirm our pathology-aware GANs' clinical relevance for diagnosis as a clinical decision support system and non-expert physician training tool \textit{via} two discussions: (\textit{i}) Conducting a questionnaire survey about our GAN projects for 9 physicians; (\textit{ii}) Holding a workshop about how to develop medical Artificial Intelligence (AI) fitting into a clinical environment in five years for 7 professionals with various AI and/or Healthcare background.

\noindent \textbf{Contributions.} Our main contributions are as follows:
\begin{itemize}
\item \textbf{Noise-to-Image GAN Applications:} We propose clinically-valuable novel noise-to-image GAN applications, medical DA and physician training, focusing on their ability to generate realistic and diverse images.
\item \textbf{Pathology-Aware GANs:} For required extrapolation, always in collaboration with physicians, we propose novel 2D/3D GANs controlling pathology (i.e., tumors and nodules) on most major modalities (i.e., brain MRI and lung CT).
\item \textbf{Clinical Validation:} After detailed discussions with many physicians and professionals with various AI and/or Healthcare background, we confirm our pathology-aware GANs' clinical relevance as a (\textit{i}) clinical decision support system and (\textit{ii}) non-expert physician training tool.
\end{itemize}

\newpage

\begin{figure}[t!]
  \centering
  \centerline{\includegraphics[width=1\linewidth]{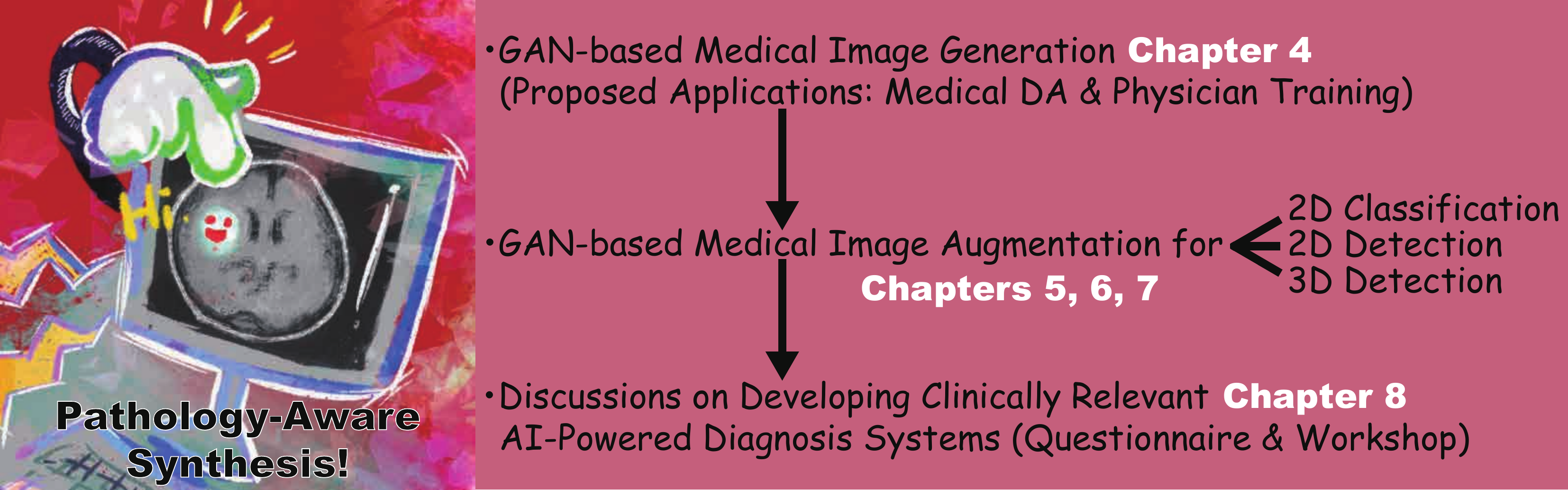}}
\caption[Conceptual scheme of this thesis]{Conceptual scheme of this thesis: inspired by their ability to generate realistic/diverse medical images, we propose novel noise-to-image GAN-based clinical applications, (\textit{i}) medical DA and (\textit{ii}) physician training; then, to present such GAN applications' technical soundness, we successfully tackle 2D classification, 2D detection, and 3D detection in collaboration with physicians---we propose novel pathology-aware GANs for effective extrapolation; lastly, we discuss how to develop clinically relevant AI-powered diagnosis systems, especially focusing on our pathology-aware GAN applications, \textit{via} a questionnaire survey and workshop.}
\label{synthesis}
\end{figure}

\section{Thesis Overview}
This Ph.D. thesis aims to present the clinical relevance of our novel pathology-aware GAN applications, medical DA and physician training, always in collaboration with physicians.

The thesis is organized as follows (Fig.~\ref{synthesis}). \textbf{Chapter 2} covers the background of Medical Image Analysis and Deep Learning, as well as methods  to address data paucity to bridge them. \textbf{Chapter 3} describes related work on the GAN-based medical DA and physician training, which emerged after our proposal to use noise-to-image GANs for those applications in \textbf{Chapter 4}. \textbf{Chapter 5} presents a two-step GAN for 2D classification that combines both noise-to-image and image-to-image GANs.  \textbf{Chapter 6} proposes CPGGANs for 2D detection that incorporates highly-rough bounding box conditions incrementally into the noise-to-image GAN. Finally, we propose 3D MCGAN for 3D detection that translates noise boxes into desired pathology in \textbf{Chapter 7}. \textbf{Chapter 8} discusses both our pathology-aware GANs' clinical relevance \textit{via} a questionnaire survey and how to develop medical AI fitting into a clinical environment in five years \textit{via} a workshop. Lastly, \textbf{Chapter 9} provides the conclusive remarks and future directions for further GAN-based extrapolation.
\chapter{\LARGE Background}
This chapter introduces basic concepts in Medical Image Analysis and Deep Learning. Afterwards, we describe methods to address data paucity because they play the greatest role in bridging the Medical Image Analysis and Deep Learning.

\section{Medical Image Analysis}
Medical Image Analysis refers to the process of increasing clinical examination/medical intervention efficiency, based on several imaging modalities and digital image processing techniques~\cite{lambin2017radiomics,rueckert2016learning}; to effectively visualize the human body's anatomical and physiological features, it covers various modalities including X-ray, CT, MRI, positron emission tomography, endoscopy, optical coherence tomography, pathology, ultrasound imaging, and fundus imaging. Its tasks are mainly classified into three groups: (\textit{i}) Early detection/diagnosis/prognosis of disease often based on pathology classification/detection/segmentation and survival prediction~\cite{Yap,lao2017deep}; (\textit{ii}) Clinical workflow enhancement often based on body part segmentation, inter-modality registration, 3D reconstruction, flow measurement, and surgery simulation~\cite{Chaitanya,yiannakopoulou2015virtual}; (\textit{iii}) Clinically impossible image analysis, such as radiogenomics that identifies the correlation between cancer imaging features and gene expression~\cite{zhu2019deep}.

Among the various modalities, this thesis focuses on the most common 3D modalities for non-invasive diagnosis, CT and MRI. To get a detailed picture inside the body, the CT merges multiple X-rays at different angles using computational tomographic reconstruction~\cite{hsieh2009computed,geyer2015state}. Since X-ray intensity is associated with the mass attenuation coefficient, higher-density tissues show higher attenuation and \textit{vice versa}. Accordingly, each voxel possesses its attenuation value following the Hounsfield scale from $-1,000$ to $+1,000$ (e.g., Hounsfield units $-1,000$ for air, 0 for water, and $+1,000$ for dense bone). The CT can provide a outstanding contrast within soft-tissue, bone, and lung while the soft-tissue contrast is poor---accordingly, it is especially performed for comprehensive lung assessment. 

The MRI uses magnetization properties of atomic nuclei~\cite{brown2014magnetic,mcrobbie2017mri}. Since different tissues show various relaxation processes when the nuclei return to their resting alignment, the tissues' proton density maps serve as both anatomical and functional images. Since the tissues possess two different relaxation times, T1 (i.e., longitudinal relaxation time) and T2 (i.e., transverse relaxation time), as MRI sequences, we can obtain both T1-weighted (T1) and T2-weighted (T2) images. Moreover, using very long repetition time and time to echo, we can obtain FLuid Attenuation Inversion Recovery (FLAIR) images. The MRI can provide a superior soft-tissue contrast to the CT---accordingly, it is especially performed for comprehensive brain assessment.

\section{Deep Learning}
Deep Learning is a kind of Machine Learning algorithms, based on Artificial Neural Networks~\cite{lecun2015deep}. The Deep Neural Networks consist of many linearly connected non-linear units whose parameters are optimized by gradient descent~\cite{Debate}; accordingly, their multiple layers can gradually grasp more-detailed features as training progresses (i.e., learning which features to place is automatic). Thanks to the good generalization ability, under large-scale data, the Deep Learning significantly outperforms classical Machine Learning algorithms relying on feature engineering. A visual cortex includes arrangements of simple and complex cells activated by a receptive field (i.e., subregions of a visual field); inspired by this biological structure~\cite{hubel1959receptive}, CNNs adopt a mathematical operation called convolution to achieve translation invariance~\cite{Russakovsky}. Since the CNNs are excellent at image/video recognition, their diverse medical applications include pathology classification/detection/segmentation and survival prediction~\cite{Yap,lao2017deep}.

Variational AutoEncoders (VAEs) often suffer from blurred samples despite easier training, due to the imperfect reconstruction using a single objective function~\cite{Kingma2013}; meanwhile, GANs have revolutionized image generation in terms of realism and diversity~\cite{Zhu}, including denoising~\cite{wolterink2017generative} and MRI-to-CT translation~\cite{emami2018generating}, based on a two-player objective function using two CNNs~\cite{Goodfellow}: a generator $G$ tries to generate realistic images to fool a discriminator $D$ while maintaining diversity; $D$ attempts to distinguish between the real and synthetic images. However, difficult GAN training from the two-player objective function accompanies artifacts and mode collapse~\cite{Gulrajani}, when generating high-resolution images (e.g., $256 \times 256$ pixels)~\cite{Radford}; to tackle this, multi-stage noise-to-image GANs have been proposed: AttnGAN generated images from text using attention-based multi-stage refinement~\cite{xu2018attngan}; PGGANs generated realistic images using low-to-high resolution multi-stage training~\cite{Karras}.

Contrarily, to obtain images with desired texture and shape, some researchers have proposed image-to-image GANs: MUNIT translated images using both GANs and VAEs~\cite{Huang}; Simulated and unsupervised learning GAN (SimGAN) translated images for DA using the self-regularization term and local adversarial loss~\cite{Shrivastava}; Isola \textit{et al.} proposed \textit{Pix2Pix} GAN to produce robust images using paired training samples~\cite{isola2017image}. Others have proposed conditional GANs: Reed \textit{et al.} proposed bounding box-based conditional GAN to control generated images' local properties~\cite{reed2016learning}; Park \textit{et al.} proposed multi-conditional GAN to refine base images based on texts describing desired position~\cite{park2018mc}.

In Healthcare, medical images have generated the largest volume of data and this trend will no doubt increase due to equipment improvement~\cite{hung2015ubiquitous,yaffe2019emergence}. Accordingly, as the Deep Learning dominates Computer Vision, Medical Image Analysis is not an exception; their combination can analyze the large-scale medical images and extract valuable insights for better clinical examination and medical intervention. However, the biggest challenge to bridge them lies in the difficulty of obtaining desired pathological images, especially for rare disease~\cite{litjens2017survey,greenspan2016guest}. Moreover, it is time-consuming and observer-dependent, even for expert physicians, to annotate them.

\section{Methods to Address Data Paucity}
So, how can we tackle the data paucity? We can either attempt to (\textit{a}) overcome the lack of generalization or (\textit{b}) overcome difficulties in optimization. The most straightforward and effective way to address the generalization is DA~\cite{perez2017effectiveness,shorten2019survey}; because the best model when given data is uncertain, we commonly increase training set size. Human perception is invariant to size, shape, brightness, and color~\cite{heeger1994representation}. Accordingly, we recognize the same objects while their such features change, and thus intentionally changing the features is plausible to obtain more data. Such classical DA include (\textit{i}) x/y/z-axis flipping and rotating, (\textit{ii}) zooming and scaling, (\textit{iii}) cropping, (\textit{iv}) translating, (\textit{v}) elastic deformation, (\textit{vi}) adding Gaussian noise (i.e, the distortion of high frequency features), and (\textit{vii}) brightness and contrast fluctuation.

Recent DA techniques focus on regularization: Mixup~\cite{zhang2017mixup} and Between-class learning~\cite{tokozume2018between} mixed two images during training, such as a dog image and a cat one, for regularization; Cutout randomly masked out square regions during training for regularization~\cite{devries2017improved}; CutMix combined the Mixup and Cutout~\cite{yun2019cutmix}. As a recent impressive DA approach, AutoAugment automatically searched for improved DA policies~\cite{cubuk2018autoaugment}. Moreover, similarly to the Mixup among all images within the same class, GAN-based DA can fill the uncovered real image distribution by generating realistic and diverse images \textit{via} many-to-many mapping~\cite{antoniou2017data}.

Along with the DA, researchers proposed many other techniques to improve the generalization: semi-supervised learning can considerably increase accuracy under limited labeled data by using pseudo labels for unlabeled data~\cite{chapelle2009semi}; unsupervised anomaly detection allows to detect out-of-distribution images from normal ones, such as disease, without any labeled data~\cite{han2019gan}; regularization techniques, such as dropout~\cite{srivastava2014dropout}, Lasso~\cite{park2008bayesian}, and elastic net~\cite{zou2005regularization}, are commonly used for reducing overfitting; similarly, ensembling multiple models \textit{via} bagging~\cite{breiman1996bagging} and boosting~\cite{freund1999short} can effectively increase the robustness; Lastly, in Medical Image Analysis, we can fuse multiple image modalities and/or sequences, such as MRI $+$ CT~\cite{hou2019brain} and T1 MRI $+$ T2 MRI~\cite{chen2018mmfnet}.

Moreover, many techniques exist for overcoming the difficulties in optimization: transfer learning can achieve better parameter initialization~\cite{shin2016deep}; problem reduction, such as inputting 2D/3D image patches instead of a whole image, can eliminate unnecessary parameters~\cite{zbontar2016stereo}; learning methods with less data, such as zero-shot learning~\cite{romera2015embarrassingly}, one-shot learning~\cite{vinyals2016matching}, and neural Turing machine~\cite{graves2014neural}, are also promising; meta-learning promotes a versatile model applicable to various tasks without requiring multiple training from scratch~\cite{finn2017model}.

\chapter{\LARGE Investigated Contexts and Applications}
In terms of interpolation, GAN-based medical image augmentation is reliable because medical modalities (e.g., X-ray, CT, MRI) can display the human body's strong anatomical consistency at fixed position while clearly reflecting inter-subject variability~\cite{hsieh2009computed,brown2014magnetic}---this is different from natural images, where various objects can appear at any position. Accordingly, we proposed to use noise-to-image GANs for (\textit{i}) medical DA and (\textit{ii}) physician training~\cite{Han1} in \textbf{Chapter 4}. Since then, research towards such clinically valuable applications has shown great promise. This chapter covers such related research works except our own works~\cite{Han2, han2019combining, han2019learning, han2019synthesizing} included in \textbf{Chapters 5-7}. Involving 9 physicians, we discuss in detail the clinical relevance of the GAN-based medical DA and physician training~\cite{han2020bridging} in \textbf{Chapter 8}.

\section{GAN-based Medical DA}
Because the lack of annotated pathological images is the greatest challenge in CAD~\cite{litjens2017survey,greenspan2016guest}, to handle various types of small/fragmented datasets from multiple scanners, researchers have actively conducted GAN-based DA studies especially in Medical Image Analysis. For better classification, some researchers adopted image-to-image GANs similarly to their conventional medical applications, such as denoising~\cite{wolterink2017generative} and MRI-to-CT translation~\cite{emami2018generating}: Wu \textit{et al.} translated $256 \times 256$ normal mammograms into lesion ones~\cite{Wu}, Gupta \textit{et al.} translated $1024 \times 512$ normal leg X-ray images into bone lesion ones~\cite{Gupta}, and Malygina \textit{et al.} translated $256 \times 256$/$512 \times 512$ normal chest X-ray images into pneumonia/pleural-thickening ones~\cite{malygina2019data}. Meanwhile, others adopted the noise-to-image GANs as we proposed, to increase image diversity for further performance boost---the diversity matters more for the DA to sufficiently fill the real image distribution: Frid-Adar \textit{et al.} augmented $64 \times 64$ liver lesion CT images~\cite{frid2018gan}, Madani \textit{et al.} augmented $128 \times 128$ chest X-ray images with cardiovascular abnormality~\cite{Madani}, and Konidaris \textit{et al.} augmented $192 \times 160$ brain MR images with Alzheimer's disease~\cite{konidaris2018generative}.

To facilitate pathology detection and segmentation, researchers conditioned the image-to-image GANs, not the noise-to-image GANs like our work in \textbf{Chapter 6}, with pathology features (e.g., position, size, and appearance) and generated realistic/diverse pathology at desired positions in medical images. In terms of extrapolation, the pathology-aware GANs are promising because common and/or desired medical priors can play a key role in the conditioning---theoretically, infinite conditioning instances, external to the training data, exist and enforcing such constraints have an extrapolation effect \textit{via} model reduction~\cite{stinis2019enforcing}. To the best of our knowledge, only Kanayama \textit{et al.} tackled bounding box-based pathology detection using the image-to-image GAN~\cite{kanayama2019gastric}; they translated normal endoscopic images with various image sizes ($458 \times 405$ on average) into gastric cancer ones by inputting both a benign image and a black image (i.e., pixel value: 0) with a specific lesion Region Of Interest (ROI) at desired position. Without conditioning the noise-to-image GAN with nodule position, Gao \textit{et al.} generated $40 \times  40 \times 18$ 3D nodule subvolumes only applicable to their subvolume-based detector using binary classification~\cite{gao2019augmenting}.

Since 3D imaging is spreading in radiology (e.g., CT and MRI), most GAN-based DA works for segmentation exploited 3D conditional image-to-image GANs. However, 3D medical image generation is more challenging than 2D one due to expensive computational cost and strong anatomical consistency; so, instead of generating a whole image including pathology, researchers only focused on a malignant Voxel Of Interest (VOI): Shin \textit{et al.} translated $128 \times 128 \times 54$ normal brain MR images into tumor ones by inputting both a benign image and a tumor-conditioning image~\cite{shin2018medical}, similarly to the Kanayama \textit{et al.}'s work~\cite{kanayama2019gastric}; Jin \textit{et al.}~ generated $64 \times 64 \times 64$ CT images of lung nodules including the surrounding tissues by only inputting a VOI centered at a lung nodule, but with a central sphere region erased~\cite{jin2018ct}. Recently, instead of generating realistic images and training classifiers on them separately, Chaitanya \textit{et al.} directly optimized segmentation results on cardiac MR images~\cite{Chaitanya}; however, it segmented body parts, instead of pathology. Since effective GAN-based medical DA generally requires much engineering effort, we also published a tutorial journal paper~\cite{han2019learning2} about tricks to boost classification/detection/segmentation performance using the GANs, based on our experience and related work.

\section{GAN-based Physician Training}
While medical students and radiology trainees must view thousands of images to become competent~\cite{wang2012competencies}, accessing such abundant medical images is often challenging due to infrastructural and legal constraints~\cite{ching2018opportunities}. Because pathology-aware GANs can generate novel medical images with desired abnormalities (e.g., position, size, and appearance)---while maintaining enough realism not to confuse the medical trainees---GAN-based physician training concept is drawing attention: Chuquicusma \textit{et al.} appreciated the GAN potential to train radiologists for educational purpose after successfully generating $56 \times 56$ CT images of lung nodules that even deceived experts~\cite{Chuquicusma}; thanks to their anonymization ability, Shin \textit{et al.} proposed to share pathology-aware GAN-generated images outside institutions after achieving considerable tumor segmentation results with only synthetic $128 \times 128 \times 54$ MR images for training~\cite{shin2018medical}; more importantly, Finlayson \textit{et al.} from Harvard Medical School are currently validating a class-conditional GANs' radiology educational efficacy after succeeding in learning features that distinguish fractures from non-fractures on $1024 \times 1024$ pelvic X-ray images~\cite{finlayson2018towards}.
\chapter{\LARGE GAN-based Medical Image Generation}

\section{Prologue to First Project}
\subsection{Project Publication}
\begin{itemize}
\item \textbf{GAN-based Synthetic Brain MR Image Generation}. \textbf{C. Han}, H. Hayashi, L. Rundo, R. Araki, Y. Furukawa, W. Shimoda, S. Muramatsu, G. Mauri, H. Nakayama, In \textit{IEEE International Symposium on Biomedical Imaging (ISBI)}, pp. 734--738, April 2018.
\end{itemize}
\subsection{Context}
Prior to this work, it remained challenging to generate realistic and diverse medical images using noise-to-image GANs, not image-to-image GANs~\cite{wolterink2017generative}, due to their unstable training. GAN architectures well-suited for medical images were unclear. Yi \textit{et al.} published results on the noise-to-image GAN-based brain MR image generation, proposing its potential for medical DA and physician training while our paper was under submission~\cite{calimeri2017biomedical}; however, they only generated single-sequence low-resolution $128 \times 64$ brain MR images without tumors.

\subsection{Contributions}
This project's main contribution is to propose to use recently developed Wasserstein Generative Adversarial Network (WGAN)~\cite{Arjovsky} for medical DA and physician training---the medical GAN applications are reliable in terms of interpolation because medical modalities can display the human body's strong anatomical consistency at fixed position while clearly reflecting inter-subject variability. We also demonstrate the noise-to-image GAN's such potential by generating multi-sequence realistic/diverse $128 \times 128$ whole brain tumor MR images~\cite{Arjovsky}; then, we confirm the superb realism \textit{via} Visual Turing Test by a physician. 

\subsection{Recent Developments}
Since proposing the GAN applications, we have successfully applied the noise-to-image GANs to improve 2D tumor classification/detection on $256 \times 256$ brain MR images~\cite{Han2, han2019combining, han2019learning} as described in \textbf{Chapters 5 and 6}. For better 3D tumor segmentation, Shin \textit{et al.} have translated $128 \times 128 \times 54$ normal brain MR images into tumor ones using the image-to-image GAN~\cite{shin2018medical}. Finlayson \textit{et al.} have generated $1024 \times 1024$ pelvic fracture/non-fracture X-ray images using a class-conditional noise-to-image GAN, also introducing ongoing work on validating such GANs' radiology educational efficacy~\cite{finlayson2018towards}. Kwon \textit{et al.} have generated realistic/diverse 3D brain MR images using the noise-to-image GAN~\cite{kwon2019generation}.


\newpage
\section{Motivation}
Along with classic methods~\cite{Rundo}, CNNs have recently revolutionized medical image analysis~\cite{Shen}, including brain MRI segmentation~\cite{Havaei}.
However, CNN training demands extensive medical data that are laborious to obtain~\cite{Ravi}.
To overcome this issue, DA techniques via reconstructing original images are common for better performance, such as geometry  and intensity transformations~\cite{Ronneberger, Milletari}.

However, those reconstructed images intrinsically resemble the original ones, leading to limited performance improvement in terms of generalization abilities; thus, generating realistic (similar to the real image distribution) but completely new images is essential. In this context, GAN-based DA has excellently performed in general computer vision tasks. It attributes to GAN's good generalization ability from matching the noise-generated distribution to the real one with a sharp value function. Especially, Shrivastava \emph{et al.} (SimGAN) outperformed the state-of-the-art with a relative 21\% improvement in eye-gaze estimation~\cite{Shrivastava}.

\begin{figure}[t!]
  \centering
  \centerline{\includegraphics[width=1\linewidth]{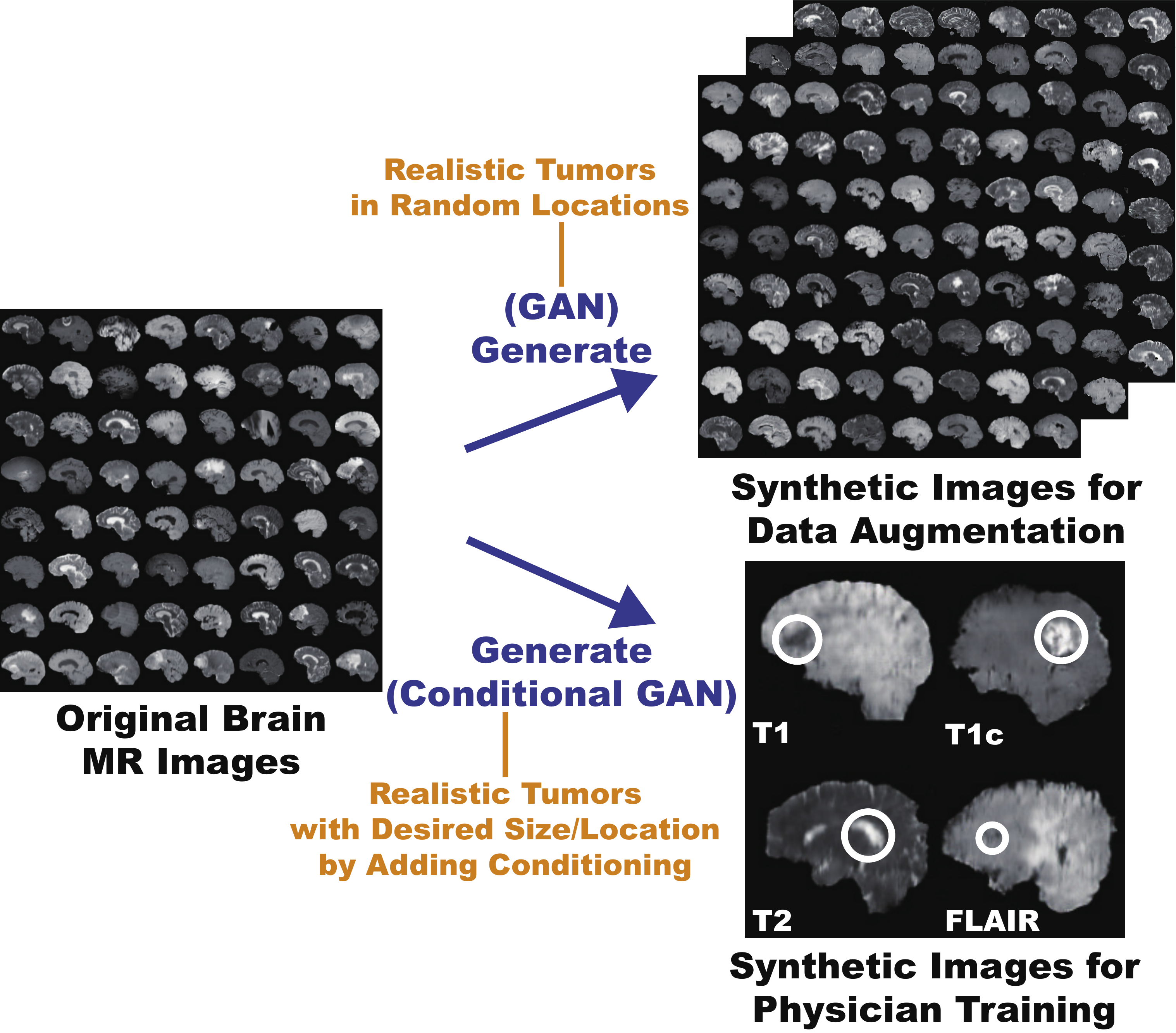}}
\caption[Potential applications of the proposed GAN-based synthetic brain MR image generation.]{Potential applications of the proposed GAN-based synthetic brain MR image generation: (1) DA for better diagnostic accuracy by generating random realistic images giving insights in classification; (2) physician training for better understanding various diseases to prevent misdiagnosis by generating desired realistic pathological images.}
\label{fig4_1}
\end{figure}

So, how can we generate realistic medical images completely different from the original samples? Our aim is to generate synthetic multi-sequence brain MR images using GANs, which is essential in medical imaging to increase diagnostic reliability, such as \textit{via} DA in CAD as well as physician training (Fig.~\ref{fig4_1})~\cite{Prastawa}. However, this is extremely challenging---MR images are characterized by low contrast, strong visual consistency in brain anatomy, and intra-sequence variability. Our novel GAN-based approach for medical DA adopts Deep Convolutional Generative Adversarial Network (DCGAN)~\cite{Radford} and WGAN~\cite{Arjovsky} to generate realistic images, and an expert physician validates them \textit{via} Visual Turing Test~\cite{Salimans}.

\noindent \textbf{Research Questions.} We mainly address two questions:
\begin{itemize}
\item \textbf{GAN Selection:} Which GAN architecture is well-suited for realistic medical image generation?
\item \textbf{Medical Image Processing:} How can we handle MR images with specific intra-sequence variability?\\
\end{itemize}
\newpage

\noindent \textbf{Contributions.} Our main contributions are as follows:
\begin{itemize}
\item \textbf{MR Image Generation:} This research shows that WGAN can generate realistic multi-sequence brain MR images, possibly leading to valuable clinical applications: DA and physician training.
\item \textbf{Medical Image Generation:} This research provides how to exploit medical images with intrinsic intra-sequence variability towards GAN-based DA for medical imaging.
\end{itemize}

\section{Materials and Methods}
Towards clinical applications utilizing realistic brain MR images, we generate synthetic brain MR images from the original samples using GANs.
Here, we compare the most used two GANs, DCGAN and WGAN, to find a well-suited GAN between them for medical image generation---it must avoid mode collapse 
and generate realistic MR images with high resolution.

\subsection{BRATS 2016 Dataset}
This project exploits a dataset of multi-sequence brain MR images to train GANs with sufficient data and resolution, which was originally produced for the Multimodal Brain Tumor Image Segmentation Benchmark (BRATS) Challenge~\cite{Menze}. In particular, the BRATS 2016 training dataset contains 220 High-Grade Glioma (HGG) and 54 Low-Grade Glioma (LGG) cases, with T1-weighted (T1), contrast enhanced T1-weighted (T1c), T2-weighted, and FLAIR sequences---they were skull stripped and resampled to isotropic $1\mbox{mm} \times 1\mbox{mm} \times 1\mbox{mm}$ resolution with $240 \times 240 \times 155$ voxels; among the different sectional planes, we use sagittal multi-sequence scans of the HGG patients to show that our GANs can generate a complete view of the whole brain anatomy (allowing for visual consistency among the different brain lobes), including also severe tumors for clinical purpose.

\subsection{DCGAN/WGAN-based Image Generation}
\begin{figure}[t!]
  \centering
  \centerline{\includegraphics[width=1\linewidth]{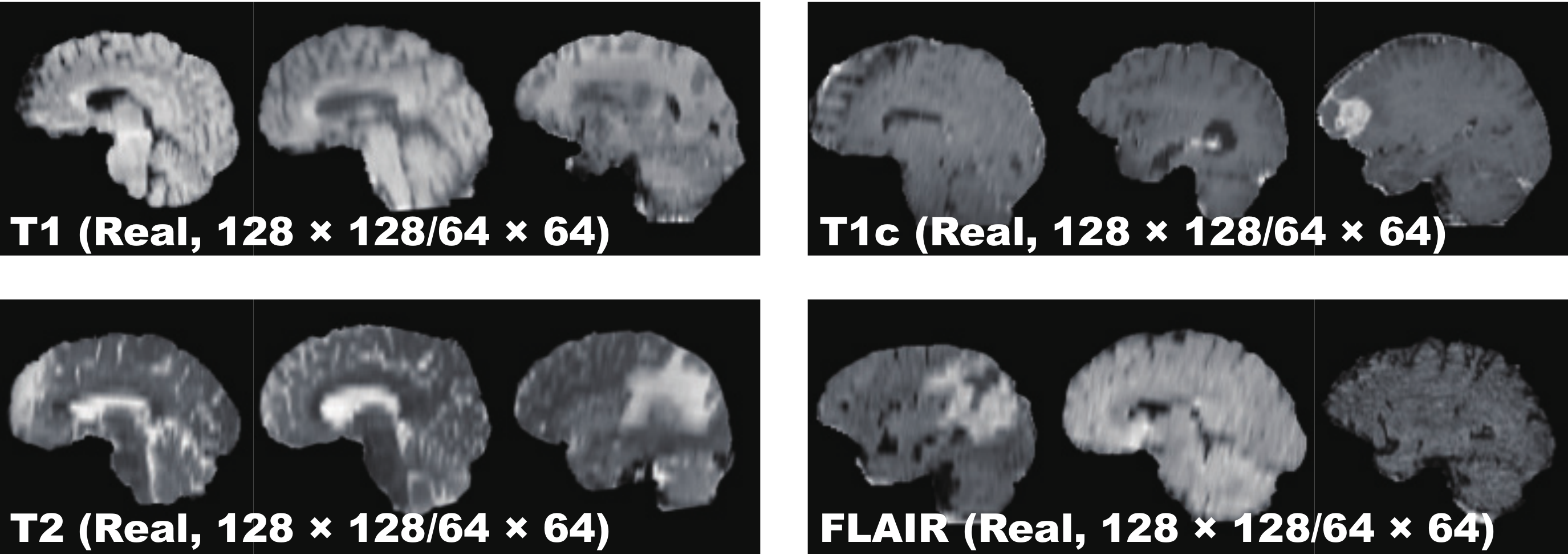}}
\caption[Example real $128\times128$/$64\times64$ MR images used for GAN training.]{Example real $128\times128$/$64\times64$ MR images used for GAN training: the resized sagittal multi-sequence brain MRI scans of patients with HGG on the BRATS 2016 training dataset~\cite{Menze}.}
\label{fig4_2}
\end{figure}

\noindent \textbf{Pre-processing} We select the slices from $\#80$ to $\#149$ among the whole $240$ slices to omit initial/final slices, since they convey a negligible amount of useful information and could affect the training. The images are resized to both $64 \times 64$/$128 \times 128$ pixels from $ 240 \times 155$ for better GAN training (DCGAN architecture results in stable training on 64 $\times$ 64 pixels~\cite{Radford}, and so $128 \times 128$ is reasonably a high-resolution). Fig.~\ref{fig4_2} shows example real MR images used for training; each sequence contains 15,400 images with 220 patients $\times$ 70 slices (61,600 in total).

\noindent \textbf{MR Image Generation} DCGAN and WGAN generate six types of images as follows:
\begin{itemize}
\item T1 sequence ($128 \times 128$) from the real T1;
\item T1c sequence ($128 \times 128$) from the real T1c;
\item T2 sequence ($128 \times 128$) from the real T2;
\item FLAIR sequence ($128 \times 128$) from the real FLAIR;
\item Concat sequence ($128 \times 128$) from concatenating the real T1, T1c, T2, and FLAIR (i.e., feeding the model with samples from all the MRI sequences);
\item Concat sequence ($64 \times 64$) from concatenating the real T1, T1c, T2, and FLAIR.
\end{itemize}
Concat sequence refers to a new ensemble sequence for an alternative DA, containing features of all four sequences. We also generate $64 \times 64$ Concat images to compare the generation performance in terms of image resolution.\\

\noindent \textbf{DCGAN}~\cite{Radford} is a standard GAN~\cite{Goodfellow} with a convolutional architecture for unsupervised learning; this generative model uses up-convolutions interleaved with Rectified Lineaer Unit (ReLU) non-linearity and batch normalization.\\
\indent Let $p_{\rm data}$ be a generating distribution over data $\bm{x}$. The generator $G(\bm{z}; \theta_{g})$ is a mapping to data space that takes a prior on input noise variables $p_{\bm{z}}(\bm{z})$, where $G$ is a neural network with parameters $\theta_{g}$. Similarly, the discriminator $D(\bm{x}; \theta_{d})$ is a neural network with parameters $\theta_{d}$ that takes either real data or synthetic data and outputs a single scalar probability that $\bm{x}$ came from the real data. The discriminator $D$ maximizes the probability of classifying both training examples and samples from $G$ correctly while the generator $G$ minimizes the likelihood; it is formulated as a minimax two-player game with value function $V(G,D)$:
\begin{equation}
\begin{split}
\min_{G} \max_{D} V(D,G)& = \E_{\bm{x} \sim p_{{\rm data}} (\bm{x})} [\log D(\bm{x})] \\
&\quad + \E_{\bm{z} \sim p_{\bm{z}} (\bm{z})} [\log (1 - D(G(\bm{z})))].
\end{split}
\end{equation}
This can be reformulated as the minimization of the Jensen-Shannon (JS) divergence between the distribution $p_{\rm data}$ and another distribution $p_{g}$ derived from $p_{\bm{z}}$ and $G$.\\

\noindent \textbf{DCGAN Implementation Details} We use the same DCGAN architecture~\cite{Radford} with no $\tanh$ in the generator, ELU as the discriminator, all filters of size $4 \times 4$, and a half channel size for DCGAN training. A batch size of $64$ and Adam optimizer with $2.0 \times 10^{-4}$ learning rate were implemented. \\

\noindent \textbf{WGAN}~\cite{Arjovsky} is an alternative to traditional GAN training, as the JS divergence is limited, such as when it is discontinuous; this novel GAN achieves stable learning with less mode collapse by replacing it to the Earth Mover (EM) distance (i.e., the Wasserstein-1 metrics):
\begin{equation}
W(p_{g},p_{r}) = \inf_{p \in \prod (p_{g},p_{r})} \E_{(\bm{x},\bm{x}') \sim p} \| \bm{x} - \bm{x}' \|,
\end{equation}
where $\prod (p_{g},p_{r})$ is the set of all joint distributions $p$ whose marginals are $p_{g}$ and $p_{r}$, respectively. In other words, $p$ implies how much mass must be transported from one distribution to another. This distance intuitively indicates the cost of the optimal transport plan.\\

\noindent \textbf{WGAN Implementation Details} We use the same DCGAN architecture~\cite{Radford} for WGAN training. A batch size of 64 and Root Mean Square Propagation (RMSprop) optimizer with $5.0 \times 10^{-5}$ learning rate were implemented. 

\subsection{Clinical Validation \textit{via} Visual Turing Test}
To quantitatively evaluate how realistic the synthetic images are, an expert physician was asked to constantly classify a random selection of $50$ real/$50$ synthetic MR images as real or synthetic shown in random order for each GAN/sequence, without previous training stages revealing which is real/synthetic; Concat images were classified together with real T1, T1c, T2, and FLAIR images in equal proportion. The so-called Visual Turing Test~\cite{Salimans} uses binary questions to probe a human ability to identify attributes and relationships in images.
For these motivations, it is commonly used to evaluate GAN-generated images, such as for SimGAN~\cite{Shrivastava}.
This applies also to medical images in clinical environments~\cite{Chuquicusma}, wherein physicians' expertise is critical.


\section{Results}
\label{sec:pb}
This section shows how DCGAN and WGAN generate synthetic brain MR images. The results include instances of synthetic images and their quantitative evaluation of the realism by an expert physician.

\subsection{MR Images Generated by DCGAN/WGAN}
\begin{figure}[t!]
  \centering
  \centerline{\includegraphics[width=1\linewidth]{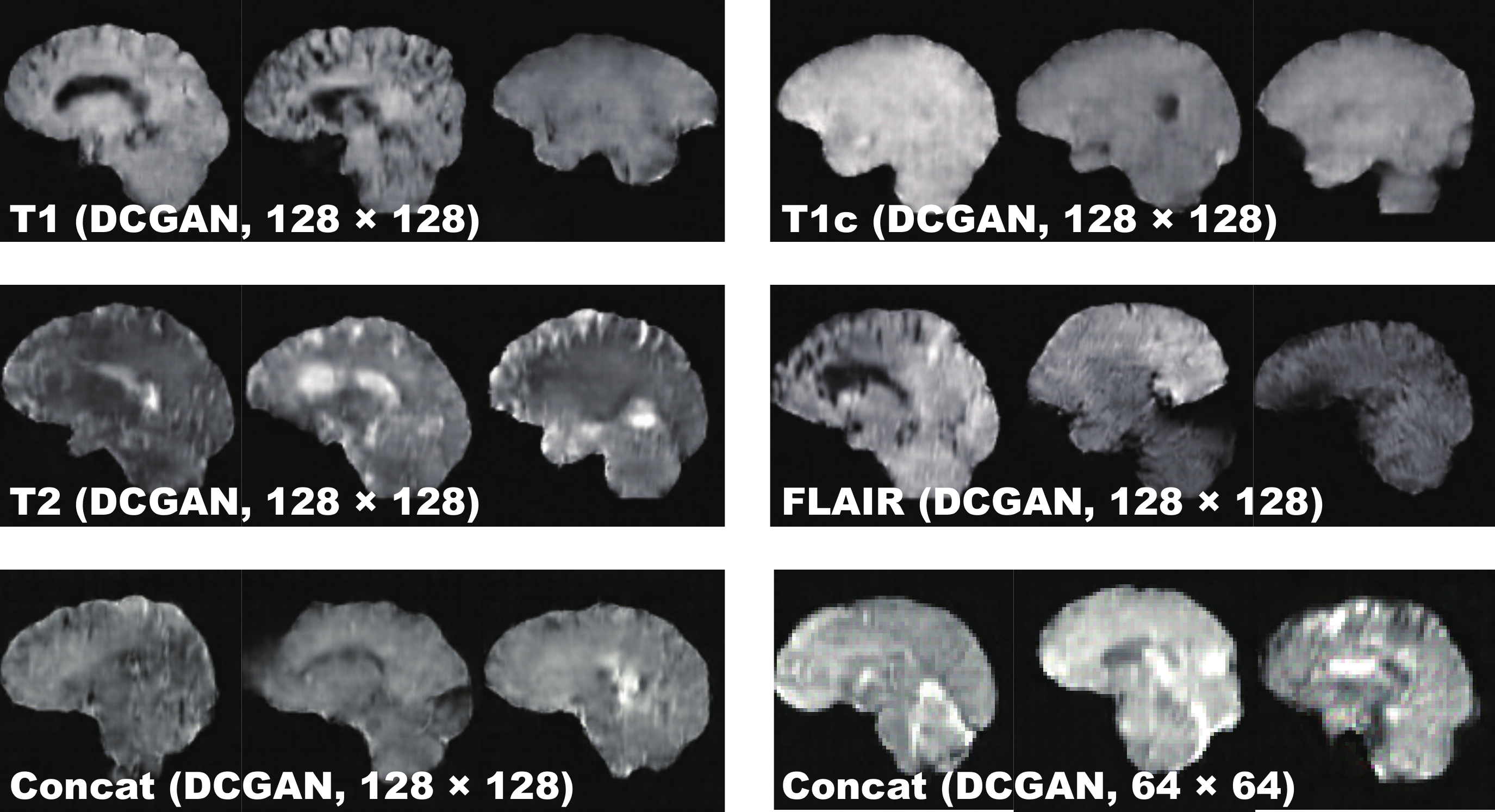}}
\caption{Example synthetic $128\times128$/$64\times64$ DCGAN-generated MR images.}
\label{fig4_3}
\end{figure}

\noindent \textbf{DCGAN} Fig.~\ref{fig4_3} illustrates examples of synthetic images by DCGAN. The images look similar to the real samples.
Concat images combine appearances and patterns from all the four sequences used in training. Since DCGAN's value function could be unstable, it often generates hyper-intense T1-like images analogous to mode collapse for 64 $\times$ 64 Concat images, while sharing the same hyper-parameters with 128 $\times$ 128.

\begin{figure}[t!]
  \centering
  \centerline{\includegraphics[width=1\linewidth]{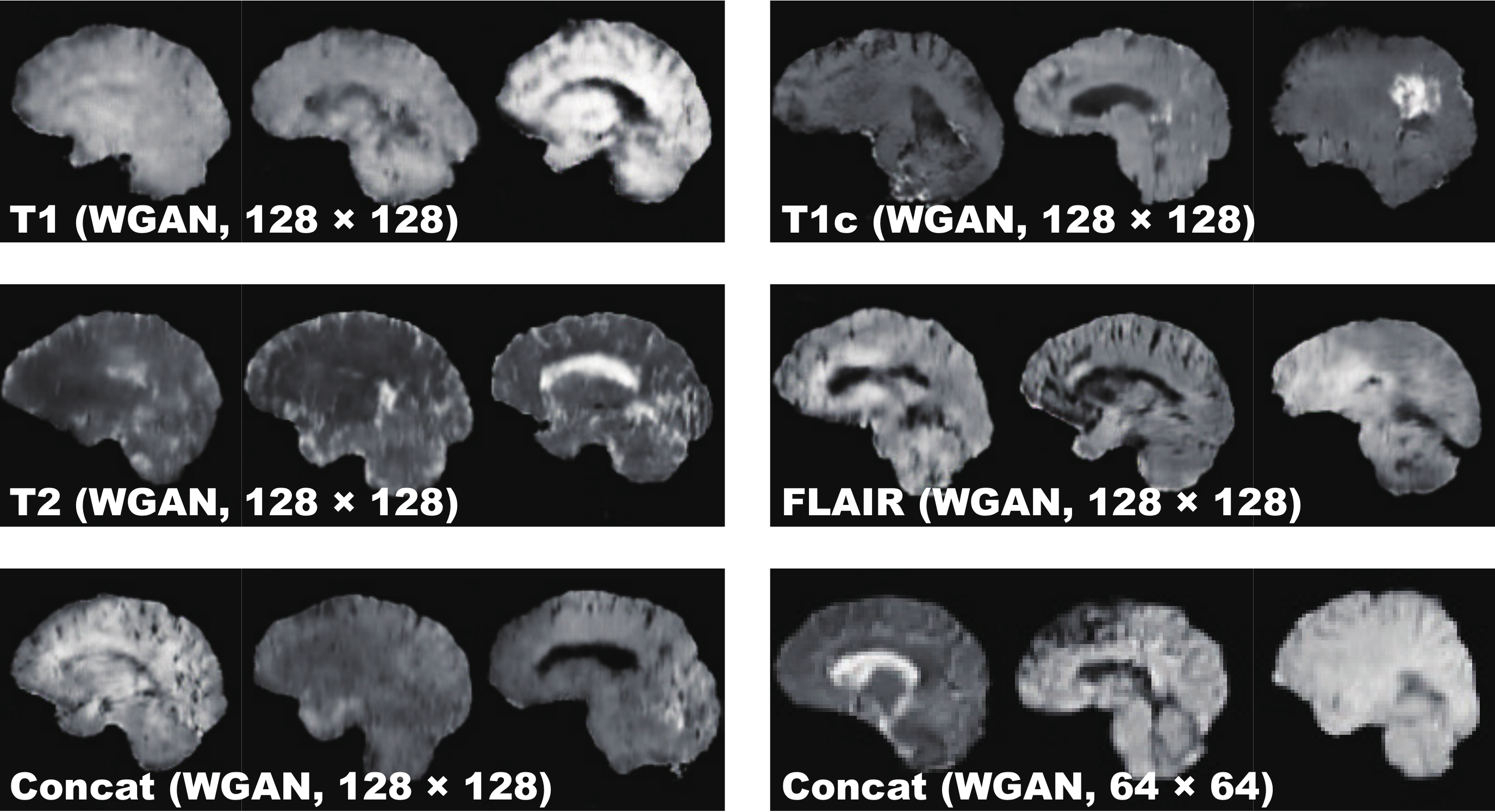}}
\caption{Example synthetic $128\times128$/$64\times64$ WGAN-generated MR images.}
\label{fig4_4}
\end{figure}

\noindent \textbf{WGAN} Fig.~\ref{fig4_4} shows the example output of WGAN in each sequence. Remarkably outperforming DCGAN, WGAN successfully captures the sequence-specific texture and tumor appearance while maintaining the realism of the original brain MR images. As expected, $128 \times 128$ Concat images tend to have more messy and unrealistic artifacts than $64 \times 64$ Concat ones, especially around boundaries of the brain, due to the introduction of unexpected intensity patterns.

\subsection{Visual Turing Test Results}
Table~\ref{tab4_1} shows the confusion matrix concerning the Visual Turing Test. Even the expert physician found classifying real and synthetic images challenging, especially in lower resolution due to their less detailed appearances unfamiliar in clinical routine, even for highly hyper-intense $64 \times 64$ Concat images by DCGAN; distinguishing Concat images was easier compared to the case of T1, T1c, T2, and FLAIR images because the physician often felt odd from the artificial sequence. WGAN succeeded to deceive the physician significantly better than DCGAN for all the MRI sequences except FLAIR images ($62\%$ to $54\%$).

\begin{table}[!t]
\caption[Visual Turing Test results by a physician for classifying $50$ real \textit{vs} $50$ synthetic images.]{Visual Turing Test results by a physician for classifying $50$ real \textit{vs} $50$ synthetic images. Accuracy denotes the physician’s successful classification ratio between the real/synthetic images and between the tumor/non-tumor images, respectively. It should be noted that proximity to 50\% of accuracy indicates superior performance (chance = 50\%).}
\label{tab4_1}
\centering
\scalebox{0.62}{
\begin{tabular}{lccccc}
\Hline\noalign{\smallskip}
\bfseries  & \bfseries Accuracy \hspace{-0.1in}  (\%) & \bfseries Real as Real \hspace{-0.1in}  (\%) & \bfseries Real as Synt \hspace{-0.1in}  (\%) & \bfseries Synt as Real \hspace{-0.1in}  (\%) & \bfseries Synt as Synt \hspace{-0.1in}  (\%) \\\noalign{\smallskip}\hline\noalign{\smallskip}
T1 (DCGAN, $128 \times 128$) & 70 & 52 & 48 & 12 & 88\\
T1c (DCGAN, $128 \times 128$) & 71 & 48 & 52 & 6 & 94\\
T2 (DCGAN, $128 \times 128$) & 64 & 44 & 56 & 16 & 84\\
FLAIR (DCGAN, $128 \times 128$) & 54 & 24 & 76 & 16 & 84\\
Concat (DCGAN, $128 \times 128$) & 77 & 68 & 32 & 14 & 86\\
Concat (DCGAN, $64 \times 64$) & 54 & 26& 74& 18&82\\
\noalign{\smallskip}\hline\noalign{\smallskip}
T1 (WGAN, $128 \times 128$) & 64 & 40 & 60 & 12 & 88\\
T1c (WGAN, $128 \times 128$) & 55 & 26 & 74 & 16 & 84\\
T2 (WGAN, $128 \times 128$) & 58 & 38 & 62 & 22 & 78\\
FLAIR (WGAN, $128 \times 128$) & 62 & 32 & 68 & 8 & 92\\
Concat (WGAN, $128 \times 128$) & 66 & 62 & 38 & 30 & 70\\
Concat (WGAN, $64 \times 64$) & 53 & 36 & 64 & 30 & 70\\
\noalign{\smallskip}\Hline\noalign{\smallskip}
\end{tabular}
}
\end{table}

\section{Conclusion}
Our preliminary results show that GANs, especially WGAN, can generate $128 \times 128$ realistic multi-sequence brain MR images that even an expert physician is unable to accurately distinguish from the real, leading to valuable clinical applications, such as DA and physician training. This attributes to WGAN's good generalization ability with a sharp value function. In this context, DCGAN might be unsuitable due to both inferior realism and mode collapse in terms of intensity. We only use slices of interest in training to obtain desired MR images and generate both original/Concat sequence images for DA in medical imaging.

This study confirms the synthetic image quality by the human expert evaluation, but a more objective computational evaluation for GANs should also follow, such as Classifier Two-Sample Tests (C2ST)~\cite{Lopez}, which assesses whether two samples are drawn from the same distribution. Currently this work uses sagittal MR images alone, so we plan to generate coronal and transverse images. As this research uniformly selects middle slices in pre-processing, better data generation demands developing a classifier to only select brain MRI slices with/without tumors.

Towards DA, whereas realistic images give more insights on geometry/intensity transformations in classification, more realistic images do not always assure better DA, so we have to find suitable image resolutions and sequences; that is why we generate both high-resolution images and Concat images, yet they looked more unrealistic for the physician. For physician training, generating desired realistic tumors by adding conditioning requires exploring latent spaces of GANs extensively.

Overall, our novel GAN-based realistic brain MR image generation approach sheds light on diagnostic and prognostic medical applications; future studies on these applications are needed to confirm our encouraging results.
\chapter{\LARGE GAN-based Medical Image Augmentation for 2D Classification}

\section{Prologue to Second Project}
\subsection{Project Publications}
\begin{itemize}
\item \textbf{Infinite Brain MR Images: PGGAN-based Data Augmentation for Tumor Detection}. \textbf{C. Han}, L. Rundo, R. Araki, Y. Furukawa, G. Mauri, H. Nakayama, H. Hayashi, In A. Esposito, M. Faundez-Zanuy, F. C. Morabito, E. Pasero (eds.) Neural Approaches to Dynamics of Signal Exchanges, Springer, pp. 291--303, September 2019.
\item \textbf{Combining Noise-to-Image and Image-to-Image GANs: Brain MR Image Augmentation for Tumor Detection}. \textbf{C. Han}, L. Rundo, R. Araki, Y. Nagano, Y. Furukawa, G. Mauri, H. Nakayama, H. Hayashi, \textit{IEEE Access}, pp. 156966--156977, October 2019.
\end{itemize}

\subsection{Context}
At the time we wrote the former paper, high-resolution (e.g.,  256 $\times$ 256) medical image generation using noise-to-image GANs had been challenging~\cite{Madani} while most CNN architectures adopt around 256 $\times$ 256 input sizes (e.g., InceptionResNetV2~\cite{Szegedy}: $299\times299$, ResNet-50~\cite{He}: $224\times224$). Moreover, prior to the latter paper, analysis had been immature on GAN-generated additional training images for better CNN-based classification.

\subsection{Contributions}
This project's core contribution is to firstly combine noise-to-image and image-to-image GANs for improved 2D classification. The former paper adopts a noise-to-image GAN called PGGANs to generate realistic/diverse original-sized 256 $\times$ 256 whole brain MR images with/without tumors separately; additionally, the latter paper exploits an image-to-image GAN called MUNIT to further refine the synthetic images' texture and shape similarly to real ones. By so doing, our two-step GAN-based DA boosts sensitivity 93.7\% to 97.5\% in tumor/non-tumor classification. Moreover, we firstly analyze how medical GAN-based DA is associated with pre-training on ImageNet and discarding weird-looking synthetic images to humans to achieve high sensitivity. A physician classifies a few synthetic images as real in Visual Turing Test despite the high resolution.

\subsection{Recent Developments}
Since the former paper's acceptance (the book chapter's publication process took more than a year), to improve 2D classification, Konidaris \textit{et al.} generated $192 \times 160$ brain MR images with Alzheimer's disease using the noise-to-image GAN~\cite{konidaris2018generative}. No more recent developments to report exist for the latter paper because it is very recent.

\newpage

\section{Motivation}
CNNs are playing a key role in Medical Image Analysis, updating the state-of-the-art in many tasks~\cite{Havaei, Rundo2, Ker} when large-scale annotated training data are available. However, preparing such massive medical data is demanding; thus, for better diagnosis, researchers generally adopt classic DA techniques, such as geometric/intensity transformations of original images~\cite{Ronneberger,Milletari}. Those augmented images, however, intrinsically have a similar distribution to the original ones, resulting in limited performance improvement. In this sense, GAN-based DA can considerably increase the performance~\cite{Goodfellow}; since the generated images are realistic but completely novel samples, they can relieve the sampling biases and fill the real image distribution uncovered by the original dataset~\textcolor{black}{\cite{Yi}}.

The main problem in CAD lies in small/fragmented medical imaging datasets from \textcolor{black}{multiple} scanners; thus, researchers have improved classification by augmenting images with noise-to-image GANs~\cite{Han1} or image-to-image GANs~\cite{Wu}. However, no research has \textcolor{black}{achieved further performance boost} by combining noise-to-image and image-to-image GANs.

\begin{figure}[t!]
  \centering
  \centerline{\includegraphics[width=1\linewidth]{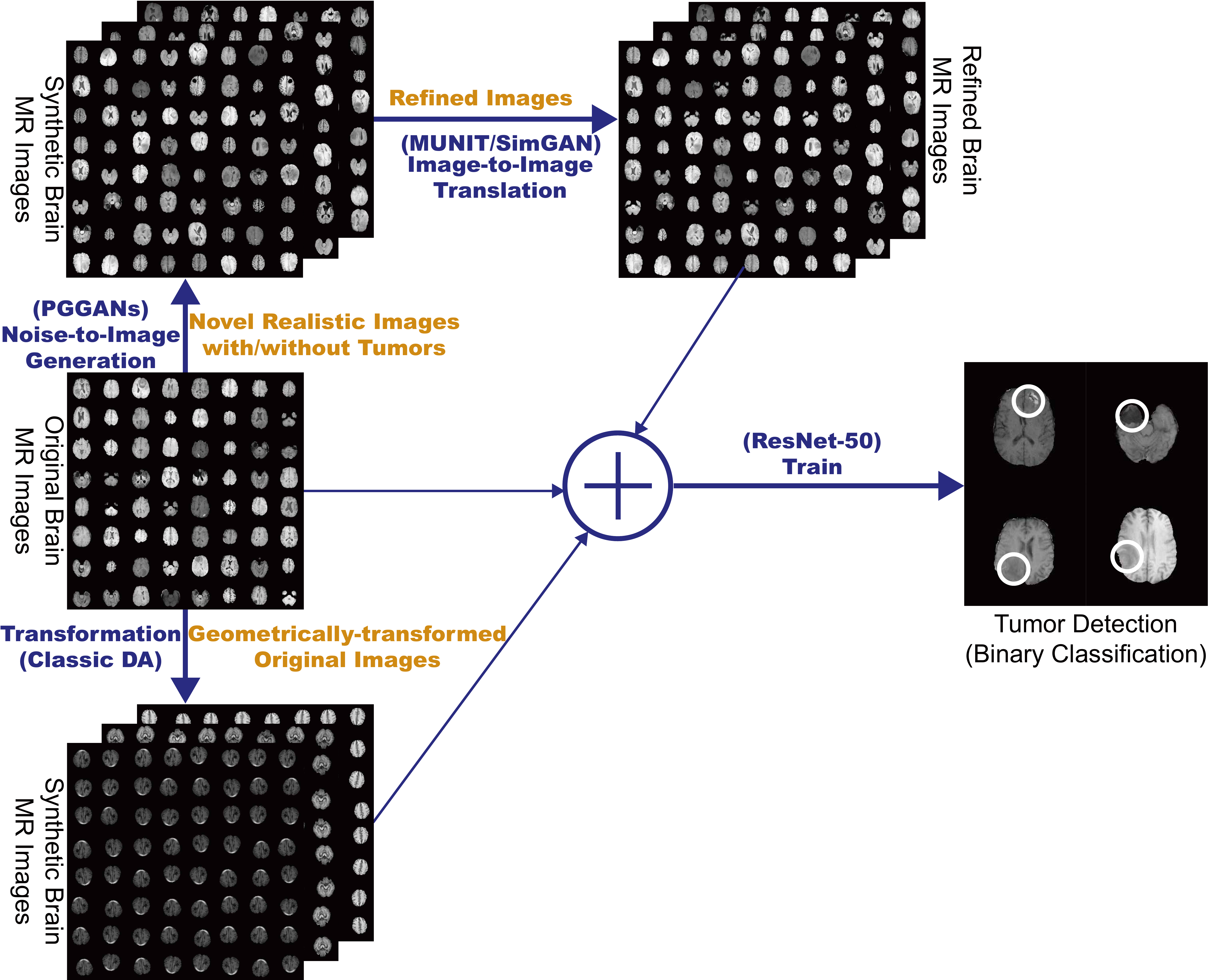}}
\caption[Combining noise-to-image and image-to-image GANs for better tumor classification.]{Combining noise-to-image and image-to-image GANs for better tumor classification: the PGGANs generates a number
of realistic brain tumor/non-tumor MR images separately, the \textcolor{black}{MUNIT}/SimGAN refines them separately, and the binary classifier uses them as additional training data.}
\label{fig5_1}
\end{figure}

So, how can we maximize the DA effect under limited training images using the GAN combinations? To generate and refine brain \textcolor{black}{MR} images with/without tumors separately (Fig.~\ref{fig5_1}), we propose a two-step GAN-based DA approach: (\textit{i}) PGGANs~\cite{Karras}, low-to-high resolution noise-to-image GAN, first generates realistic/diverse $256\times256$ images---the PGGANs \textcolor{black}{helps DA} since most CNN architectures adopt around $256\times256$ input sizes (e.g., InceptionResNetV2~\cite{Szegedy}: $299\times299$, ResNet-50~\cite{He}: $224\times224$); (\textit{ii}) \textcolor{black}{MUNIT~\cite{Huang} that combines GANs/VAEs~\cite{Kingma2013} or SimGAN~\cite{Shrivastava} that uses a DA-focused GAN loss}, further refines the texture and shape of the PGGAN-generated images to fit them into the real image distribution. \textcolor{black}{Since training a single sophisticated GAN system is already difficult, instead of end-to-end training, we adopt a two-step approach for performance boost \textit{via} an ensemble generation process from those state-of-the-art GANs' different algorithms.}

\newpage

We thoroughly investigate CNN-based tumor classification results, also considering the influence of pre-training on ImageNet~\cite{Russakovsky} and discarding weird-looking GAN-generated images. Moreover, we evaluate the synthetic images' \textcolor{black}{appearance} \textit{via} Visual Turing Test~\cite{Salimans} by an expert physician, and visualize the data distribution of real/synthetic images \textit{via} t-Distributed Stochastic Neighbor Embedding (t-SNE)~\cite{Maaten}. When combined with classic DA, our two-step GAN-based DA approach significantly outperforms the classic DA alone, boosting sensitivity \textcolor{black}{$93.67\%$ to $97.48\%$}.

\noindent \textbf{Research Questions.} We mainly address two questions:
\begin{itemize}
\item \textbf{GAN Selection:} Which GAN architectures are well-suited for realistic/diverse medical image generation?

\item \textbf{Medical DA:} How to use GAN-generated images as additional training data for better CNN-based diagnosis?

\end{itemize}

\noindent \textbf{Contributions.} Our main contributions are as follows:
\begin{itemize}
\item \textbf{Whole Image Generation:} This research shows that PGGANs can generate realistic/diverse $256 \times 256$ whole medical images\textcolor{black}{---not only small pathological sub-areas---and \textcolor{black}{MUNIT} can further refine their texture and shape similarly to real ones}.

\item \textbf{Two-step GAN-based DA:} This novel two-step approach, combining for the first time noise-to-image and image-to-image GANs, \textcolor{black}{significantly} boosts tumor classification \textcolor{black}{sensitivity}.

\item \textbf{Misdiagnosis Prevention:} This study firstly analyzes how medical GAN-based DA is associated with pre-training on ImageNet and discarding weird-looking synthetic images to achieve high sensitivity with small and fragmented datasets.

\end{itemize}


\section{Materials and Methods}
\subsection{BRATS 2016 Dataset}
We use a dataset of $240 \times 240$ T1c brain axial MR images of $220$ HGG cases from BRATS 2016~\cite{Menze}. T1c is the most common sequence in tumor classification thanks to its high-contrast~\cite{Koley}.

\begin{figure}[t!]
  \centering
  \centerline{\includegraphics[width=1\linewidth]{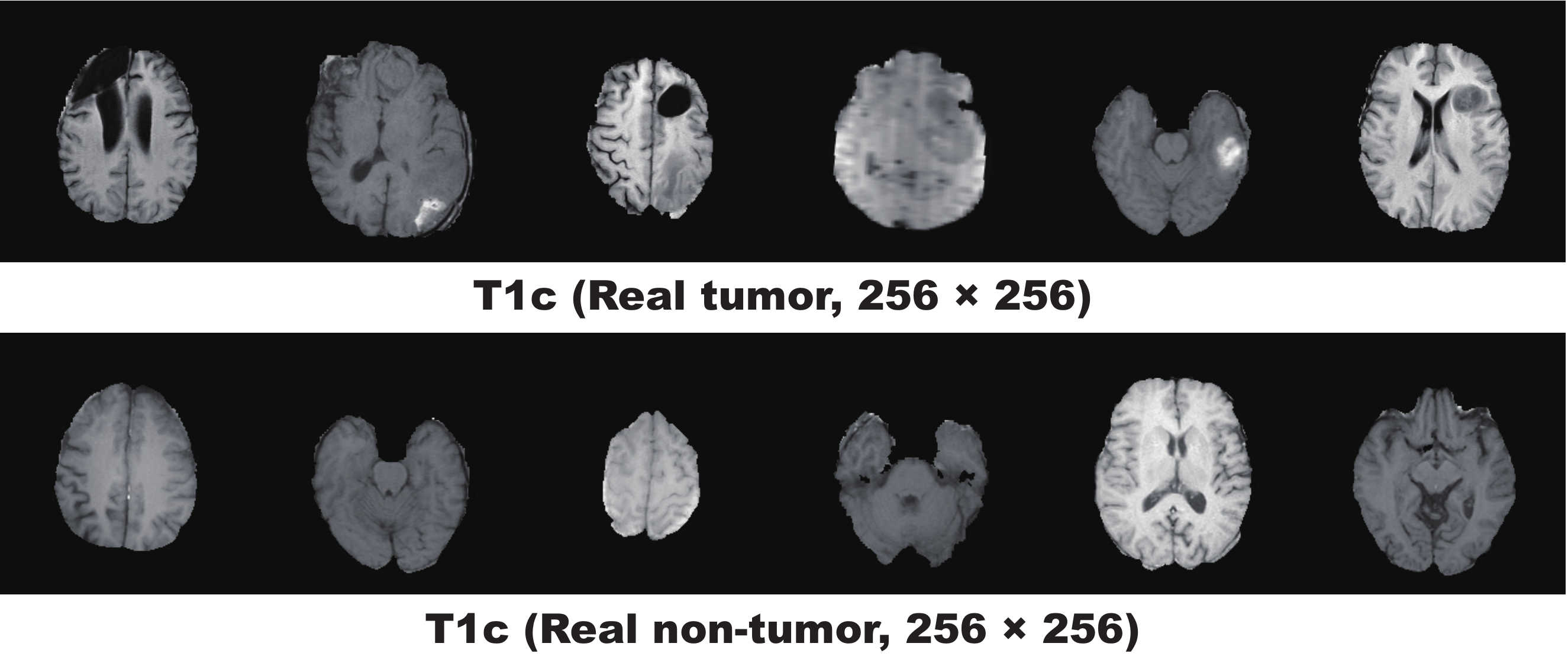}}
\caption{Example real $256\times256$ MR images used for PGGAN training.}
\label{fig5_2}
\end{figure}
\subsection{PGGAN-based Image Generation}
\noindent \textbf{Pre-processing}
For better GAN/ResNet-50 training, we select the slices from $\#30$ to $\#130$ among the whole $155$ slices to omit initial/final slices, which convey negligible useful information; also, since tumor/non-tumor annotation in the BRATS 2016 dataset, based on 3D volumes, is highly incorrect/ambiguous on 2D slices, we exclude ($i$) tumor images tagged as non-tumor, ($ii$) non-tumor images tagged as tumor, ($iii$) borderline images with unclear tumor/non-tumor appearance, and ($iv$) images with missing brain parts due to the skull-stripping procedure. For tumor classification, we divide the whole dataset ($220$ patients) into:


\begin{itemize}
\item Training set \\($154$ patients/$4,679$ tumor/$3,750$ non-tumor images);
\item Validation set \\($44$ patients/$750$ tumor/$608$ non-tumor images);
\item Test set \\($22$ patients/$1,232$ tumor/$1,013$ non-tumor images).
\end{itemize}

During the GAN training, we only use the training set to be fair; for better \textcolor{black}{PGGAN} training, the training set images are zero-padded to reach a power of $2$: $256 \times 256$ pixels from $240 \times 240$. Fig.~\ref{fig5_2} shows example real MR images.

\begin{figure}[t!]
  \centering
  \centerline{\includegraphics[width=1\linewidth]{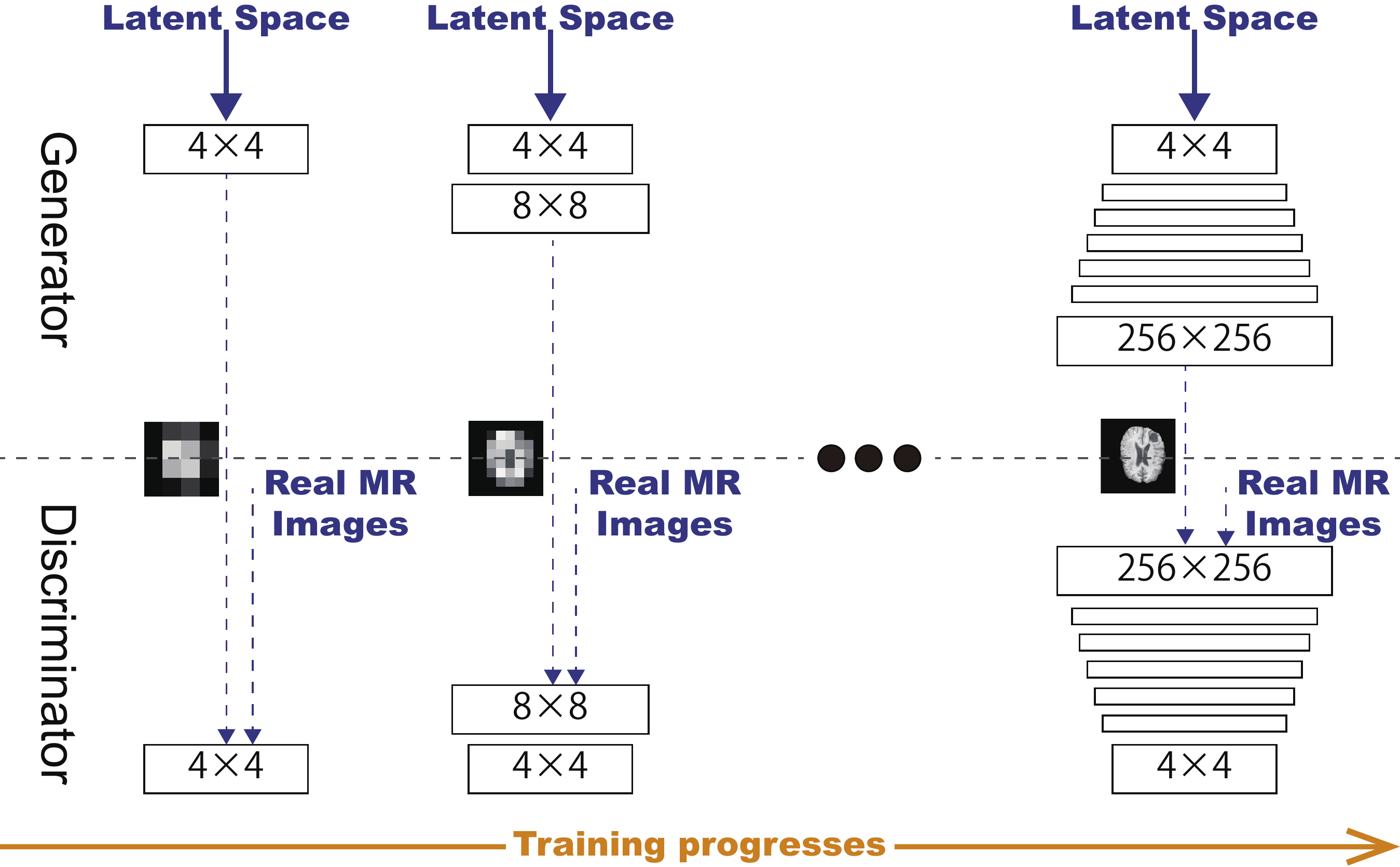}}
\caption[PGGAN architecture for $256 \times 256$ brain MR image generation.]{PGGAN architecture for $256 \times 256$ brain MR image generation. \textcolor{black}{$N \times N$ refers to convolutional layers operating on $N \times N$ spatial resolution.}}
\label{fig5_3}
\end{figure}

\noindent \textbf{PGGANs}~\cite{Karras} is a GAN training method that progressively grows a generator and discriminator: starting from low resolution, new layers model details as training progresses. This study adopts the PGGANs to synthesize realistic/diverse $256 \times 256$ brain MR images (Fig.~\ref{fig5_3}); we train and generate tumor/non-tumor images separately.

\noindent \textbf{PGGAN Implementation Details} The PGGAN architecture adopts the Wasserstein loss with Gradient Penalty (WGAN-GP)~\cite{Gulrajani}:
\begin{eqnarray}\label{eq:wgan_gp}
\underset{{\tilde{\mathbf{y}}\sim{\mathbb{P}_g}}}{\mathbb{E}}[D(\tilde{\mathbf{y}})]-\underset{{\mathbf{y}\sim{\mathbb{P}_r}}}{\mathbb{E}}[D(\mathbf{y})] +
\lambda_\text{gp}\underset{{{\hat{\mathbf{y}}}\sim{\mathbb{P}_{\hat{\mathbf{y}}}}}} {\mathbb{E}}[(\left \| \nabla_{\hat{\mathbf{y}}}{D({\hat{\mathbf{y}}})} \right \|_2-1)^2],
\end{eqnarray}
\textcolor{black}{where $\mathbb{E}[\cdot]$ denotes the expected value, the discriminator $D \in \mathcal{D}$ (i.e., the set of $1$-Lipschitz functions)}, $\mathbb{P}_r$ is the data distribution defined by the true data sample $\mathbf{y}$, and $\mathbb{P}_g$ is the model distribution defined by the generated sample \textcolor{black}{${\tilde{\mathbf{y}} = G(\mathbf{z})}$ ($\mathbf{z} \sim p(\mathbf{z})$ is the input noise $\mathbf{z}$ to the generator sampled from a Gaussian distribution)}. A gradient penalty is added for the random sample ${\hat{\mathbf{y}}}\sim{\mathbb{P}_{\hat{\mathbf{y}}}}$, \textcolor{black}{where $\nabla_{\hat{\mathbf{y}}}$ is the gradient operator towards the generated samples} and $\lambda_\text{gp}$ is the gradient penalty coefficient.


We train the model \textcolor{black}{(Table~\ref{tab5_1})} for $100$ epochs with a batch size of $16$ and $1.0 \times 10^{-3}$ learning rate for the Adam optimizer \textcolor{black}{(the exponential decay rates $\beta_{1} = 0, \beta_{2} = 0.99$)}~\cite{Kingma2015}. \textcolor{black}{All experiments use $\lambda_\text{gp} = 10$ with $1$ critic iteration per generator iteration.} During training, we apply random cropping in $0$-$15$ pixels \textcolor{black}{as DA.}

\begin{table*}[t!]
\caption[PGGAN architecture details for the generator/discriminator.]{PGGAN architecture details for the generator/discriminator. Pixelwise feature vector normalization~\cite{Krizhevsky} is applied in the generator after each convolutional layer except for the final output layer as in the original paper~\cite{Karras}. LReLU denotes Leaky ReLU with leakiness $0.2$.}
\small
  \centering
\scalebox{0.84}{
\begin{tabular}{lcc}
\Hline
\textbf{\textcolor{black}{Generator}} & \textbf{\textcolor{black}{Activation}} & \textbf{\textcolor{black}{Output Shape}}\\
\hline
\textcolor{black}{Latent vector} & \textcolor{black}{--} & \textcolor{black}{$\makebox[\widthof{512}][c]{512}\times\makebox[\widthof{1024}][c]{1}\times\makebox[\widthof{1024}][c]{1}$} \\
\textcolor{black}{Conv $4\times4$} & \textcolor{black}{LReLU} & \textcolor{black}{$\makebox[\widthof{512}][c]{512}\times\makebox[\widthof{1024}][c]{4}\times\makebox[\widthof{1024}][c]{4}$} \\
\textcolor{black}{Conv $3\times3$} &  \textcolor{black}{LReLU} & \textcolor{black}{$\makebox[\widthof{512}][c]{512}\times\makebox[\widthof{1024}][c]{4}\times\makebox[\widthof{1024}][c]{4}$} \\
\hline
\textcolor{black}{Upsample} & \textcolor{black}{--} & \textcolor{black}{$\makebox[\widthof{512}][c]{512}\times\makebox[\widthof{1024}][c]{8}\times\makebox[\widthof{1024}][c]{8}$} \\
\textcolor{black}{Conv $3\times3$} & \textcolor{black}{LReLU} & \textcolor{black}{$\makebox[\widthof{512}][c]{512}\times\makebox[\widthof{1024}][c]{8}\times\makebox[\widthof{1024}][c]{8}$} \\
\textcolor{black}{Conv $3\times3$} & \textcolor{black}{LReLU} & \textcolor{black}{$\makebox[\widthof{512}][c]{512}\times\makebox[\widthof{1024}][c]{8}\times\makebox[\widthof{1024}][c]{8}$} \\
\hline
\textcolor{black}{Upsample} & \textcolor{black}{--} & \textcolor{black}{$\makebox[\widthof{512}][c]{512}\times\makebox[\widthof{1024}][c]{16}\times\makebox[\widthof{1024}][c]{16}$}  \\
\textcolor{black}{Conv $3\times3$} & \textcolor{black}{LReLU} & \textcolor{black}{$\makebox[\widthof{512}][c]{256}\times\makebox[\widthof{1024}][c]{16}\times\makebox[\widthof{1024}][c]{16}$} \\
\textcolor{black}{Conv $3\times3$} & \textcolor{black}{LReLU} & \textcolor{black}{$\makebox[\widthof{512}][c]{256}\times\makebox[\widthof{1024}][c]{16}\times\makebox[\widthof{1024}][c]{16}$} \\
\hline
\textcolor{black}{Upsample} & \textcolor{black}{--} & \textcolor{black}{$\makebox[\widthof{512}][c]{256}\times\makebox[\widthof{1024}][c]{32}\times\makebox[\widthof{1024}][c]{32}$}  \\
\textcolor{black}{Conv $3\times3$} & \textcolor{black}{LReLU} & \textcolor{black}{$\makebox[\widthof{512}][c]{128}\times\makebox[\widthof{1024}][c]{32}\times\makebox[\widthof{1024}][c]{32}$} \\
\textcolor{black}{Conv $3\times3$} & \textcolor{black}{LReLU} & \textcolor{black}{$\makebox[\widthof{512}][c]{128}\times\makebox[\widthof{1024}][c]{32}\times\makebox[\widthof{1024}][c]{32}$} \\
\hline
\textcolor{black}{Upsample} & \textcolor{black}{--} & \textcolor{black}{$\makebox[\widthof{512}][c]{128}\times\makebox[\widthof{1024}][c]{64}\times\makebox[\widthof{1024}][c]{64}$} \\ 
\textcolor{black}{Conv $3\times3$} & \textcolor{black}{LReLU} & \textcolor{black}{$\makebox[\widthof{512}][c]{64}\times\makebox[\widthof{1024}][c]{64}\times\makebox[\widthof{1024}][c]{64}$} \\
\textcolor{black}{Conv $3\times3$} & \textcolor{black}{LReLU} & \textcolor{black}{$\makebox[\widthof{512}][c]{64}\times\makebox[\widthof{1024}][c]{64}\times\makebox[\widthof{1024}][c]{64}$} \\
\hline
\textcolor{black}{Upsample} & \textcolor{black}{--} & \textcolor{black}{$\makebox[\widthof{512}][c]{64}\times\makebox[\widthof{1024}][c]{128}\times\makebox[\widthof{1024}][c]{128}$} \\
\textcolor{black}{Conv $3\times3$} & \textcolor{black}{LReLU} & \textcolor{black}{$\makebox[\widthof{512}][c]{32}\times\makebox[\widthof{1024}][c]{128}\times\makebox[\widthof{1024}][c]{128}$}  \\
\textcolor{black}{Conv $3\times3$} & \textcolor{black}{LReLU} & \textcolor{black}{$\makebox[\widthof{512}][c]{32}\times\makebox[\widthof{1024}][c]{128}\times\makebox[\widthof{1024}][c]{128}$}  \\
\hline
\textcolor{black}{Upsample} & \textcolor{black}{--} & \textcolor{black}{$\makebox[\widthof{512}][c]{32}\times\makebox[\widthof{1024}][c]{256}\times\makebox[\widthof{1024}][c]{256}$} \\
\textcolor{black}{Conv $3\times3$} & \textcolor{black}{LReLU} & \textcolor{black}{$\makebox[\widthof{512}][c]{16}\times\makebox[\widthof{1024}][c]{256}\times\makebox[\widthof{1024}][c]{256}$}  \\
\textcolor{black}{Conv $3\times3$} & \textcolor{black}{LReLU} & \textcolor{black}{$\makebox[\widthof{512}][c]{16}\times\makebox[\widthof{1024}][c]{256}\times\makebox[\widthof{1024}][c]{256}$}  \\
\hline
\textcolor{black}{Conv $1\times1$} & \textcolor{black}{Linear} & \textcolor{black}{$\makebox[\widthof{512}][c]{1}\times\makebox[\widthof{1024}][c]{256}\times\makebox[\widthof{1024}][c]{256}$}  \\
\Hline
\end{tabular}
\begin{tabular}{lcc}
\Hline
\textcolor{black}{\textbf{Discriminator}} & \textcolor{black}{\textbf{Activation}} & \textcolor{black}{\textbf{Output Shape}}\\
\hline
\textcolor{black}{Input image} & \textcolor{black}{--} & \textcolor{black}{$\makebox[\widthof{512}][c]{1}\times\makebox[\widthof{1024}][c]{256}\times\makebox[\widthof{1024}][c]{256}$} \\
\textcolor{black}{Conv $1\times1$} & \textcolor{black}{LReLU} & \textcolor{black}{$\makebox[\widthof{512}][c]{16}\times\makebox[\widthof{1024}][c]{256}\times\makebox[\widthof{1024}][c]{256}$} \\
\textcolor{black}{Conv $3\times3$} &  \textcolor{black}{LReLU} & \textcolor{black}{$\makebox[\widthof{512}][c]{16}\times\makebox[\widthof{1024}][c]{256}\times\makebox[\widthof{1024}][c]{256}$} \\
\textcolor{black}{Conv $3\times3$} &  \textcolor{black}{LReLU} & \textcolor{black}{$\makebox[\widthof{512}][c]{32}\times\makebox[\widthof{1024}][c]{256}\times\makebox[\widthof{1024}][c]{256}$} \\
\textcolor{black}{Downsample} & \textcolor{black}{--} & \textcolor{black}{$\makebox[\widthof{512}][c]{32}\times\makebox[\widthof{1024}][c]{128}\times\makebox[\widthof{1024}][c]{128}$} \\
\hline
\textcolor{black}{Conv $3\times3$} & \textcolor{black}{LReLU} & \textcolor{black}{$\makebox[\widthof{512}][c]{32}\times\makebox[\widthof{1024}][c]{128}\times\makebox[\widthof{1024}][c]{128}$} \\
\textcolor{black}{Conv $3\times3$} & \textcolor{black}{LReLU}& \textcolor{black}{$\makebox[\widthof{512}][c]{64}\times\makebox[\widthof{1024}][c]{128}\times\makebox[\widthof{1024}][c]{128}$} \\
\textcolor{black}{Downsample} & \textcolor{black}{--} & \textcolor{black}{$\makebox[\widthof{512}][c]{64}\times\makebox[\widthof{1024}][c]{64}\times\makebox[\widthof{1024}][c]{64}$}  \\
\hline
\textcolor{black}{Conv $3\times3$} & \textcolor{black}{LReLU} & \textcolor{black}{$\makebox[\widthof{512}][c]{64}\times\makebox[\widthof{1024}][c]{64}\times\makebox[\widthof{1024}][c]{64}$} \\
\textcolor{black}{Conv $3\times3$} & \textcolor{black}{LReLU} & \textcolor{black}{$\makebox[\widthof{512}][c]{128}\times\makebox[\widthof{1024}][c]{64}\times\makebox[\widthof{1024}][c]{64}$} \\
\textcolor{black}{Downsample} & \textcolor{black}{--} & \textcolor{black}{$\makebox[\widthof{512}][c]{128}\times\makebox[\widthof{1024}][c]{32}\times\makebox[\widthof{1024}][c]{32}$}  \\
\hline
\textcolor{black}{Conv $3\times3$} & \textcolor{black}{LReLU} & \textcolor{black}{$\makebox[\widthof{512}][c]{128}\times\makebox[\widthof{1024}][c]{32}\times\makebox[\widthof{1024}][c]{32}$} \\
\textcolor{black}{Conv $3\times3$} & \textcolor{black}{LReLU} & \textcolor{black}{$\makebox[\widthof{512}][c]{256}\times\makebox[\widthof{1024}][c]{32}\times\makebox[\widthof{1024}][c]{32}$} \\
\textcolor{black}{Downsample} & \textcolor{black}{--} & \textcolor{black}{$\makebox[\widthof{512}][c]{256}\times\makebox[\widthof{1024}][c]{16}\times\makebox[\widthof{1024}][c]{16}$} \\ 
\hline
\textcolor{black}{Conv $3\times3$} & \textcolor{black}{LReLU} & \textcolor{black}{$\makebox[\widthof{512}][c]{256}\times\makebox[\widthof{1024}][c]{16}\times\makebox[\widthof{1024}][c]{16}$} \\
\textcolor{black}{Conv $3\times3$} & \textcolor{black}{LReLU} & \textcolor{black}{$\makebox[\widthof{512}][c]{512}\times\makebox[\widthof{1024}][c]{16}\times\makebox[\widthof{1024}][c]{16}$} \\
\textcolor{black}{Downsample} & \textcolor{black}{--} & \textcolor{black}{$\makebox[\widthof{512}][c]{512}\times\makebox[\widthof{1024}][c]{8}\times\makebox[\widthof{1024}][c]{8}$} \\
\hline
\textcolor{black}{Conv $3\times3$} & \textcolor{black}{LReLU} & \textcolor{black}{$\makebox[\widthof{512}][c]{512}\times\makebox[\widthof{1024}][c]{8}\times\makebox[\widthof{1024}][c]{8}$}  \\
\textcolor{black}{Conv $3\times3$} & \textcolor{black}{LReLU} & \textcolor{black}{$\makebox[\widthof{512}][c]{512}\times\makebox[\widthof{1024}][c]{8}\times\makebox[\widthof{1024}][c]{8}$}  \\
\textcolor{black}{Downsample} & \textcolor{black}{--} & \textcolor{black}{$\makebox[\widthof{512}][c]{512}\times\makebox[\widthof{1024}][c]{4}\times\makebox[\widthof{1024}][c]{4}$} \\
\hline
\textcolor{black}{Minibatch stddev} & \textcolor{black}{--} & \textcolor{black}{$\makebox[\widthof{512}][c]{513}\times\makebox[\widthof{1024}][c]{4}\times\makebox[\widthof{1024}][c]{4}$}  \\
\textcolor{black}{Conv $3\times3$} & \textcolor{black}{LReLU} & \textcolor{black}{$\makebox[\widthof{512}][c]{512}\times\makebox[\widthof{1024}][c]{4}\times\makebox[\widthof{1024}][c]{4}$} \\
\textcolor{black}{Conv $4\times4$} & \textcolor{black}{LReLU} & \textcolor{black}{$\makebox[\widthof{512}][c]{512}\times\makebox[\widthof{1024}][c]{1}\times\makebox[\widthof{1024}][c]{1}$} \\
\textcolor{black}{Fully-connected} & \textcolor{black}{Linear} & \textcolor{black}{$\makebox[\widthof{512}][c]{1}\times\makebox[\widthof{1024}][c]{1}\times\makebox[\widthof{1024}][c]{1}$}  \\
\Hline
\label{tab5_1}
\end{tabular}
}
\end{table*}

\subsection{MUNIT/SimGAN-based Image Refinement}
\noindent \textbf{Refinement}
\textcolor{black}{Using resized $224\times224$ images for ResNet-50,} we further refine the texture and shape of PGGAN-generated tumor/non-tumor images separately to fit them into the real image distribution using \textcolor{black}{MUNIT}~\cite{Huang} or SimGAN~\cite{Shrivastava}. SimGAN remarkably improved eye gaze estimation results after refining non-GAN-based synthetic images from the UnityEyes simulator $via$ image-to-image translation; thus, we also expect such performance improvement after refining synthetic images from a noise-to-image GAN (i.e., PGGANs) $via$ an image-to-image GAN (i.e., \textcolor{black}{MUNIT}/SimGAN) with considerably different GAN algorithms.

We randomly select $3,000$ real/$3,000$ PGGAN-generated tumor images for tumor image training, and we perform the same for non-tumor image training. To find suitable refining steps for each architecture, we pick the \textcolor{black}{MUNIT}/SimGAN models with the highest accuracy on tumor classification validation, when pre-trained and combined with classic DA, among $20,000$/$50,000$/$100,000$ steps, respectively.

\noindent \textbf{MUNIT}~\cite{Huang} \textcolor{black}{is an image-to-image GAN based on both auto-encoding/translation; it extends UNIT~\cite{Liu} to increase the generated images' realism/diversity \textit{via} a stochastic model representing continuous output distributions.}


\noindent \textbf{MUNIT Implementation Details}
The \textcolor{black}{MUNIT} architecture adopts the following loss:
\begin{eqnarray}\label{eq:unit_loss}
\min_{E_1,E_2,G_1,G_2} \max_{D_1,D_2} 
&&\mathcal{L}_{\text{\tiny VAE}_1} +\mathcal{L}_{\text{\tiny GAN}_1} + \mathcal{L}_{\text{\tiny CC}_1} + \mathcal{L}_{\text{\tiny VGG}_1}\nonumber\\
&+&\mathcal{L}_{\text{\tiny VAE}_2} + \mathcal{L}_{\text{\tiny GAN}_2} + \mathcal{L}_{\text{\tiny CC}_2} + \mathcal{L}_{\text{\tiny VGG}_2},
\end{eqnarray}
\textcolor{black}{where $\mathcal{L}(\cdot)$ denotes the loss function.}
Using the multiple encoders $E_1$/$E_2$, generators $G_1$/$G_2$, discriminators $D_1$/$D_2$, cycle-consistencies CC$_1$/CC$_2$\textcolor{black}{, and domain-invariant perceptions VGG$_1$/VGG$_2$~\cite{Simonyan}}, this framework jointly solves learning problems of the VAE$_1$/VAE$_2$ and GAN$_1$/GAN$_2$ for the image reconstruction streams, image translation streams, cycle-consistency reconstruction streams, \textcolor{black}{and domain-invariant perception streams. Since we do not need the style loss for our experiments, instead of the MUNIT loss, we use the UNIT loss with the perceptual loss for the MUNIT architecture (as in the UNIT authors' GitHub repository).}
The MUNIT architecture adopts the following loss:

\begin{eqnarray}\label{eq:munit_loss}
\min_{\epsilon,\tilde{\epsilon},\delta} \max_{\tilde{\delta},D,\tilde{D}} 
\mathcal{L}^{\mathbf{x}}_{\text{GAN}} + \mathcal{L}^{\tilde{\mathbf{x}}}_{\text{GAN}}
+ \lambda_{x}(\mathcal{L}^{\mathbf{x}}_{\text{recon}}+\mathcal{L}^{\tilde{\mathbf{x}}}_{\text{recon}})\nonumber\\
\vspace{0.1mm} + \lambda_{c}(\mathcal{L}^{\mathbf{c}}_{\text{recon}}+\mathcal{L}^{\tilde{\mathbf{c}}}_{\text{recon}}),
\end{eqnarray}



%

We train \textcolor{black}{the model (Table~\ref{tab5_2})} for $100,000$ steps with a batch size of 1 and $1.0 \times 10^{-4}$ learning rate for the Adam optimizer \textcolor{black}{($\beta_{1} = 0.5, \beta_{2} = 0.999$)}~\cite{Kingma2015}. The learning rate is reduced by half every $20,000$ steps. \textcolor{black}{We use the following MUNIT weights: the adversarial loss weight $= 1$; the image reconstruction loss weight $ = 10$; the Kullback-Leibler (KL) divergence loss weight for reconstruction $= 0.01$; the cycle consistency loss weight $ = 10$; the KL divergence loss weight for cycle consistency $= 0.01$; the domain-invariant perceptual loss weight $= 1$; \textcolor{black}{the Least Squares GAN objective function for the discriminators~\cite{mao2017least}}.} During training, we apply horizontal flipping as DA.

\begin{table*}[t!]
\caption[MUNIT architecture details for the generator/discriminator.]{MUNIT architecture details for the generator/discriminator. We input color images (i.e., 3 channels) to use ImageNet initialization. Instance normalization~\cite{Ulyanov}/adaptive instance normalization~\cite{Huang2} are applied in the content encoder/decoder after each convolutional layer respectively except for the final decoder output layer as in the original paper~\cite{Huang}. LReLU denotes Leaky ReLU with leakiness $0.2$.}
\small
  \centering
\scalebox{0.80}{
\begin{tabular}{lcc}
\Hline
\textbf{\textcolor{black}{Generator}} & \textbf{\textcolor{black}{Activation}} & \textbf{\textcolor{black}{Output Shape}}\\
\textbf{\textcolor{black}{Content Encoder}} &  & \\
\hline
\textcolor{black}{Input image} & \textcolor{black}{--} & \textcolor{black}{$\makebox[\widthof{512}][c]{3}\times\makebox[\widthof{1024}][c]{224}\times\makebox[\widthof{1024}][c]{224}$} \\
\textcolor{black}{Conv $7\times7$} & \textcolor{black}{ReLU} & \textcolor{black}{$\makebox[\widthof{512}][c]{64}\times\makebox[\widthof{1024}][c]{224}\times\makebox[\widthof{1024}][c]{224}$} \\
\textcolor{black}{Conv $4\times4$} & \textcolor{black}{ReLU} & \textcolor{black}{$\makebox[\widthof{512}][c]{128}\times\makebox[\widthof{1024}][c]{112}\times\makebox[\widthof{1024}][c]{112}$} \\
\textcolor{black}{Conv $4\times4$} & \textcolor{black}{ReLU} & \textcolor{black}{$\makebox[\widthof{512}][c]{256}\times\makebox[\widthof{1024}][c]{56}\times\makebox[\widthof{1024}][c]{56}$} \\
\hline
\textcolor{black}{\blockc{4}} & \textcolor{black}{ReLU} & \textcolor{black}{$\makebox[\widthof{512}][c]{256}\times\makebox[\widthof{1024}][c]{56}\times\makebox[\widthof{1024}][c]{56}$} \\
& \textcolor{black}{--} & \textcolor{black}{$\makebox[\widthof{512}][c]{256}\times\makebox[\widthof{1024}][c]{56}\times\makebox[\widthof{1024}][c]{56}$} \\
\hline
\textbf{\textcolor{black}{Decoder}} &  & \\
\hline
\textcolor{black}{\blockc{4}} & \textcolor{black}{ReLU} & \textcolor{black}{$\makebox[\widthof{512}][c]{256}\times\makebox[\widthof{1024}][c]{56}\times\makebox[\widthof{1024}][c]{56}$} \\
& \textcolor{black}{--} & \textcolor{black}{$\makebox[\widthof{512}][c]{256}\times\makebox[\widthof{1024}][c]{56}\times\makebox[\widthof{1024}][c]{56}$} \\
\hline

\textcolor{black}{Upsample} & \textcolor{black}{--} & \textcolor{black}{$\makebox[\widthof{512}][c]{256}\times\makebox[\widthof{1024}][c]{112}\times\makebox[\widthof{1024}][c]{112}$} \\
\textcolor{black}{Conv $5\times5$} & \textcolor{black}{ReLU} & \textcolor{black}{$\makebox[\widthof{512}][c]{128}\times\makebox[\widthof{1024}][c]{112}\times\makebox[\widthof{1024}][c]{112}$} \\
\hline

\textcolor{black}{Upsample} & \textcolor{black}{--} & \textcolor{black}{$\makebox[\widthof{512}][c]{128}\times\makebox[\widthof{1024}][c]{224}\times\makebox[\widthof{1024}][c]{224}$} \\
\textcolor{black}{Conv $5\times5$} & \textcolor{black}{ReLU} & \textcolor{black}{$\makebox[\widthof{512}][c]{64}\times\makebox[\widthof{1024}][c]{224}\times\makebox[\widthof{1024}][c]{224}$} \\
\hline
\textcolor{black}{Conv $7\times7$} & \textcolor{black}{Tanh} & \textcolor{black}{$\makebox[\widthof{512}][c]{3}\times\makebox[\widthof{1024}][c]{224}\times\makebox[\widthof{1024}][c]{224}$}  \\
\Hline
\end{tabular}
\begin{tabular}{lcc}
\Hline
\textcolor{black}{\textbf{Discriminator}} & \textcolor{black}{\textbf{Activation}} & \textcolor{black}{\textbf{Output Shape}}\\
\hline
\textcolor{black}{Input image} & \textcolor{black}{--} & \textcolor{black}{$\makebox[\widthof{512}][c]{3}\times\makebox[\widthof{1024}][c]{224}\times\makebox[\widthof{1024}][c]{224}$} \\
\textcolor{black}{Conv $4\times4$} & \textcolor{black}{LReLU} & \textcolor{black}{$\makebox[\widthof{512}][c]{64}\times\makebox[\widthof{1024}][c]{112}\times\makebox[\widthof{1024}][c]{112}$} \\
\textcolor{black}{Conv $4\times4$} & \textcolor{black}{LReLU} & \textcolor{black}{$\makebox[\widthof{512}][c]{128}\times\makebox[\widthof{1024}][c]{56}\times\makebox[\widthof{1024}][c]{56}$} \\
\textcolor{black}{Conv $4\times4$} & \textcolor{black}{LReLU} & \textcolor{black}{$\makebox[\widthof{512}][c]{256}\times\makebox[\widthof{1024}][c]{28}\times\makebox[\widthof{1024}][c]{28}$} \\
\textcolor{black}{Conv $4\times4$} & \textcolor{black}{LReLU} & \textcolor{black}{$\makebox[\widthof{512}][c]{512}\times\makebox[\widthof{1024}][c]{14}\times\makebox[\widthof{1024}][c]{14}$} \\
\textcolor{black}{Conv $4\times4$} & \textcolor{black}{--} & \textcolor{black}{$\makebox[\widthof{512}][c]{1}\times\makebox[\widthof{1024}][c]{14}\times\makebox[\widthof{1024}][c]{14}$} \\
\textcolor{black}{AveragePool} & \textcolor{black}{--} & \textcolor{black}{$\makebox[\widthof{512}][c]{3}\times\makebox[\widthof{1024}][c]{112}\times\makebox[\widthof{1024}][c]{112}$} \\
\hline
\textcolor{black}{Conv $4\times4$} & \textcolor{black}{LReLU} & \textcolor{black}{$\makebox[\widthof{512}][c]{64}\times\makebox[\widthof{1024}][c]{56}\times\makebox[\widthof{1024}][c]{56}$} \\
\textcolor{black}{Conv $4\times4$} & \textcolor{black}{LReLU} & \textcolor{black}{$\makebox[\widthof{512}][c]{128}\times\makebox[\widthof{1024}][c]{28}\times\makebox[\widthof{1024}][c]{28}$} \\
\textcolor{black}{Conv $4\times4$} & \textcolor{black}{LReLU} & \textcolor{black}{$\makebox[\widthof{512}][c]{256}\times\makebox[\widthof{1024}][c]{14}\times\makebox[\widthof{1024}][c]{14}$} \\
\textcolor{black}{Conv $4\times4$} & \textcolor{black}{LReLU} & \textcolor{black}{$\makebox[\widthof{512}][c]{512}\times\makebox[\widthof{1024}][c]{7}\times\makebox[\widthof{1024}][c]{7}$} \\
\textcolor{black}{Conv $4\times4$} & \textcolor{black}{--} & \textcolor{black}{$\makebox[\widthof{512}][c]{1}\times\makebox[\widthof{1024}][c]{7}\times\makebox[\widthof{1024}][c]{7}$} \\
\textcolor{black}{AveragePool} & \textcolor{black}{--} & \textcolor{black}{$\makebox[\widthof{512}][c]{3}\times\makebox[\widthof{1024}][c]{56}\times\makebox[\widthof{1024}][c]{56}$} \\
\hline
\textcolor{black}{Conv $4\times4$} & \textcolor{black}{LReLU} & \textcolor{black}{$\makebox[\widthof{512}][c]{64}\times\makebox[\widthof{1024}][c]{28}\times\makebox[\widthof{1024}][c]{28}$} \\
\textcolor{black}{Conv $4\times4$} & \textcolor{black}{LReLU} & \textcolor{black}{$\makebox[\widthof{512}][c]{128}\times\makebox[\widthof{1024}][c]{14}\times\makebox[\widthof{1024}][c]{14}$} \\
\textcolor{black}{Conv $4\times4$} & \textcolor{black}{LReLU} & \textcolor{black}{$\makebox[\widthof{512}][c]{256}\times\makebox[\widthof{1024}][c]{7}\times\makebox[\widthof{1024}][c]{7}$} \\
\textcolor{black}{Conv $4\times4$} & \textcolor{black}{LReLU} & \textcolor{black}{$\makebox[\widthof{512}][c]{512}\times\makebox[\widthof{1024}][c]{3}\times\makebox[\widthof{1024}][c]{3}$} \\
\textcolor{black}{Conv $4\times4$} & \textcolor{black}{--} & \textcolor{black}{$\makebox[\widthof{512}][c]{1}\times\makebox[\widthof{1024}][c]{3}\times\makebox[\widthof{1024}][c]{3}$} \\
\textcolor{black}{AveragePool} & \textcolor{black}{--} & \textcolor{black}{$\makebox[\widthof{512}][c]{3}\times\makebox[\widthof{1024}][c]{28}\times\makebox[\widthof{1024}][c]{28}$} \\
\Hline
\label{tab5_2}
\end{tabular}
}
\end{table*}

\noindent \textbf{SimGAN}~\cite{Shrivastava} is an image-to-image GAN designed for DA that adopts the self-regularization term/local adversarial loss; it updates a discriminator with a history of refined images.

\noindent \textbf{SimGAN Implementation Details}
The SimGAN architecture \textcolor{black}{(i.e., a refiner)} \textcolor{black}{uses} the following loss:

\begin{eqnarray}\label{eq:simgan_loss}
\sum_i  \mathcal{L}_{\text{real}} (\boldsymbol \theta; \mathbf x_i, \mathcal Y ) + \lambda_{\text{reg}} \mathcal{L}_{\text{reg}} (\boldsymbol \theta;  {\mathbf x_i}),
\end{eqnarray}
where \textcolor{black}{$\mathcal{L}(\cdot)$ denotes the loss function,} $\boldsymbol {\theta} $ is the function parameters, $\mathbf x_i$ is the $i^{\text th}$ PGGAN-generated training image, and $\mathcal Y$ is the set of the real images $\mathbf y_j$. The first part $ \mathcal{L}_{\text{real}}$ adds realism to the synthetic images \textcolor{black}{using a discriminator}, while the second part $\mathcal{L}_{\text{reg}}$ preserves the tumor/non-tumor features.


We train \textcolor{black}{the model (Table~\ref{tab5_3})} for $20,000$ steps with a batch size of 10 and $1.0 \times 10^{-4}$ learning rate for the Stochastic Gradient Descent (SGD) optimizer~\cite{Bottou} \textcolor{black}{without momentum}. The learning rate is reduced by half at 15,000 steps. \textcolor{black}{We train the refiner first with just the self-regularization loss with $\lambda_\text{reg} = 5 \times 10^{-5}$ for $500$ steps; then, for each update of the discriminator, we update the refiner $5$ times. } During training, we apply horizontal flipping as DA.

\begin{table*}[t!]
\caption[SimGAN architecture details for the refiner/discriminator.]{SimGAN architecture details for the refiner/discriminator. Batch normalization is applied both in the refiner/discriminator after each convolutional layer except for the final output layers respectively as in the original paper~\cite{Shrivastava}.}
\small
  \centering
\scalebox{0.79}{
\begin{tabular}{lcc}
\Hline
\textbf{\textcolor{black}{Refiner}} & \textbf{\textcolor{black}{Activation}} & \textbf{\textcolor{black}{Output Shape}}\\
\hline
\textcolor{black}{Input image} & \textcolor{black}{--} & \textcolor{black}{$\makebox[\widthof{512}][c]{1}\times\makebox[\widthof{1024}][c]{224}\times\makebox[\widthof{1024}][c]{224}$} \\
\textcolor{black}{Conv $9\times9$} & \textcolor{black}{ReLU} & \textcolor{black}{$\makebox[\widthof{512}][c]{64}\times\makebox[\widthof{1024}][c]{224}\times\makebox[\widthof{1024}][c]{224}$} \\
\hline
\textcolor{black}{\blockc{12}} & \textcolor{black}{ReLU} & \textcolor{black}{$\makebox[\widthof{512}][c]{64}\times\makebox[\widthof{1024}][c]{224}\times\makebox[\widthof{1024}][c]{224}$} \\
& \textcolor{black}{--} & \textcolor{black}{$\makebox[\widthof{512}][c]{64}\times\makebox[\widthof{1024}][c]{224}\times\makebox[\widthof{1024}][c]{224}$} \\
\hline
\textcolor{black}{Conv $1\times1$} & \textcolor{black}{Tanh} & \textcolor{black}{$\makebox[\widthof{512}][c]{1}\times\makebox[\widthof{1024}][c]{224}\times\makebox[\widthof{1024}][c]{224}$}  \\
\Hline
\end{tabular}
\begin{tabular}{lcc}
\Hline
\textcolor{black}{\textbf{Discriminator}} & \textcolor{black}{\textbf{Activation}} & \textcolor{black}{\textbf{Output Shape}}\\
\hline
\textcolor{black}{Input image} & \textcolor{black}{--} & \textcolor{black}{$\makebox[\widthof{512}][c]{1}\times\makebox[\widthof{1024}][c]{224}\times\makebox[\widthof{1024}][c]{224}$} \\
\textcolor{black}{Conv $9\times9$} & \textcolor{black}{ReLU} & \textcolor{black}{$\makebox[\widthof{512}][c]{96}\times\makebox[\widthof{1024}][c]{72}\times\makebox[\widthof{1024}][c]{72}$} \\
\textcolor{black}{Conv $5\times5$} & \textcolor{black}{ReLU} & \textcolor{black}{$\makebox[\widthof{512}][c]{64}\times\makebox[\widthof{1024}][c]{68}\times\makebox[\widthof{1024}][c]{68}$} \\
\textcolor{black}{Maxpool} &  \textcolor{black}{--} & \textcolor{black}{$\makebox[\widthof{512}][c]{64}\times\makebox[\widthof{1024}][c]{34}\times\makebox[\widthof{1024}][c]{34}$} \\
\hline
\textcolor{black}{Conv $5\times5$} & \textcolor{black}{ReLU} & \textcolor{black}{$\makebox[\widthof{512}][c]{64}\times\makebox[\widthof{1024}][c]{15}\times\makebox[\widthof{1024}][c]{15}$} \\
\textcolor{black}{Conv $3\times3$} & \textcolor{black}{ReLU} & \textcolor{black}{$\makebox[\widthof{512}][c]{32}\times\makebox[\widthof{1024}][c]{13}\times\makebox[\widthof{1024}][c]{13}$} \\
\textcolor{black}{Maxpool} & \textcolor{black}{--} & \textcolor{black}{$\makebox[\widthof{512}][c]{32}\times\makebox[\widthof{1024}][c]{7}\times\makebox[\widthof{1024}][c]{7}$} \\
\hline
\textcolor{black}{Conv $1\times1$} & \textcolor{black}{ReLU} & \textcolor{black}{$\makebox[\widthof{512}][c]{32}\times\makebox[\widthof{1024}][c]{7}\times\makebox[\widthof{1024}][c]{7}$} \\
\textcolor{black}{Conv $1\times1$} & \textcolor{black}{ReLU} & \textcolor{black}{$\makebox[\widthof{512}][c]{2}\times\makebox[\widthof{1024}][c]{7}\times\makebox[\widthof{1024}][c]{7}$}  \\
\Hline
\label{tab5_3}
\end{tabular}
}
\end{table*}

\subsection{ResNet-50-based Tumor Classification}
\noindent \textbf{Pre-processing} As ResNet-50's input size is $224 \times 224$ pixels, we resize the whole real images from $240 \times 240$ and whole PGGAN-generated images from $256 \times 256$.

\noindent \textbf{ResNet-50}~\cite{He} is a $50$-layer residual learning-based CNN. We adopt it to conduct tumor/non-tumor binary classification on MR images due to its outstanding performance in image classification tasks~\cite{Bianco}\textcolor{black}{, including binary classification~\cite{Yap}}. \textcolor{black}{Chang \textit{et al.}\cite{Chang} also used a similar $34$-layer residual convolutional network for the binary classification of brain tumors (i.e., determining the Isocitrate Dehydrogenase status in LGG/HGG).}

\textcolor{black}{\noindent \textbf{DA Setups}} To confirm the effect of PGGAN-based DA and its refinement using MUNIT/SimGAN, we compare the following $10$ DA setups under sufficient images both with/without ImageNet~\cite{Russakovsky} pre-training (i.e., 20 DA setups):

\begin{enumerate}
\item $8,429$ real images;
\item + $200$k classic DA;
\item + $400$k classic DA;
\item + $200$k PGGAN-based DA;
\item + $200$k PGGAN-based DA w/o clustering/discarding;
\item + $200$k classic DA \& $200$k PGGAN-based DA;
\item + $200$k \textcolor{black}{MUNIT}-refined DA;
\item + $200$k classic DA \& $200$k \textcolor{black}{MUNIT}-refined DA;
\item + $200$k SimGAN-refined DA;
\item + $200$k classic DA \& $200$k SimGAN-refined DA.
\end{enumerate}

\textcolor{black}{Due to the risk of overlooking the tumor diagnosis, higher sensitivity matters much more than higher specificity~\cite{Mazurowski}; thus, we aim to achieve higher sensitivity, using the additional synthetic training images. We perform McNemar's test on paired tumor classification results~\cite{McNemar} to confirm our two-step GAN-based DA's statistically-significant sensitivity improvement; since this statistical analysis involves multiple comparison tests, we adjust their $p$-values using the Holm-Bonferroni method~\cite{Holm}.}

\begin{figure}[t!]
  \centering
  \centerline{\includegraphics[width=1\linewidth]{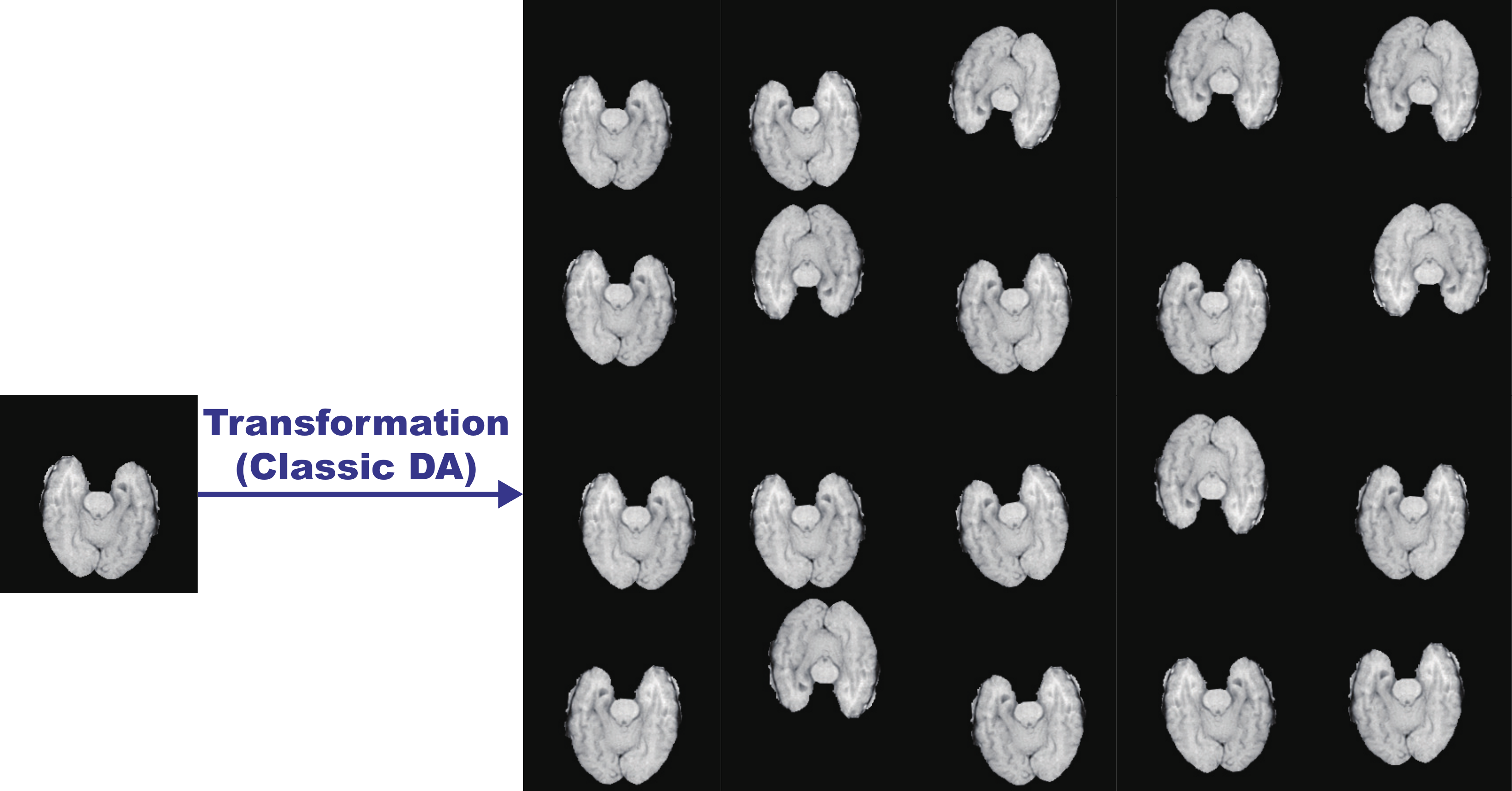}}
\caption{Example real $256\times256$ MR image and its geometrically-transformed images.}
\label{fig5_4}
\end{figure}

\begin{figure}[t!]
  \centering
  \centerline{\includegraphics[width=1\linewidth]{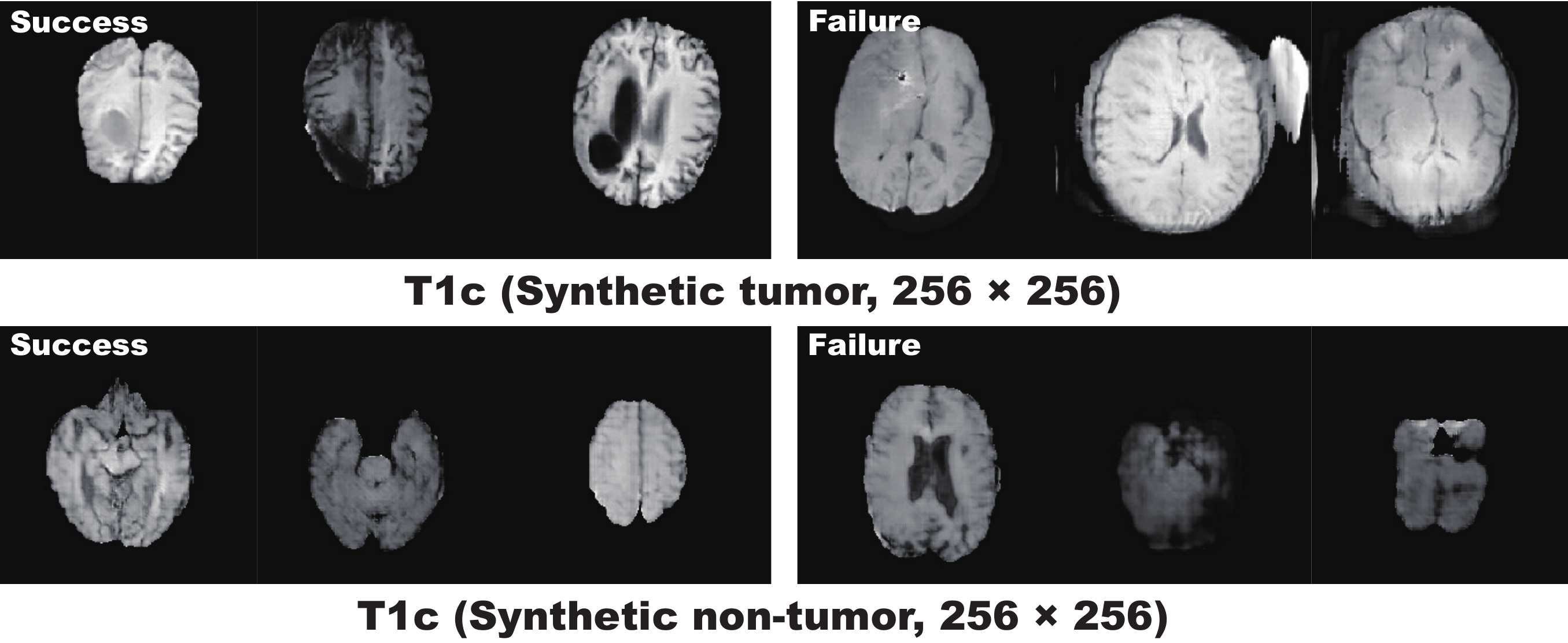}}
\caption[Example PGGAN-generated $256\times256$ MR images.]{Example PGGAN-generated MR images: (a) Success cases; (b) Failure cases.}
\label{fig5_5}
\end{figure}

Whereas medical imaging researchers widely use the ImageNet initialization despite different textures of natural/medical images, recent study found that such ImageNet-trained CNNs are biased towards recognizing texture rather than shape~\cite{Geirhos}; thus, we aim to investigate how the medical GAN-based DA affects classification performance with/without the pre-training. As the classic DA, we adopt a random combination of horizontal/vertical flipping, rotation up to $10$ degrees, width/height shift up to $8\%$, shearing up to $8\%$, zooming up to $8\%$, and constant filling of points outside the input boundaries (Fig.~\ref{fig5_4}). For the PGGAN-based DA and its refinement, we only use success cases after discarding weird-looking synthetic images (Fig.~\ref{fig5_5}); DenseNet-169~\cite{Iandola} extracts image features and k-means++~\cite{Arthur} clusters the features into $200$ groups, and then we manually discard each cluster containing similar weird-looking images. To verify its effect, we also conduct \textcolor{black}{a} PGGAN-based DA experiment without the discarding step. Additionally, to confirm the effect of changing training data set sizes, we compare classification results with pre-training on $8,429$/$4,183$/$1,646$/$834$ real images \textit{vs} real images + $200$k classic DA \textit{vs} real images + $200$k classic DA \& $200$k PGGAN-based DA (i.e., $4 \times 3 = 12$ setups).

\noindent \textbf{ResNet-50 Implementation Details}
The ResNet-50 architecture adopts the binary cross-entropy loss for binary classification both with/without ImageNet pre-training. \textcolor{black}{As shown in Table~\ref{tab5_4},} for robust training, before the final sigmoid layer, we \textcolor{black}{introduce} a $0.5$ dropout~\cite{srivastava2014dropout}, linear dense, and batch normalization~\cite{ioffe2015batch} layers---training with GAN-based DA tends to be unstable especially without the batch normalization layer. We use a batch size of $96$, $1.0 \times 10^{-2}$ learning rate for the SGD optimizer~\cite{Bottou} with $0.9$ momentum, and early stopping of $20$ epochs. The learning rate was multiplied by $0.1$ every $20$ epochs for the training from scratch and by $0.5$ every $5$ epochs for the ImageNet pre-training.

\begin{table}[t!]
\caption[ResNet-$50$ architecture details without/with pre-training.]{ResNet-$50$ architecture details without/with pre-training. We input grayscale images (i.e., $1$ channel) for experiments without pre-training, whereas we input color images (i.e., $3$ channels) for experiments with pre-training to use ImageNet initialization. Batch normalization is applied after each convolutional layer as in the original paper~\cite{He}.}
\small
  \centering
\begin{tabular}{lcc}
\Hline
\textcolor{black}{\textbf{Classifier}} & \textcolor{black}{\textbf{Activation}} & \textcolor{black}{\textbf{Output Shape}}\\
\hline
\textcolor{black}{Input image} & \textcolor{black}{--} & \textcolor{black}{$\makebox[\widthof{512}][c]{1 (3)}\times\makebox[\widthof{1024}][c]{224}\times\makebox[\widthof{1024}][c]{224}$} \\
\textcolor{black}{Conv $7\times7$} & \textcolor{black}{ReLU} & \textcolor{black}{$\makebox[\widthof{512}][c]{64}\times\makebox[\widthof{1024}][c]{112}\times\makebox[\widthof{1024}][c]{112}$} \\
\textcolor{black}{Maxpool} &  \textcolor{black}{--} & \textcolor{black}{$\makebox[\widthof{512}][c]{64}\times\makebox[\widthof{1024}][c]{55}\times\makebox[\widthof{1024}][c]{55}$} \\
\hline
\textcolor{black}{\blockb{3}} & \textcolor{black}{ReLU} & \textcolor{black}{$\makebox[\widthof{512}][c]{64}\times\makebox[\widthof{1024}][c]{55}\times\makebox[\widthof{1024}][c]{55}$} \\
& \textcolor{black}{ReLU} & \textcolor{black}{$\makebox[\widthof{512}][c]{64}\times\makebox[\widthof{1024}][c]{55}\times\makebox[\widthof{1024}][c]{55}$} \\
& \textcolor{black}{ReLU} &\textcolor{black}{$\makebox[\widthof{512}][c]{256}\times\makebox[\widthof{1024}][c]{55}\times\makebox[\widthof{1024}][c]{55}$} \\
\hline
\textcolor{black}{\blockb{4}} & \textcolor{black}{ReLU} & \textcolor{black}{$\makebox[\widthof{512}][c]{128}\times\makebox[\widthof{1024}][c]{28}\times\makebox[\widthof{1024}][c]{28}$} \\
& \textcolor{black}{ReLU} & \textcolor{black}{$\makebox[\widthof{512}][c]{128}\times\makebox[\widthof{1024}][c]{28}\times\makebox[\widthof{1024}][c]{28}$} \\
& \textcolor{black}{ReLU} &\textcolor{black}{$\makebox[\widthof{512}][c]{512}\times\makebox[\widthof{1024}][c]{28}\times\makebox[\widthof{1024}][c]{28}$} \\
\hline
\textcolor{black}{\blockb{6}} & \textcolor{black}{ReLU} & \textcolor{black}{$\makebox[\widthof{512}][c]{256}\times\makebox[\widthof{1024}][c]{14}\times\makebox[\widthof{1024}][c]{14}$} \\
& \textcolor{black}{ReLU} & \textcolor{black}{$\makebox[\widthof{512}][c]{256}\times\makebox[\widthof{1024}][c]{14}\times\makebox[\widthof{1024}][c]{14}$} \\
& \textcolor{black}{ReLU} &\textcolor{black}{$\makebox[\widthof{512}][c]{1024}\times\makebox[\widthof{1024}][c]{14}\times\makebox[\widthof{1024}][c]{14}$} \\
\hline
\textcolor{black}{\blockb{3}} & \textcolor{black}{ReLU} & \textcolor{black}{$\makebox[\widthof{512}][c]{512}\times\makebox[\widthof{1024}][c]{7}\times\makebox[\widthof{1024}][c]{7}$} \\
& \textcolor{black}{ReLU} & \textcolor{black}{$\makebox[\widthof{512}][c]{512}\times\makebox[\widthof{1024}][c]{7}\times\makebox[\widthof{1024}][c]{7}$} \\
& \textcolor{black}{ReLU} &\textcolor{black}{$\makebox[\widthof{512}][c]{2048}\times\makebox[\widthof{1024}][c]{7}\times\makebox[\widthof{1024}][c]{7}$} \\
\hline
\textcolor{black}{AveragePool} & \textcolor{black}{--} & \textcolor{black}{$\makebox[\widthof{512}][c]{2048}\times\makebox[\widthof{1024}][c]{1}\times\makebox[\widthof{1024}][c]{1}$}  \\
\textcolor{black}{Flatten} & \textcolor{black}{--} & \textcolor{black}{$\makebox[\widthof{1024}][c]{2048}$}  \\
\textcolor{black}{$0.5$ Dropout} & \textcolor{black}{--} & \textcolor{black}{$\makebox[\widthof{1024}][c]{2048}$}  \\
\textcolor{black}{Dense} & \textcolor{black}{--} & \textcolor{black}{$\makebox[\widthof{1024}][c]{2}$}  \\
\textcolor{black}{BatchNorm} & \textcolor{black}{Sigmoid} & \textcolor{black}{$\makebox[\widthof{1024}][c]{2}$}  \\
\Hline
\label{tab5_4}
\end{tabular}
\end{table}

\subsection{Clinical Validation \textit{via} Visual Turing Test}
To \textcolor{black}{quantify} the (\textit{i}) realism of \textcolor{black}{$224 \times 224$} synthetic images \textcolor{black}{by PGGANs, \textcolor{black}{MUNIT}, and SimGAN against real images respectively (i.e., 3 setups)} and (\textit{ii}) clearness of their tumor/non-tumor features, we supply, in random order, to an expert physician a random selection of:
\begin{itemize}
\item $50$ real tumor images;
\item $50$ real non-tumor images;
\item $50$ synthetic tumor images;
\item $50$ synthetic non-tumor images.
\end{itemize}

Then, the physician is asked to classify them as both (\textit{i}) real/synthetic and (\textit{ii}) tumor/non-tumor, without previously knowing which is real/synthetic and tumor/non-tumor.

\subsection{Visualization \textit{via} t-SNE}
To \textcolor{black}{visualize} distributions of geometrically-transformed and each GAN-based \textcolor{black}{$224 \times 224$} images by PGGANs, \textcolor{black}{MUNIT}, and SimGAN against real images \textcolor{black}{respectively} (i.e., 4 setups), we adopt t-SNE~\cite{Maaten} on a random selection of:
\begin{itemize}
\item $300$ real tumor images;
\item $300$ real non-tumor images;
\item $300$ geometrically-transformed or each GAN-based tumor images;
\item $300$ geometrically-transformed or each GAN-based non-tumor images.
\end{itemize}

We select only $300$ images per each category for better visualization. The t-SNE method reduces the dimensionality to represent high-dimensional data into a lower-dimensional (2D/3D) space; it non-linearly balances between the input data's local and global aspects using perplexity.

\noindent \textbf{T-SNE Implementation Details}
The t-SNE uses a perplexity of $100$ for $1,000$ iterations to visually represent a 2D space.
\textcolor{black}{We input the images after normalizing pixel values to $[0, 1]$. For point locations of the real images, we compress all the images simultaneously and plot each setup (i.e., the geometrically-transformed or each GAN-based images against the real ones) separately; we maintain their locations by projecting all the data onto the same subspace.}

\section{Results}
This section shows how PGGANs generates synthetic brain MR images and how \textcolor{black}{MUNIT} and SimGAN refine them. The results include instances of synthetic images, their quantitative evaluation by a physician, their t-SNE visualization, and their influence on tumor classification.

\subsection{MR Images Generated by PGGANs}
Fig.~\ref{fig5_5} illustrates examples of synthetic MR images by PGGANs. We visually confirm that, for about $75\%$ of cases, it successfully captures the T1c-specific texture and tumor appearance, while maintaining the realism of the original brain MR images; but, for the rest $25\%$, the generated images lack clear tumor/non-tumor features or contain unrealistic features (i.e., hyper-intensity, gray contours, and odd artifacts).

\subsection{MR Images Refined by MUNIT/SimGAN}
\begin{figure}[t!]
  \centering
  \centerline{\includegraphics[width=1\linewidth]{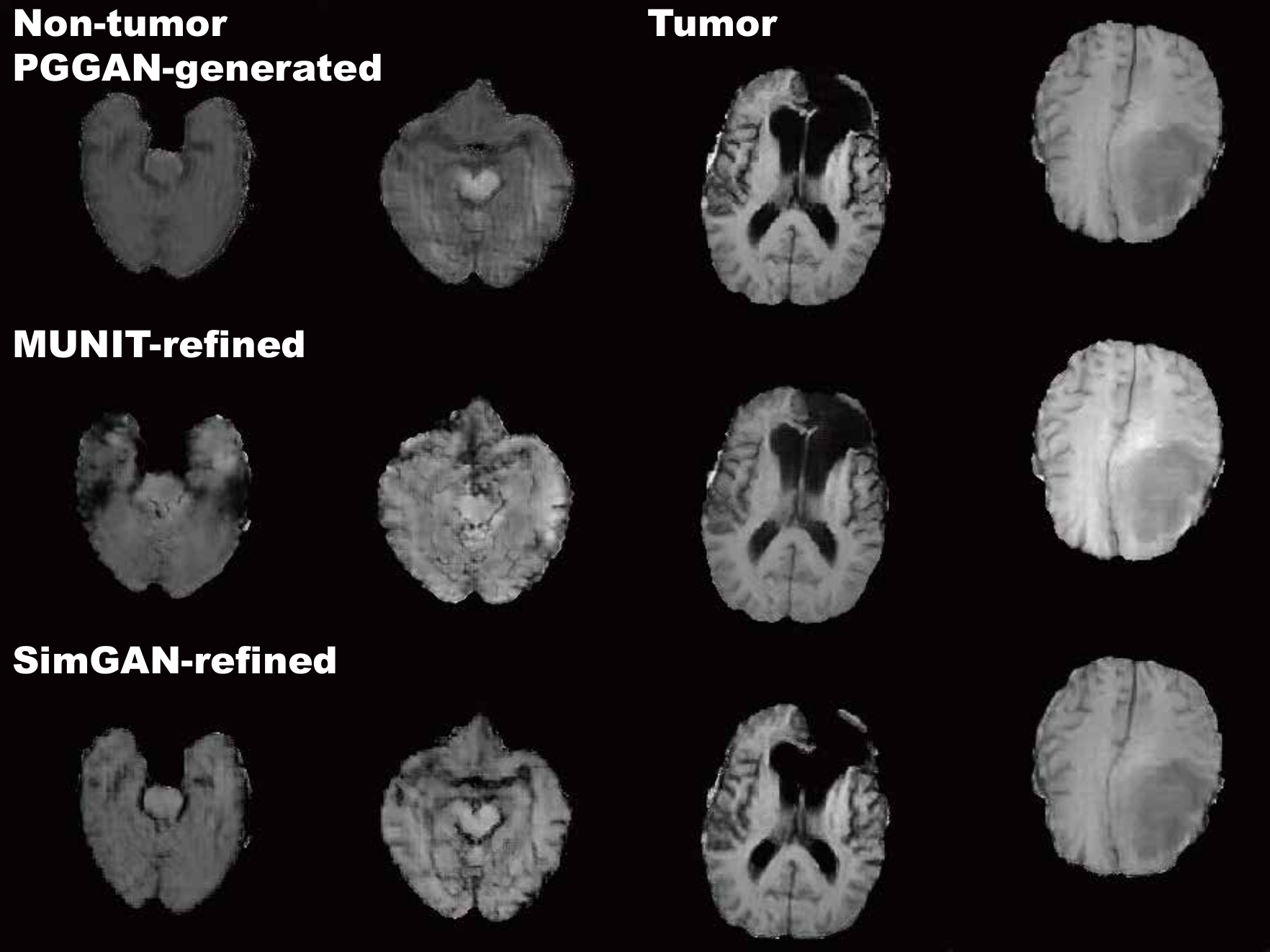}}
\caption{Example PGGAN-generated $256\times256$ MR images and their refined versions by \textcolor{black}{MUNIT}/SimGAN.}
\label{fig5_6}
\end{figure}

\textcolor{black}{MUNIT} and SimGAN differently refine PGGAN-generated images---they render the texture and contours while maintaining the overall shape (Fig.~\ref{fig5_6}). Non-tumor images change more remarkably than tumor images for both \textcolor{black}{MUNIT} and SimGAN; it probably derives from unsupervised image translation's loss for consistency to avoid image collapse, resulting in conservative change for more complicated images.

\newpage

\subsection{Tumor Classification Results}

Table~\ref{tab5_5} shows the brain tumor classification results with/without DA \textcolor{black}{while Table~\ref{tab5_6} indicates their pairwise comparison ($p$-values between our two-step GAN-based DA setups and the other DA setups) using McNemar's test.} ImageNet pre-training generally outperforms training from scratch despite different image domains (i.e., natural images to medical images). As expected, classic DA remarkably improves classification, while no clear difference exists between the $200,000$/$400,000$ classic DA under sufficient geometrically-transformed training images. When pre-trained, each GAN-based DA (i.e., PGGANs/\textcolor{black}{MUNIT}/SimGAN) alone helps classification due to the robustness from GAN-generated images; but, without pre-training, it harms classification due to the biased initialization from the GAN-overwhelming data distribution. Similarly, without pre-training, PGGAN-based DA without clustering/discarding causes poor classification due to the synthetic images with severe artifacts, unlike the PGGAN-based DA's comparable results with/without the discarding step when pre-trained.

\begin{table*}[t!]
\caption[ResNet-$50$ tumor classification results of $20$ DA setups, with (without) ImageNet pre-training.]{ResNet-$50$ tumor results of $20$ DA setups, with (without) ImageNet pre-training. \textcolor{black}{Sensitivity and specificity consider the slight tumor/non-tumor class imbalance (about $6:5$) in the test set. Boldface indicates the best performance.}}
\centering
\begin{small}
\scalebox{0.8}{
\begin{tabular}{lrrr}
\Hline\noalign{\smallskip}
\multicolumn{1}{l}{\bfseries DA Setups}  & \multicolumn{1}{c}{\bfseries Accuracy} \hspace{-0.1in} (\%) & \multicolumn{1}{c}{\bfseries Sensitivity} \hspace{-0.1in} (\%) & \multicolumn{1}{c}{\bfseries Specificity} \hspace{-0.1in} (\%)  \\\noalign{\smallskip}\hline\noalign{\smallskip}
\textcolor{black}{(1)} 8,429 real images & \textcolor{black}{93.1} (\textcolor{black}{86.3}) \hspace{-0.1in} & \textcolor{black}{90.9} (\textcolor{black}{88.9}) \hspace{-0.1in} & \textcolor{black}{95.9} (\textcolor{black}{83.2}) \hspace{-0.1in} \\
\textcolor{black}{(2)} + 200k classic DA & \textcolor{black}{95.0} (\textcolor{black}{92.2}) \hspace{-0.1in} & \textcolor{black}{93.7} (\textcolor{black}{89.9}) \hspace{-0.1in} & \textcolor{black}{96.6} (\textcolor{black}{95.0}) \hspace{-0.1in} \\
\textcolor{black}{(3)} + 400k classic DA & \textcolor{black}{94.8} (\textcolor{black}{93.2}) \hspace{-0.1in} & \textcolor{black}{91.9} (\textcolor{black}{90.9}) \hspace{-0.1in} & \textcolor{black}{98.4} (\textcolor{black}{96.1}) \hspace{-0.1in} \\\noalign{\smallskip}\hline\noalign{\smallskip}
\textcolor{black}{(4)} + 200k PGGAN-based DA & \textcolor{black}{93.9} (\textcolor{black}{86.2}) \hspace{-0.1in} & \textcolor{black}{92.6} (\textcolor{black}{87.3}) \hspace{-0.1in} & \textcolor{black}{95.6} (\textcolor{black}{84.9}) \hspace{-0.1in} \\
\textcolor{black}{(5)} + 200k PGGAN-based DA w/o clustering/discarding & \textcolor{black}{94.8} (\textcolor{black}{80.7}) \hspace{-0.1in} & \textcolor{black}{91.9} (\textcolor{black}{80.2}) \hspace{-0.1in} & \textcolor{black}{98.4} (\textcolor{black}{81.2}) \hspace{-0.1in} \\
\textcolor{black}{(6)} + 200k classic DA \& 200k PGGAN-based DA & \textcolor{black}{96.2} (\textcolor{black}{95.6}) \hspace{-0.1in} & \textcolor{black}{94.0} (\textcolor{black}{94.2}) \hspace{-0.1in} & \textbf{\textcolor{black}{98.8}} (\textcolor{black}{97.3}) \hspace{-0.1in} \\\noalign{\smallskip}\hline\noalign{\smallskip}
\textcolor{black}{(7)} + 200k \textcolor{black}{MUNIT}-refined DA & \textcolor{black}{94.3} (\textcolor{black}{83.7}) \hspace{-0.1in} & \textcolor{black}{93.0} (\textcolor{black}{87.8}) \hspace{-0.1in} & \textcolor{black}{95.8} (\textcolor{black}{78.5}) \hspace{-0.1in} \\
\textcolor{black}{(8)} + 200k classic DA \& 200k \textcolor{black}{MUNIT}-refined DA & \textbf{\textcolor{black}{96.7}} (\textcolor{black}{96.3}) \hspace{-0.1in} & \textcolor{black}{95.4} (\textbf{\textcolor{black}{97.5}}) \hspace{-0.1in} & \textcolor{black}{98.2} (\textcolor{black}{95.0}) \hspace{-0.1in} \\
\textcolor{black}{(9)} + 200k SimGAN-refined DA & \textcolor{black}{94.5} (\textcolor{black}{77.6}) \hspace{-0.1in} & \textcolor{black}{92.3} (\textcolor{black}{82.3}) \hspace{-0.1in} & \textcolor{black}{97.1} (\textcolor{black}{72.0}) \hspace{-0.1in} \\
\textcolor{black}{(10)} + 200k classic DA \& 200k SimGAN-refined DA & \textcolor{black}{96.4} (\textcolor{black}{95.0}) \hspace{-0.1in} & \textcolor{black}{95.1} (\textcolor{black}{95.1}) \hspace{-0.1in} & \textcolor{black}{97.9} (\textcolor{black}{95.0}) \hspace{-0.1in} \\

\noalign{\smallskip}\Hline\noalign{\smallskip}
\label{tab5_5}
\end{tabular}
}
\end{small}
\end{table*}

\begin{table*}[t!]
\caption[McNemar's test $p$-values for the pairwise comparison of the ResNet-50 tumor classification results by accuracy, sensitivity, and specificity.]{McNemar's test $p$-values for the pairwise comparison of the ResNet-50 tumor classification results by accuracy, sensitivity, and specificity. We compare our two-step GAN-based DA setups and all the other DA setups. All numbers within parentheses refer to DA setups on Table~\ref{tab5_5} and PT denotes pre-training. Boldface indicates statistical significance (threshold $p$-value $< 0.05$).}
\centering
\begin{small}
\scalebox{0.48}{
\begin{tabular}{lrrrlrrrlrrr}
\Hline\noalign{\smallskip}
\multicolumn{1}{l}{\bfseries DA Setup Comparison} & \multicolumn{1}{c}{\bfseries Accu} \hspace{-0.1in} & \multicolumn{1}{c}{\bfseries Sens} \hspace{-0.1in}  & \multicolumn{1}{c}{\bfseries Spec} \hspace{0.0in} & \multicolumn{1}{l}{\bfseries DA Setup Comparison} & \multicolumn{1}{c}{\bfseries Accu} \hspace{-0.1in} & \multicolumn{1}{c}{\bfseries Sens} \hspace{-0.1in}  & \multicolumn{1}{c}{\bfseries Spec} \hspace{0.0in} & \multicolumn{1}{l}{\bfseries DA Setup Comparison} & \multicolumn{1}{c}{\bfseries Accu} \hspace{-0.1in} & \multicolumn{1}{c}{\bfseries Sens} \hspace{-0.1in}  & \multicolumn{1}{c}{\bfseries Spec} \hspace{-0.1in} \\\noalign{\smallskip}\hline\noalign{\smallskip}
\textcolor{black}{(7) w/ PT \textit{vs} (1) w/ PT} & \textcolor{black}{0.693} \hspace{-0.1in} & \textcolor{black}{0.206} \hspace{-0.1in} & \textcolor{black}{1} \hspace{0.0in} &

\textcolor{black}{(7) w/ PT \textit{vs} (1) w/o PT} & \textcolor{black}{\textbf{$<$ 0.001}} \hspace{-0.1in} & \textcolor{black}{\textbf{0.002}} \hspace{-0.1in} & \textcolor{black}{\textbf{$<$ 0.001}} \hspace{0.0in} &

\textcolor{black}{(7) w/ PT \textit{vs} (2) w/ PT} & \textcolor{black}{1} \hspace{-0.1in} & \textcolor{black}{1} \hspace{-0.1in} & \textcolor{black}{1} \hspace{-0.1in}
\\\noalign{\smallskip}\hline\noalign{\smallskip}

\textcolor{black}{(7) w/ PT \textit{vs} (2) w/o PT} & \textcolor{black}{\textbf{0.034}} \hspace{-0.1in} & \textcolor{black}{\textbf{0.024}} \hspace{-0.1in} & \textcolor{black}{1} \hspace{0.0in} &

\textcolor{black}{(7) w/ PT \textit{vs} (3) w/ PT} & \textcolor{black}{1} \hspace{-0.1in} & \textcolor{black}{1} \hspace{-0.1in} & \textcolor{black}{\textbf{0.035}} \hspace{0.0in} &

\textcolor{black}{(7) w/ PT \textit{vs} (3) w/o PT} & \textcolor{black}{1} \hspace{-0.1in} & \textcolor{black}{0.468} \hspace{-0.1in} & \textcolor{black}{1} \hspace{-0.1in}
\\\noalign{\smallskip}\hline\noalign{\smallskip}

\textcolor{black}{(7) w/ PT \textit{vs} (4) w/ PT} & \textcolor{black}{1} \hspace{-0.1in} & \textcolor{black}{1} \hspace{-0.1in} & \textcolor{black}{1} \hspace{0.0in} &

\textcolor{black}{(7) w/ PT \textit{vs} (4) w/o PT} & \textcolor{black}{\textbf{$<$ 0.001}} \hspace{-0.1in} & \textcolor{black}{\textbf{$<$ 0.001}} \hspace{-0.1in} & \textcolor{black}{\textbf{$<$ 0.001}} \hspace{0.0in} &

\textcolor{black}{(7) w/ PT \textit{vs} (5) w/ PT} & \textcolor{black}{1} \hspace{-0.1in} & \textcolor{black}{1} \hspace{-0.1in} & \textcolor{black}{\textbf{0.003}} \hspace{-0.1in}
\\\noalign{\smallskip}\hline\noalign{\smallskip}

\textcolor{black}{(7) w/ PT \textit{vs} (5) w/o PT} & \textcolor{black}{\textbf{$<$ 0.001}} \hspace{-0.1in} & \textcolor{black}{\textbf{$<$ 0.001}} \hspace{-0.1in} & \textcolor{black}{\textbf{$<$ 0.001}} \hspace{0.0in} &

\textcolor{black}{(7) w/ PT \textit{vs} (6) w/ PT} & \textcolor{black}{\textbf{0.009}} \hspace{-0.1in} & \textcolor{black}{1} \hspace{-0.1in} & \textcolor{black}{\textbf{$<$ 0.001}} \hspace{0.0in} &

\textcolor{black}{(7) w/ PT \textit{vs} (6) w/o PT} & \textcolor{black}{0.397} \hspace{-0.1in} & \textcolor{black}{1} \hspace{-0.1in} & \textcolor{black}{1} \hspace{-0.1in}
\\\noalign{\smallskip}\hline\noalign{\smallskip}

\textcolor{black}{(7) w/ PT \textit{vs} (7) w/o PT} & \textcolor{black}{\textbf{$<$ 0.001}} \hspace{-0.1in} & \textcolor{black}{\textbf{$<$ 0.001}} \hspace{-0.1in} & \textcolor{black}{\textbf{$<$ 0.001}} \hspace{0.0in} &

\textcolor{black}{(7) w/ PT \textit{vs} (8) w/ PT} & \textcolor{black}{\textbf{$<$ 0.001}} \hspace{-0.1in} & \textcolor{black}{\textbf{0.025}} \hspace{-0.1in} & \textcolor{black}{\textbf{0.045}} \hspace{0.0in} &

\textcolor{black}{(7) w/ PT \textit{vs} (8) w/o PT} & \textcolor{black}{\textbf{0.008}} \hspace{-0.1in} & \textcolor{black}{\textbf{$<$ 0.001}} \hspace{-0.1in} & \textcolor{black}{1} \hspace{-0.1in}
\\\noalign{\smallskip}\hline\noalign{\smallskip}

\textcolor{black}{(7) w/ PT \textit{vs} (9) w/ PT} & \textcolor{black}{1} \hspace{-0.1in} & \textcolor{black}{1} \hspace{-0.1in} & \textcolor{black}{1} \hspace{0.0in} &

\textcolor{black}{(7) w/ PT \textit{vs} (9) w/o PT} & \textcolor{black}{\textbf{$<$ 0.001}} \hspace{-0.1in} & \textcolor{black}{\textbf{$<$ 0.001}} \hspace{-0.1in} & \textcolor{black}{\textbf{$<$ 0.001}} \hspace{0.0in} &

\textcolor{black}{(7) w/ PT \textit{vs} (10) w/ PT} & \textcolor{black}{\textbf{$<$ 0.001}} \hspace{-0.1in} & \textcolor{black}{0.077} \hspace{-0.1in} & \textcolor{black}{0.108} \hspace{-0.1in}
\\\noalign{\smallskip}\hline\noalign{\smallskip}

\textcolor{black}{(7) w/ PT \textit{vs} (10) w/o PT} & \textcolor{black}{1} \hspace{-0.1in} & \textcolor{black}{0.206} \hspace{-0.1in} & \textcolor{black}{1} \hspace{0.0in} &

\textcolor{black}{(7) w/o PT \textit{vs} (1) w/ PT} & \textcolor{black}{\textbf{$<$ 0.001}} \hspace{-0.1in} & \textcolor{black}{0.135} \hspace{-0.1in} & \textcolor{black}{\textbf{$<$ 0.001}} \hspace{0.0in} &

\textcolor{black}{(7) w/o PT \textit{vs} (1) w/o PT} & \textcolor{black}{\textbf{0.026}} \hspace{-0.1in} & \textcolor{black}{1} \hspace{-0.1in} & \textcolor{black}{\textbf{0.014}} \hspace{-0.1in}
\\\noalign{\smallskip}\hline\noalign{\smallskip}

\textcolor{black}{(7) w/o PT \textit{vs} (2) w/ PT} & \textcolor{black}{\textbf{$<$ 0.001}} \hspace{-0.1in} & \textcolor{black}{\textbf{$<$ 0.001}} \hspace{-0.1in} & \textcolor{black}{\textbf{$<$ 0.001}} \hspace{0.0in} &

\textcolor{black}{(7) w/o PT \textit{vs} (2) w/o PT} & \textcolor{black}{\textbf{$<$ 0.001}} \hspace{-0.1in} & \textcolor{black}{1} \hspace{-0.1in} & \textcolor{black}{\textbf{$<$ 0.001}} \hspace{0.0in} &

\textcolor{black}{(7) w/o PT \textit{vs} (3) w/ PT} & \textcolor{black}{\textbf{$<$ 0.001}} \hspace{-0.1in} & \textcolor{black}{\textbf{0.020}} \hspace{-0.1in} & \textcolor{black}{\textbf{$<$ 0.001}} \hspace{-0.1in}
\\\noalign{\smallskip}\hline\noalign{\smallskip}

\textcolor{black}{(7) w/o PT \textit{vs} (3) w/o PT} & \textcolor{black}{\textbf{$<$ 0.001}} \hspace{-0.1in} & \textcolor{black}{0.147} \hspace{-0.1in} & \textcolor{black}{\textbf{$<$ 0.001}} \hspace{0.0in} &

\textcolor{black}{(7) w/o PT \textit{vs} (4) w/ PT} & \textcolor{black}{\textbf{$<$ 0.001}} \hspace{-0.1in} & \textcolor{black}{\textbf{0.002}} \hspace{-0.1in} & \textcolor{black}{\textbf{$<$ 0.001}} \hspace{0.0in} &

\textcolor{black}{(7) w/o PT \textit{vs} (4) w/o PT} & \textcolor{black}{\textbf{0.044}} \hspace{-0.1in} & \textcolor{black}{1} \hspace{-0.1in} & \textcolor{black}{\textbf{$<$ 0.001}} \hspace{-0.1in}
\\\noalign{\smallskip}\hline\noalign{\smallskip}

\textcolor{black}{(7) w/o PT \textit{vs} (5) w/ PT} & \textcolor{black}{\textbf{$<$ 0.001}} \hspace{-0.1in} & \textcolor{black}{\textbf{0.015}} \hspace{-0.1in} & \textcolor{black}{\textbf{$<$ 0.001}} \hspace{0.0in} &

\textcolor{black}{(7) w/o PT \textit{vs} (5) w/o PT} & \textcolor{black}{\textbf{0.011}} \hspace{-0.1in} & \textcolor{black}{\textbf{$<$ 0.001}} \hspace{-0.1in} & \textcolor{black}{1} \hspace{0.0in} &

\textcolor{black}{(7) w/o PT \textit{vs} (6) w/ PT} & \textcolor{black}{\textbf{$<$ 0.001}} \hspace{-0.1in} & \textcolor{black}{\textbf{$<$ 0.001}} \hspace{-0.1in} & \textcolor{black}{\textbf{$<$ 0.001}} \hspace{-0.1in}
\\\noalign{\smallskip}\hline\noalign{\smallskip}

\textcolor{black}{(7) w/o PT \textit{vs} (6) w/o PT} & \textcolor{black}{\textbf{$<$ 0.001}} \hspace{-0.1in} & \textcolor{black}{\textbf{$<$ 0.001}} \hspace{-0.1in} & \textcolor{black}{\textbf{$<$ 0.001}} \hspace{0.0in} &

\textcolor{black}{(7) w/o PT \textit{vs} (8) w/ PT} & \textcolor{black}{\textbf{$<$ 0.001}} \hspace{-0.1in} & \textcolor{black}{\textbf{$<$ 0.001}} \hspace{-0.1in} & \textcolor{black}{\textbf{$<$ 0.001}} \hspace{0.0in} &

\textcolor{black}{(7) w/o PT \textit{vs} (8) w/o PT} & \textcolor{black}{\textbf{$<$ 0.001}} \hspace{-0.1in} & \textcolor{black}{\textbf{$<$ 0.001}} \hspace{-0.1in} & \textcolor{black}{\textbf{$<$ 0.001}} \hspace{-0.1in}
\\\noalign{\smallskip}\hline\noalign{\smallskip}

\textcolor{black}{(7) w/o PT \textit{vs} (9) w/ PT} & \textcolor{black}{\textbf{$<$ 0.001}} \hspace{-0.1in} & \textcolor{black}{\textbf{0.004}} \hspace{-0.1in} & \textcolor{black}{\textbf{$<$ 0.001}} \hspace{0.0in} &

\textcolor{black}{(7) w/o PT \textit{vs} (9) w/o PT} & \textcolor{black}{\textbf{$<$ 0.001}} \hspace{-0.1in} & \textcolor{black}{\textbf{$<$ 0.001}} \hspace{-0.1in} & \textcolor{black}{\textbf{$<$ 0.001}} \hspace{0.0in} &

\textcolor{black}{(7) w/o PT \textit{vs} (10) w/ PT} & \textcolor{black}{\textbf{$<$ 0.001}} \hspace{-0.1in} & \textcolor{black}{\textbf{$<$ 0.001}} \hspace{-0.1in} & \textcolor{black}{\textbf{$<$ 0.001}} \hspace{-0.1in}
\\\noalign{\smallskip}\hline\noalign{\smallskip}

\textcolor{black}{(7) w/o PT \textit{vs} (10) w/o PT} & \textcolor{black}{\textbf{$<$ 0.001}} \hspace{-0.1in} & \textcolor{black}{\textbf{$<$ 0.001}} \hspace{-0.1in} & \textcolor{black}{\textbf{$<$ 0.001}} \hspace{0.0in} &

\textcolor{black}{(8) w/ PT \textit{vs} (1) w PT} & \textcolor{black}{\textbf{$<$ 0.001}} \hspace{-0.1in} & \textcolor{black}{\textbf{$<$ 0.001}} \hspace{-0.1in} & \textcolor{black}{\textbf{0.010}} \hspace{0.0in} &

\textcolor{black}{(8) w/ PT \textit{vs} (1) w/o PT} & \textcolor{black}{\textbf{$<$ 0.001}} \hspace{-0.1in} & \textcolor{black}{\textbf{$<$ 0.001}} \hspace{-0.1in} & \textcolor{black}{\textbf{$<$ 0.001}} \hspace{-0.1in}
\\\noalign{\smallskip}\hline\noalign{\smallskip}

\textcolor{black}{(8) w/ PT \textit{vs} (2) w/ PT} & \textcolor{black}{\textbf{$<$ 0.001}} \hspace{-0.1in} & \textcolor{black}{0.074} \hspace{-0.1in} & \textcolor{black}{0.206} \hspace{0.0in} &

\textcolor{black}{(8) w/ PT \textit{vs} (2) w/o PT} & \textcolor{black}{\textbf{$<$ 0.001}} \hspace{-0.1in} & \textcolor{black}{\textbf{$<$ 0.001}} \hspace{-0.1in} & \textcolor{black}{\textbf{$<$ 0.001}} \hspace{0.0in} &

\textcolor{black}{(8) w/ PT \textit{vs} (3) w/ PT} & \textcolor{black}{\textbf{0.002}} \hspace{-0.1in} & \textcolor{black}{\textbf{$<$ 0.001}} \hspace{-0.1in} & \textcolor{black}{1} \hspace{-0.1in}
\\\noalign{\smallskip}\hline\noalign{\smallskip}

\textcolor{black}{(8) w/ PT \textit{vs} (3) w/o PT} & \textcolor{black}{\textbf{$<$ 0.001}} \hspace{-0.1in} & \textcolor{black}{\textbf{$<$ 0.001}} \hspace{-0.1in} & \textcolor{black}{0.112} \hspace{0.0in} &

\textcolor{black}{(8) w/ PT \textit{vs} (4) w/ PT} & \textcolor{black}{\textbf{$<$ 0.001}} \hspace{-0.1in} & \textcolor{black}{\textbf{$<$ 0.001}} \hspace{-0.1in} & \textcolor{black}{\textbf{0.006}} \hspace{0.0in} &

\textcolor{black}{(8) w/ PT \textit{vs} (4) w/o PT} & \textcolor{black}{\textbf{$<$ 0.001}} \hspace{-0.1in} & \textcolor{black}{\textbf{$<$ 0.001}} \hspace{-0.1in} & \textcolor{black}{\textbf{$<$ 0.001}} \hspace{-0.1in}
\\\noalign{\smallskip}\hline\noalign{\smallskip}

\textcolor{black}{(8) w/ PT \textit{vs} (5) w/ PT} & \textcolor{black}{\textbf{0.002}} \hspace{-0.1in} & \textcolor{black}{\textbf{$<$ 0.001}} \hspace{-0.1in} & \textcolor{black}{1} \hspace{0.0in} &

\textcolor{black}{(8) w/ PT \textit{vs} (5) w/o PT} & \textcolor{black}{\textbf{$<$ 0.001}} \hspace{-0.1in} & \textcolor{black}{\textbf{$<$ 0.001}} \hspace{-0.1in} & \textcolor{black}{\textbf{$<$ 0.001}} \hspace{0.0in} &

\textcolor{black}{(8) w/ PT \textit{vs} (6) w/ PT} & \textcolor{black}{1} \hspace{-0.1in} & \textcolor{black}{0.128} \hspace{-0.1in} & \textcolor{black}{1} \hspace{-0.1in}
\\\noalign{\smallskip}\hline\noalign{\smallskip}

\textcolor{black}{(8) w/ PT \textit{vs} (6) w/o PT} & \textcolor{black}{0.222} \hspace{-0.1in} & \textcolor{black}{0.760} \hspace{-0.1in} & \textcolor{black}{1} \hspace{0.0in} &

\textcolor{black}{(8) w/ PT \textit{vs} (8) w/o PT} & \textcolor{black}{1} \hspace{-0.1in} & \textcolor{black}{\textbf{0.008}} \hspace{-0.1in} & \textcolor{black}{\textbf{$<$ 0.001}} \hspace{0.0in} &

\textcolor{black}{(8) w/ PT \textit{vs} (9) w/ PT} & \textcolor{black}{\textbf{$<$ 0.001}} \hspace{-0.1in} & \textcolor{black}{\textbf{$<$ 0.001}} \hspace{-0.1in} & \textcolor{black}{1} \hspace{-0.1in}
\\\noalign{\smallskip}\hline\noalign{\smallskip}

\textcolor{black}{(8) w/ PT \textit{vs} (9) w/o PT} & \textcolor{black}{\textbf{$<$ 0.001}} \hspace{-0.1in} & \textcolor{black}{\textbf{$<$ 0.001}} \hspace{-0.1in} & \textcolor{black}{\textbf{$<$ 0.001}} \hspace{0.0in} &

\textcolor{black}{(8) w/ PT \textit{vs} (10) w/ PT} & \textcolor{black}{1} \hspace{-0.1in} & \textcolor{black}{1} \hspace{-0.1in} & \textcolor{black}{1} \hspace{0.0in} &

\textcolor{black}{(8) w/ PT \textit{vs} (10) w/o PT} & \textcolor{black}{\textbf{0.007}} \hspace{-0.1in} & \textcolor{black}{1} \hspace{-0.1in} & \textcolor{black}{0} \hspace{-0.1in}
\\\noalign{\smallskip}\hline\noalign{\smallskip}

\textcolor{black}{(8) w/o PT \textit{vs} (1) w/ PT} & \textcolor{black}{\textbf{$<$ 0.001}} \hspace{-0.1in} & \textcolor{black}{\textbf{$<$ 0.001}} \hspace{-0.1in} & \textcolor{black}{1} \hspace{0.0in} &

\textcolor{black}{(8) w/o PT \textit{vs} (1) w/o PT} & \textcolor{black}{\textbf{$<$ 0.001}} \hspace{-0.1in} & \textcolor{black}{\textbf{$<$ 0.001}} \hspace{-0.1in} & \textcolor{black}{\textbf{$<$ 0.001}} \hspace{0.0in} &

\textcolor{black}{(8) w/o PT \textit{vs} (2) w/ PT} & \textcolor{black}{0.179} \hspace{-0.1in} & \textcolor{black}{\textbf{$<$ 0.001}} \hspace{-0.1in} & \textcolor{black}{0.588} \hspace{-0.1in}
\\\noalign{\smallskip}\hline\noalign{\smallskip}

\textcolor{black}{(8) w/o PT \textit{vs} (2) w/o PT} & \textcolor{black}{\textbf{$<$ 0.001}} \hspace{-0.1in} & \textcolor{black}{\textbf{$<$ 0.001}} \hspace{-0.1in} & \textcolor{black}{1} \hspace{0.0in} &

\textcolor{black}{(8) w/o PT \textit{vs} (3) w/ PT} & \textcolor{black}{0.101} \hspace{-0.1in} & \textcolor{black}{\textbf{$<$ 0.001}} \hspace{-0.1in} & \textcolor{black}{\textbf{$<$ 0.001}} \hspace{0.0in} &

\textcolor{black}{(8) w/o PT \textit{vs} (3) w/o PT} & \textcolor{black}{\textbf{$<$ 0.001}} \hspace{-0.1in} & \textcolor{black}{\textbf{$<$ 0.001}} \hspace{-0.1in} & \textcolor{black}{1} \hspace{-0.1in}
\\\noalign{\smallskip}\hline\noalign{\smallskip}

\textcolor{black}{(8) w/o PT \textit{vs} (4) w/ PT} & \textcolor{black}{\textbf{$<$ 0.001}} \hspace{-0.1in} & \textcolor{black}{\textbf{$<$ 0.001}} \hspace{-0.1in} & \textcolor{black}{1} \hspace{0.0in} &

\textcolor{black}{(8) w/o PT \textit{vs} (4) w/o PT} & \textcolor{black}{\textbf{$<$ 0.001}} \hspace{-0.1in} & \textcolor{black}{\textbf{$<$ 0.001}} \hspace{-0.1in} & \textcolor{black}{\textbf{$<$ 0.001}} \hspace{0.0in} &

\textcolor{black}{(8) w/o PT \textit{vs} (5) w/ PT} & \textcolor{black}{0.197} \hspace{-0.1in} & \textcolor{black}{\textbf{$<$ 0.001}} \hspace{-0.1in} & \textcolor{black}{\textbf{$<$ 0.001}} \hspace{-0.1in}
\\\noalign{\smallskip}\hline\noalign{\smallskip}

\textcolor{black}{(8) w/o PT \textit{vs} (5) w/o PT} & \textcolor{black}{\textbf{$<$ 0.001}} \hspace{-0.1in} & \textcolor{black}{\textbf{$<$ 0.001}} \hspace{-0.1in} & \textcolor{black}{\textbf{$<$ 0.001}} \hspace{0.0in} &

\textcolor{black}{(8) w/o PT \textit{vs} (6) w/ PT} & \textcolor{black}{1} \hspace{-0.1in} & \textcolor{black}{\textbf{$<$ 0.001}} \hspace{-0.1in} & \textcolor{black}{\textbf{$<$ 0.001}} \hspace{0.0in} &

\textcolor{black}{(8) w/o PT \textit{vs} (6) w/o PT} & \textcolor{black}{1} \hspace{-0.1in} & \textcolor{black}{\textbf{$<$ 0.001}} \hspace{-0.1in} & \textcolor{black}{\textbf{0.007}} \hspace{-0.1in}
\\\noalign{\smallskip}\hline\noalign{\smallskip}

\textcolor{black}{(8) w/o PT \textit{vs} (9) w/ PT} & \textcolor{black}{\textbf{0.023}} \hspace{-0.1in} & \textcolor{black}{\textbf{$<$ 0.001}} \hspace{-0.1in} & \textcolor{black}{0.256} \hspace{0.0in} &

\textcolor{black}{(8) w/o PT \textit{vs} (9) w/o PT} & \textcolor{black}{\textbf{$<$ 0.001}} \hspace{-0.1in} & \textcolor{black}{\textbf{$<$ 0.001}} \hspace{-0.1in} & \textcolor{black}{\textbf{$<$ 0.001}} \hspace{0.0in} &

\textcolor{black}{(8) w/o PT \textit{vs} (10) w/ PT} & \textcolor{black}{1} \hspace{-0.1in} & \textcolor{black}{\textbf{0.002}} \hspace{-0.1in} & \textcolor{black}{\textbf{$<$ 0.001}} \hspace{-0.1in}
\\\noalign{\smallskip}\hline\noalign{\smallskip}

\textcolor{black}{(8) w/o PT \textit{vs} (10) w/o PT} & \textcolor{black}{0.143} \hspace{-0.1in} & \textcolor{black}{\textbf{0.005}} \hspace{-0.1in} & \textcolor{black}{1} \hspace{0.0in} &

\textcolor{black}{(9) w/ PT \textit{vs} (1) w/ PT} & \textcolor{black}{0.387} \hspace{-0.1in} & \textcolor{black}{1} \hspace{-0.1in} & \textcolor{black}{1} \hspace{0.0in} &

\textcolor{black}{(9) w/ PT \textit{vs} (1) w/o PT} & \textcolor{black}{\textbf{$<$ 0.001}} \hspace{-0.1in} & \textcolor{black}{\textbf{0.046}} \hspace{-0.1in} & \textcolor{black}{\textbf{$<$ 0.001}} \hspace{-0.1in}
\\\noalign{\smallskip}\hline\noalign{\smallskip}

\textcolor{black}{(9) w/ PT \textit{vs} (2) w/ PT} & \textcolor{black}{1} \hspace{-0.1in} & \textcolor{black}{1} \hspace{-0.1in} & \textcolor{black}{1} \hspace{0.0in} &

\textcolor{black}{(9) w/ PT \textit{vs} (2) w/o PT} & \textcolor{black}{\textbf{0.008}} \hspace{-0.1in} & \textcolor{black}{0.262} \hspace{-0.1in} & \textcolor{black}{0.321} \hspace{0.0in} &

\textcolor{black}{(9) w/ PT \textit{vs} (3) w/ PT} & \textcolor{black}{1} \hspace{-0.1in} & \textcolor{black}{1} \hspace{-0.1in} & \textcolor{black}{0.931} \hspace{-0.1in}
\\\noalign{\smallskip}\hline\noalign{\smallskip}

\textcolor{black}{(9) w/ PT \textit{vs} (3) w/o PT} & \textcolor{black}{0.910} \hspace{-0.1in} & \textcolor{black}{1} \hspace{-0.1in} & \textcolor{black}{1} \hspace{0.0in} &

\textcolor{black}{(9) w/ PT \textit{vs} (4) w/ PT} & \textcolor{black}{1} \hspace{-0.1in} & \textcolor{black}{1} \hspace{-0.1in} & \textcolor{black}{0.764} \hspace{0.0in} &

\textcolor{black}{(9) w/ PT \textit{vs} (4) w/o PT} & \textcolor{black}{\textbf{$<$ 0.001}} \hspace{-0.1in} & \textcolor{black}{\textbf{$<$ 0.001}} \hspace{-0.1in} & \textcolor{black}{\textbf{$<$ 0.001}} \hspace{-0.1in}
\\\noalign{\smallskip}\hline\noalign{\smallskip}

\textcolor{black}{(9) w/ PT \textit{vs} (5) w/ PT} & \textcolor{black}{1} \hspace{-0.1in} & \textcolor{black}{1} \hspace{-0.1in} & \textcolor{black}{0.639} \hspace{0.0in} &

\textcolor{black}{(9) w/ PT \textit{vs} (5) w/o PT} & \textcolor{black}{\textbf{$<$ 0.001}} \hspace{-0.1in} & \textcolor{black}{\textbf{$<$ 0.001}} \hspace{-0.1in} & \textcolor{black}{\textbf{$<$ 0.001}} \hspace{0.0in} &

\textcolor{black}{(9) w/ PT \textit{vs} (6) w/ PT} & \textcolor{black}{\textbf{0.014}} \hspace{-0.1in} & \textcolor{black}{0.660} \hspace{-0.1in} & \textcolor{black}{0.066} \hspace{-0.1in}
\\\noalign{\smallskip}\hline\noalign{\smallskip}

\textcolor{black}{(9) w/ PT \textit{vs} (6) w/o PT} & \textcolor{black}{0.716} \hspace{-0.1in} & \textcolor{black}{0.365} \hspace{-0.1in} & \textcolor{black}{1} \hspace{0.0in} &

\textcolor{black}{(9) w/ PT \textit{vs} (9) w/o PT} & \textcolor{black}{\textbf{$<$ 0.001}} \hspace{-0.1in} & \textcolor{black}{\textbf{$<$ 0.001}} \hspace{-0.1in} & \textcolor{black}{\textbf{$<$ 0.001}} \hspace{0.0in} &

\textcolor{black}{(9) w/ PT \textit{vs} (10) w/ PT} & \textcolor{black}{\textbf{0.004}} \hspace{-0.1in} & \textcolor{black}{\textbf{0.006}} \hspace{-0.1in} & \textcolor{black}{1} \hspace{-0.1in}
\\\noalign{\smallskip}\hline\noalign{\smallskip}

\textcolor{black}{(9) w/ PT \textit{vs} (10) w/o PT} & \textcolor{black}{1} \hspace{-0.1in} & \textcolor{black}{\textbf{0.017}} \hspace{-0.1in} & \textcolor{black}{0.256} \hspace{0.0in} &

\textcolor{black}{(9) w/o PT \textit{vs} (1) w/ PT} & \textcolor{black}{\textbf{$<$ 0.001}} \hspace{-0.1in} & \textcolor{black}{\textbf{$<$ 0.001}} \hspace{-0.1in} & \textcolor{black}{\textbf{$<$ 0.001}} \hspace{0.0in} &

\textcolor{black}{(9) w/o PT \textit{vs} (1) w/o PT} & \textcolor{black}{\textbf{$<$ 0.001}} \hspace{-0.1in} & \textcolor{black}{\textbf{$<$ 0.001}} \hspace{-0.1in} & \textcolor{black}{\textbf{$<$ 0.001}} \hspace{-0.1in}
\\\noalign{\smallskip}\hline\noalign{\smallskip}

\textcolor{black}{(9) w/o PT \textit{vs} (2) w/ PT} & \textcolor{black}{\textbf{$<$ 0.001}} \hspace{-0.1in} & \textcolor{black}{\textbf{$<$ 0.001}} \hspace{-0.1in} & \textcolor{black}{\textbf{$<$ 0.001}} \hspace{0.0in} &

\textcolor{black}{(9) w/o PT \textit{vs} (2) w/o PT} & \textcolor{black}{\textbf{$<$ 0.001}} \hspace{-0.1in} & \textcolor{black}{\textbf{$<$ 0.001}} \hspace{-0.1in} & \textcolor{black}{\textbf{$<$ 0.001}} \hspace{0.0in} &

\textcolor{black}{(9) w/o PT \textit{vs} (3) w/ PT} & \textcolor{black}{\textbf{$<$ 0.001}} \hspace{-0.1in} & \textcolor{black}{\textbf{$<$ 0.001}} \hspace{-0.1in} & \textcolor{black}{\textbf{$<$ 0.001}} \hspace{-0.1in}
\\\noalign{\smallskip}\hline\noalign{\smallskip}

\textcolor{black}{(9) w/o PT \textit{vs} (3) w/o PT} & \textcolor{black}{\textbf{$<$ 0.001}} \hspace{-0.1in} & \textcolor{black}{\textbf{$<$ 0.001}} \hspace{-0.1in} & \textcolor{black}{\textbf{$<$ 0.001}} \hspace{0.0in} &

\textcolor{black}{(9) w/o PT \textit{vs} (4) w/ PT} & \textcolor{black}{\textbf{$<$ 0.001}} \hspace{-0.1in} & \textcolor{black}{\textbf{$<$ 0.001}} \hspace{-0.1in} & \textcolor{black}{\textbf{$<$ 0.001}} \hspace{0.0in} &

\textcolor{black}{(9) w/o PT \textit{vs} (4) w/o PT} & \textcolor{black}{\textbf{$<$ 0.001}} \hspace{-0.1in} & \textcolor{black}{\textbf{$<$ 0.001}} \hspace{-0.1in} & \textcolor{black}{\textbf{$<$ 0.001}} \hspace{-0.1in}
\\\noalign{\smallskip}\hline\noalign{\smallskip}

\textcolor{black}{(9) w/o PT \textit{vs} (5) w/ PT} & \textcolor{black}{\textbf{$<$ 0.001}} \hspace{-0.1in} & \textcolor{black}{\textbf{$<$ 0.001}} \hspace{-0.1in} & \textcolor{black}{\textbf{$<$ 0.001}} \hspace{0.0in} &

\textcolor{black}{(9) w/o PT \textit{vs} (5) w/o PT} & \textcolor{black}{\textbf{0.022}} \hspace{-0.1in} & \textcolor{black}{1} \hspace{-0.1in} & \textcolor{black}{\textbf{$<$ 0.001}} \hspace{0.0in} &

\textcolor{black}{(9) w/o PT \textit{vs} (6) w/ PT} & \textcolor{black}{\textbf{$<$ 0.001}} \hspace{-0.1in} & \textcolor{black}{\textbf{$<$ 0.001}} \hspace{-0.1in} & \textcolor{black}{\textbf{$<$ 0.001}} \hspace{-0.1in}
\\\noalign{\smallskip}\hline\noalign{\smallskip}

\textcolor{black}{(9) w/o PT \textit{vs} (6) w/o PT} & \textcolor{black}{\textbf{$<$ 0.001}} \hspace{-0.1in} & \textcolor{black}{\textbf{$<$ 0.001}} \hspace{-0.1in} & \textcolor{black}{\textbf{$<$ 0.001}} \hspace{0.0in} &

\textcolor{black}{(9) w/o PT \textit{vs} (10) w/PT} & \textcolor{black}{\textbf{$<$ 0.001}} \hspace{-0.1in} & \textcolor{black}{\textbf{$<$ 0.001}} \hspace{-0.1in} & \textcolor{black}{\textbf{$<$ 0.001}} \hspace{0.0in} &

\textcolor{black}{(9) w/o PT \textit{vs} (10) w/o PT} & \textcolor{black}{\textbf{$<$ 0.001}} \hspace{-0.1in} & \textcolor{black}{\textbf{$<$ 0.001}} \hspace{-0.1in} & \textcolor{black}{\textbf{$<$ 0.001}} \hspace{-0.1in}
\\\noalign{\smallskip}\hline\noalign{\smallskip}

\textcolor{black}{(10) w/ PT \textit{vs} (1) w/ PT} & \textcolor{black}{\textbf{$<$ 0.001}} \hspace{-0.1in} & \textcolor{black}{\textbf{$<$ 0.001}} \hspace{-0.1in} & \textcolor{black}{\textbf{0.049}} \hspace{0.0in} &

\textcolor{black}{(10) w/ PT \textit{vs} (1) w/o PT} & \textcolor{black}{\textbf{$<$ 0.001}} \hspace{-0.1in} & \textcolor{black}{\textbf{$<$ 0.001}} \hspace{-0.1in} & \textcolor{black}{\textbf{$<$ 0.001}} \hspace{0.0in} &

\textcolor{black}{(10) w/ PT \textit{vs} (2) w/ PT} & \textcolor{black}{\textbf{0.039}} \hspace{-0.1in} & \textcolor{black}{0.515} \hspace{-0.1in} & \textcolor{black}{1} \hspace{-0.1in}
\\\noalign{\smallskip}\hline\noalign{\smallskip}

\textcolor{black}{(10) w/ PT \textit{vs} (2) w/o PT} & \textcolor{black}{\textbf{$<$ 0.001}} \hspace{-0.1in} & \textcolor{black}{\textbf{$<$ 0.001}} \hspace{-0.1in} & \textcolor{black}{\textbf{0.002}} \hspace{0.0in} &

\textcolor{black}{(10) w/ PT \textit{vs} (3) w/ PT} & \textcolor{black}{\textbf{0.017}} \hspace{-0.1in} & \textcolor{black}{\textbf{$<$ 0.001}} \hspace{-0.1in} & \textcolor{black}{1} \hspace{0.0in} &

\textcolor{black}{(10) w/ PT \textit{vs} (3) w/o PT} & \textcolor{black}{\textbf{$<$ 0.001}} \hspace{-0.1in} & \textcolor{black}{\textbf{$<$ 0.001}} \hspace{-0.1in} & \textcolor{black}{0.415} \hspace{-0.1in}
\\\noalign{\smallskip}\hline\noalign{\smallskip}

\textcolor{black}{(10) w/ PT \textit{vs} (4) w/ PT} & \textcolor{black}{\textbf{$<$ 0.001}} \hspace{-0.1in} & \textcolor{black}{\textbf{0.019}} \hspace{-0.1in} & \textcolor{black}{\textbf{0.028}} \hspace{0.0in} &

\textcolor{black}{(10) w/ PT \textit{vs} (4) w/o PT} & \textcolor{black}{\textbf{$<$ 0.001}} \hspace{-0.1in} & \textcolor{black}{\textbf{$<$ 0.001}} \hspace{-0.1in} & \textcolor{black}{\textbf{$<$ 0.001}} \hspace{0.0in} &

\textcolor{black}{(10) w/ PT \textit{vs} (5) w/ PT} & \textcolor{black}{\textbf{0.015}} \hspace{-0.1in} & \textcolor{black}{\textbf{$<$ 0.001}} \hspace{-0.1in} & \textcolor{black}{1} \hspace{-0.1in}
\\\noalign{\smallskip}\hline\noalign{\smallskip}

\textcolor{black}{(10) w/ PT \textit{vs} (5) w/o PT} & \textcolor{black}{\textbf{$<$ 0.001}} \hspace{-0.1in} & \textcolor{black}{\textbf{$<$ 0.001}} \hspace{-0.1in} & \textcolor{black}{\textbf{$<$ 0.001}} \hspace{0.0in} &

\textcolor{black}{(10) w/ PT \textit{vs} (6) w/ PT} & \textcolor{black}{1} \hspace{-0.1in} & \textcolor{black}{1} \hspace{-0.1in} & \textcolor{black}{1} \hspace{0.0in} &

\textcolor{black}{(10) w/ PT \textit{vs} (6) w/o PT} & \textcolor{black}{0.981} \hspace{-0.1in} & \textcolor{black}{1} \hspace{-0.1in} & \textcolor{black}{1} \hspace{-0.1in}
\\\noalign{\smallskip}\hline\noalign{\smallskip}

\textcolor{black}{(10) w/ PT \textit{vs} (10) w/o PT} & \textcolor{black}{0.054} \hspace{-0.1in} & \textcolor{black}{1} \hspace{-0.1in} & \textcolor{black}{\textbf{0.002}} \hspace{0.0in} &

\textcolor{black}{(10) w/o PT \textit{vs} (1) w/ PT} & \textcolor{black}{\textbf{0.039}} \hspace{-0.1in} & \textcolor{black}{\textbf{$<$ 0.001}} \hspace{-0.1in} & \textcolor{black}{1} \hspace{0.0in} &

\textcolor{black}{(10) w/o PT \textit{vs} (1) w/o PT} & \textcolor{black}{\textbf{$<$ 0.001}} \hspace{-0.1in} & \textcolor{black}{\textbf{$<$ 0.001}} \hspace{-0.1in} & \textcolor{black}{\textbf{$<$ 0.001}} \hspace{-0.1in}
\\\noalign{\smallskip}\hline\noalign{\smallskip}

\textcolor{black}{(10) w/o PT \textit{vs} (2) w/ PT} & \textcolor{black}{1} \hspace{-0.1in} & \textcolor{black}{0.727} \hspace{-0.1in} & \textcolor{black}{0.649} \hspace{0.0in} &

\textcolor{black}{(10) w/o PT \textit{vs} (2) w/o PT} & \textcolor{black}{\textbf{$<$ 0.001}} \hspace{-0.1in} & \textcolor{black}{\textbf{$<$ 0.001}} \hspace{-0.1in} & \textcolor{black}{1} \hspace{0.0in} &

\textcolor{black}{(10) w/o PT \textit{vs} (3) w/ PT} & \textcolor{black}{1} \hspace{-0.1in} & \textcolor{black}{\textbf{0.002}} \hspace{-0.1in} & \textcolor{black}{\textbf{$<$ 0.001}} \hspace{-0.1in}
\\\noalign{\smallskip}\hline\noalign{\smallskip}

\textcolor{black}{(10) w/o PT \textit{vs} (3) w/o PT} & \textcolor{black}{\textbf{0.039}} \hspace{-0.1in} & \textcolor{black}{\textbf{$<$ 0.001}} \hspace{-0.1in} & \textcolor{black}{1} \hspace{0.0in} &

\textcolor{black}{(10) w/o PT \textit{vs} (4) w/ PT} & \textcolor{black}{1} \hspace{-0.1in} & \textcolor{black}{\textbf{0.019}} \hspace{-0.1in} & \textcolor{black}{1} \hspace{0.0in} &

\textcolor{black}{(10) w/o PT \textit{vs} (4) w/o PT} & \textcolor{black}{\textbf{$<$ 0.001}} \hspace{-0.1in} & \textcolor{black}{\textbf{$<$ 0.001}} \hspace{-0.1in} & \textcolor{black}{\textbf{$<$ 0.001}} \hspace{-0.1in}
\\\noalign{\smallskip}\hline\noalign{\smallskip}

\textcolor{black}{(10) w/o PT \textit{vs} (5) w/ PT} & \textcolor{black}{1} \hspace{-0.1in} & \textcolor{black}{\textbf{0.002}} \hspace{-0.1in} & \textcolor{black}{\textbf{$<$ 0.001}} \hspace{0.0in} &

\textcolor{black}{(10) w/o PT \textit{vs} (5) w/o PT} & \textcolor{black}{\textbf{$<$ 0.001}} \hspace{-0.1in} & \textcolor{black}{\textbf{$<$ 0.001}} \hspace{-0.1in} & \textcolor{black}{\textbf{$<$ 0.001}} \hspace{0.0in} &

\textcolor{black}{(10) w/o PT \textit{vs} (6) w/ PT} & \textcolor{black}{0.308} \hspace{-0.1in} & \textcolor{black}{1} \hspace{-0.1in} & \textcolor{black}{\textbf{$<$ 0.001}} \hspace{-0.1in}
\\\noalign{\smallskip}\hline\noalign{\smallskip}

\textcolor{black}{(10) w/o PT \textit{vs} (6) w/o PT} & \textcolor{black}{1} \hspace{-0.1in} & \textcolor{black}{1} \hspace{-0.1in} & \textcolor{black}{\textbf{0.035}} \hspace{0.0in} &
\\\noalign{\smallskip}\Hline\noalign{\smallskip}

\label{tab5_6}
\end{tabular}
}
\end{small}
\end{table*}

When combined with the classic DA, each GAN-based DA remarkably outperforms the GAN-based DA or classic DA alone \textcolor{black}{in terms of sensitivity since they are mutually-complementary: the former learns the non-linear manifold of the real images to generate novel local tumor features (since we train tumor/non-tumor images separately) strongly associated with sensitivity; the latter learns the geometrically-transformed manifold of the real images to cover global features and provide the robustness on training for most cases. We confirm that test samples, originally-misclassified but correctly classified after DA, are obviously different for the GAN-based DA and classic DA;} here, both image-to-image GAN-based DA, especially \textcolor{black}{MUNIT}, produce remarkably higher sensitivity than the PGGAN-based DA after refinement. Specificity is higher than sensitivity for every DA setup with pre-training, probably due to the training data imbalance; but interestingly, without pre-training, sensitivity is higher than specificity for both image-to-image GAN-based DA \textcolor{black}{since our tumor classification-oriented two-step GAN-based DA can fill the real tumor image distribution uncovered by the original dataset under no ImageNet initialization. Accordingly,} when combined with the classic DA, the \textcolor{black}{MUNIT}-based DA \textcolor{black}{based on both GANs/VAEs} achieves the highest sensitivity \textcolor{black}{$97.48\%$ against the best performing classic DA's $93.67\%$}, allowing to significantly alleviate the risk of overlooking the tumor diagnosis; \textcolor{black}{in terms of sensitivity, it outperforms all the other DA setups, including two-step DA setups, with statistical significance.}

\newpage


\begin{figure*}[t!]
  \centering
  \centerline{\includegraphics[width=1\linewidth]{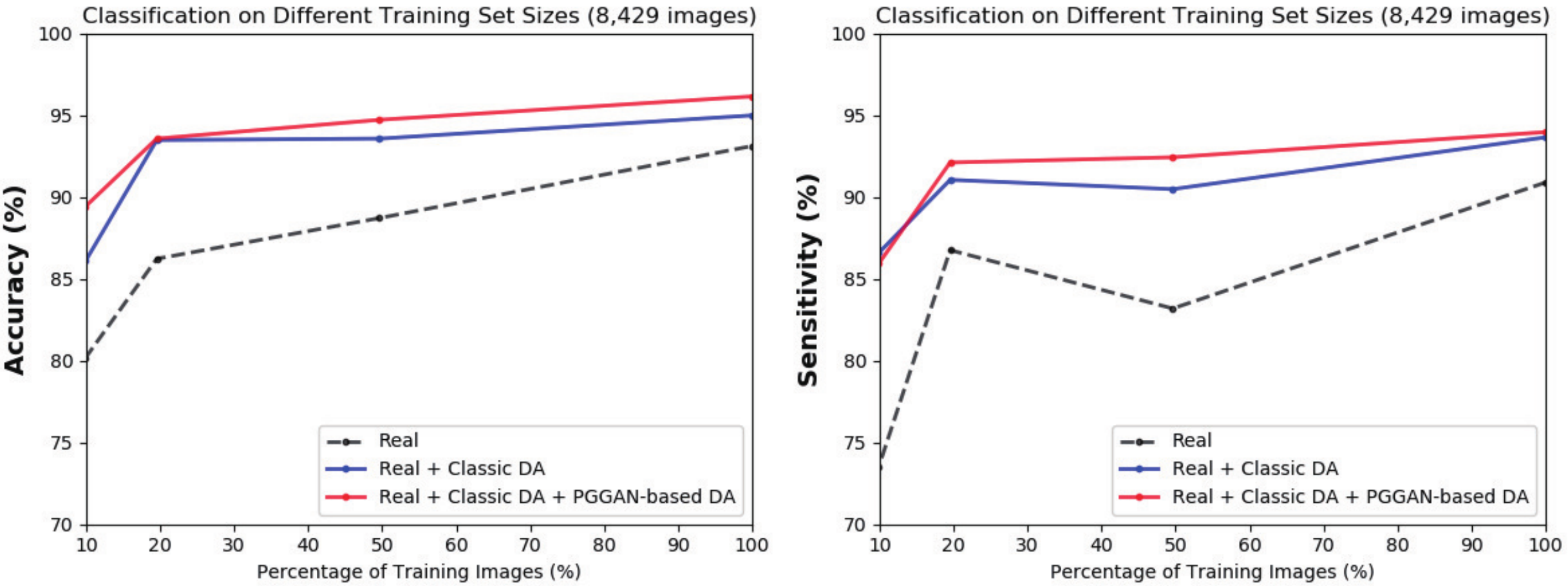}}
\caption[ResNet-$50$ tumor classification results under different training dataset sizes of $12$ DA setups, with ImageNet pre-training.]{ResNet-$50$ tumor classification results under different training dataset sizes of $12$ DA setups, with ImageNet pre-training: (a) $8,429$/$4,183$/$1,646$/$834$ real images \textit{vs} (b) real images + $200$k Classic DA \textit{vs} (c) real images + $200$k Classic DA \& $200$k PGGAN-based DA.}
\label{new_exp}
\end{figure*}

Figure~\ref{new_exp} shows that the PGGAN-based DA, even without further refinement, could moderately increase both accuracy/sensitivity on top of the classic DA in tumor classification; it achieves considerably high sensitivity with only $20\%$/$50\%$ of the real training images. However, it should be noted that the MUNIT-based DA could outperform the PGGAN-based DA in return for more computational power.

\subsection{Visual Turing Test Results}
Table~\ref{tab5_7} indicates the confusion matrix for the Visual Turing Test. The expert physician classifies a few PGGAN-generated images as real\textcolor{black}{, thanks to their realism,} despite high resolution (i.e., \textcolor{black}{$224 \times 224$} pixels); \textcolor{black}{meanwhile, the expert classifies less GAN-refined images as real due to slight artifacts induced during refinement.} The synthetic images successfully capture tumor/non-tumor features; unlike the non-tumor images, the expert recognizes a considerable number of the mild/modest tumor images as non-tumor for both real/synthetic cases. It derives from clinical tumor diagnosis relying on a full 3D volume, instead of a single 2D slice.

\begin{table*}[t!]
\caption[Visual Turing Test results by an expert physician for classifying 100 Real ($\mathsf{R}$) \textit{vs} 100 Synthetic ($\mathsf{S}$) images and 100 Tumor ($\mathsf{T}$) \textit{vs} 100 Non-tumor ($\mathsf{N}$) images.]{Visual Turing Test results by an expert physician for classifying Real ($\mathsf{R}$) \textit{vs} Synthetic ($\mathsf{S}$) images and Tumor ($\mathsf{T}$) \textit{vs} Non-tumor ($\mathsf{N}$) images. \textcolor{black}{Accuracy denotes the physician's successful classification ratio between the real/synthetic images and between the tumor/non-tumor images, respectively. It should be noted that proximity to 50\% of accuracy indicates superior performance (chance = 50\%).}}
\label{tab5_7}
\centering
\begin{small}
\scalebox{0.83}{
\begin{tabular}{p{1.4em}ccccc}
\Hline\noalign{\smallskip}
 & \textbf{Accuracy} (\%) & \textbf{Accuracy} (\%) & \textbf{Accuracy} (\%) & \textbf{Accuracy} (\%) & \textbf{Accuracy} (\%)\\
\noalign{\smallskip}\hline\noalign{\smallskip}
\parbox[t]{2mm}{\multirow{4}{*}{\rotatebox[origin=c]{270}{\textbf{\shortstack{\textcolor{black}{PGGAN}}}}}} & \multicolumn{1}{c}{\textcolor{black}{Real \textit{vs} Synthetic}}  \ \ \ &  $\mathsf{R}$ as $\mathsf{R}$ \ \ \ &  $\mathsf{R}$ as $\mathsf{S}$ \ \ \ &  $\mathsf{S}$ as $\mathsf{R}$ \ \ \ &  $\mathsf{S}$ as $\mathsf{S}$\\ 
& $79.5$ \ \ \ & $73\textcolor{black}{}$ \ \ \ & $27\textcolor{black}{}$ \ \ \ & $14\textcolor{black}{}$ \ \ \ & $86\textcolor{black}{}$ \\
& \multicolumn{1}{c}{\textcolor{black}{Tumor \textit{vs} Non-tumor}} \ \ \ &  $\mathsf{T}$ as $\mathsf{T}$ \ \ \ &  $\mathsf{T}$ as $\mathsf{N}$ \ \ \ &  $\mathsf{N}$ as $\mathsf{T}$ \ \ \ &  $\mathsf{N}$ as $\mathsf{N}$\\
& $87.5$ \ \ \ & $77\textcolor{black}{}$ \ \ \ & $23\textcolor{black}{}$ ($\mathsf{R}: 
11$, $\mathsf{S}: 12$) \ \ \ & $2\textcolor{black}{}$ ($\mathsf{S}: 2$) \ \ \ & $98\textcolor{black}{}$\\
\noalign{\smallskip}\hline\noalign{\smallskip}
\parbox[t]{2mm}{\multirow{4}{*}{\rotatebox[origin=c]{270}{\textbf{\shortstack{\textcolor{black}{MUNIT}}}}}} & \multicolumn{1}{c}{\textcolor{black}{Real \textit{vs} Synthetic}}  \ \ \ &  \textcolor{black}{$\mathsf{R}$ as $\mathsf{R}$} \ \ \ &  \textcolor{black}{$\mathsf{R}$ as $\mathsf{S}$} \ \ \ &  \textcolor{black}{$\mathsf{S}$ as $\mathsf{R}$} \ \ \ &  \textcolor{black}{$\mathsf{S}$ as $\mathsf{S}$}\\ 

& \textcolor{black}{$77.0$} \ \ \ & \textcolor{black}{$58$} \ \ \ & \textcolor{black}{$42$} \ \ \ & \textcolor{black}{$4$} \ \ \ & \textcolor{black}{$96$} \\
& \multicolumn{1}{c}{\textcolor{black}{Tumor \textit{vs} Non-tumor}} \ \ \ &  \textcolor{black}{$\mathsf{T}$ as $\mathsf{T}$} \ \ \ &  \textcolor{black}{$\mathsf{T}$ as $\mathsf{N}$} \ \ \ &  \textcolor{black}{$\mathsf{N}$ as $\mathsf{T}$} \ \ \ &  \textcolor{black}{$\mathsf{N}$ as $\mathsf{N}$}\\
& \textcolor{black}{$92.5$} \ \ \ & \textcolor{black}{$88$} \ \ \ & \textcolor{black}{$12$ ($\mathsf{R}: 
6$, $\mathsf{S}: 6$)} \ \ \ & \textcolor{black}{$3$ ($\mathsf{R}: 1$, $\mathsf{S}: 2$)} \ \ \ & \textcolor{black}{$97$}\\
\noalign{\smallskip}\hline\noalign{\smallskip}
\parbox[t]{2mm}{\multirow{4}{*}{\rotatebox[origin=c]{270}{\textbf{\shortstack{\textcolor{black}{SimGAN}}}}}} & \multicolumn{1}{c}{\textcolor{black}{Real \textit{vs} Synthetic}}  \ \ \ &  \textcolor{black}{$\mathsf{R}$ as $\mathsf{R}$} \ \ \ &  \textcolor{black}{$\mathsf{R}$ as $\mathsf{S}$} \ \ \ &  \textcolor{black}{$\mathsf{S}$ as $\mathsf{R}$} \ \ \ &  \textcolor{black}{$\mathsf{S}$ as $\mathsf{S}$}\\ 

& \textcolor{black}{$76.0$} \ \ \ & \textcolor{black}{$53$} \ \ \ & \textcolor{black}{$47$} \ \ \ & \textcolor{black}{$1$} \ \ \ & \textcolor{black}{$99$} \\
& \multicolumn{1}{c}{\textcolor{black}{Tumor \textit{vs} Non-tumor}} \ \ \ &  \textcolor{black}{$\mathsf{T}$ as $\mathsf{T}$} \ \ \ &  \textcolor{black}{$\mathsf{T}$ as $\mathsf{N}$} \ \ \ &  \textcolor{black}{$\mathsf{N}$ as $\mathsf{T}$} \ \ \ &  \textcolor{black}{$\mathsf{N}$ as $\mathsf{N}$}\\
& \textcolor{black}{$94.0$} \ \ \ & \textcolor{black}{$91$} \ \ \ & \textcolor{black}{$9$ ($\mathsf{R}: 
2$, $\mathsf{S}: 7$)} \ \ \ & \textcolor{black}{$3$ ($\mathsf{R}: 3$)} \ \ \ & \textcolor{black}{$97$}\\\noalign{\smallskip}\Hline\noalign{\smallskip}
\end{tabular}
}
\end{small}
\end{table*}

\begin{figure*}[t!]
  \centering
  \centerline{\includegraphics[width=1\linewidth]{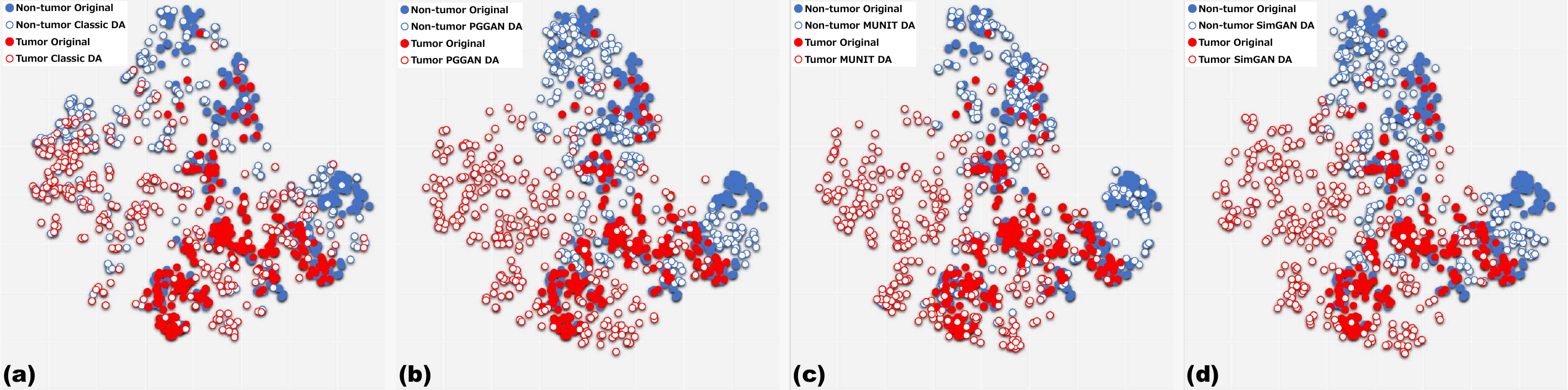}}
\caption[T-SNE plots with 300 $256\times256$ tumor/non-tumor MR images per each category.]{T-SNE plots with 300 tumor/non-tumor images per each category: Real images \textit{vs} (a) Geometrically-transformed images; (b) PGGAN-generated images; (c) \textcolor{black}{MUNIT}-refined images; (d) SimGAN-refined images.}
\label{fig5_7}
\end{figure*}

\subsection{T-SNE Results}
As Fig.~\ref{fig5_7} represents, the real tumor/non-tumor image distributions largely overlap while the non-tumor images distribute wider. The geometrically-transformed tumor/non-tumor image distributions also often overlap, and both images distribute wider than the real ones. All GAN-based synthetic images by PGGANs, MUNIT, and SimGAN distribute widely, while their tumor/non-tumor images overlap much less than the geometrically-transformed ones \textcolor{black}{(i.e., a high discrimination ability associated with sensitivity improvement)}; the \textcolor{black}{MUNIT}-refined images show \textcolor{black}{better tumor/non-tumor discrimination and} a more similar distribution to the real ones than the PGGAN-based and SimGAN-based images. \textcolor{black}{This trend derives from} the \textcolor{black}{MUNIT}'s loss function adopting both GANs/VAEs \textcolor{black}{that further fits the PGGAN-generated images into the real image distribution by refining their texture and shape; contrarily, this refinement could also induce slight human-recognizable but DA-irrelevant artifacts}. Overall, the GAN-based images, especially the \textcolor{black}{MUNIT}-refined images, fill the distribution uncovered by the real or geometrically-transformed ones with less tumor/non-tumor overlap; \textcolor{black}{this demonstrates the superiority of combining classic DA and GAN-based DA}.

\section{Conclusion}
Visual Turing Test and t-SNE results show that PGGANs, multi-stage noise-to-image GAN, can generate realistic/diverse $256 \times 256$ brain MR images with/without tumors separately. \textcolor{black}{Unlike classic DA that geometrically covers global features and provides the robustness on training for most cases, the GAN-generated images can non-linearly cover local tumor features with much less tumor/non-tumor overlap; thus, combining them can significantly boost tumor classification sensitivity}---especially after refining them with \textcolor{black}{MUNIT} or SimGAN, image-to-image GANs; thanks to an ensemble \textcolor{black}{generation process} from those GANs' different algorithms, the \textcolor{black}{texture/shape-}refined images can replace missing data points of the training set with less tumor/non-tumor overlap, and thus handle the data imbalance \textcolor{black}{by regularizing the model (i.e., improved generalization). Notably, MUNIT remarkably} outperforms SimGAN \textcolor{black}{in terms of sensitivity}, probably due to the effect of combining both GANs/VAEs.

Regarding better medical GAN-based DA, ImageNet pre-training generally improves classification despite different textures of natural/medical images; but, without pre-training, the GAN-refined images may help achieve better sensitivity, allowing to alleviate the risk of overlooking the tumor diagnosis\textcolor{black}{---this attributes to our tumor classification-oriented two-step GAN-based DA's high discrimination ability to fill the real tumor image distribution under no ImageNet initialization.} GAN-generated images typically include odd artifacts; however, only without pre-training, discarding them boosts DA performance.

Overall, by minimizing the number of annotated images required for medical imaging tasks, the two-step GAN-based DA can shed light not only on classification, but also on object detection~\cite{han2019learning} and segmentation~\cite{shin2018medical}. Moreover, other potential medical applications exist: (\textit{i}) A data anonymization tool to share patients' data outside their institution for training without losing classification performance~\cite{shin2018medical}; (\textit{ii}) A physician training tool to show random pathological images for medical students/radiology trainees despite infrastructural/legal constraints~\cite{finlayson2018towards}. As future work, we plan to define a new \textcolor{black}{end-to-end} GAN loss function that explicitly \textcolor{black}{optimizes} the classification results, instead of \textcolor{black}{optimizing} visual realism \textcolor{black}{while maintaining diversity by combining the state-of-the-art noise-to-image and image-to-image GANs; towards this, we might extend a preliminary work on a three-player GAN for classification~\cite{vandenhende2019three} to generate only hard-to-classify samples to improve classification; we could also (\textit{i}) explicitly model deformation fields/intensity transformations and (\textit{ii}) leverage unlabeled data during the generative process~\cite{Chaitanya} to effectively fill the real image distribution.}
\chapter{\LARGE GAN-based Medical Image Augmentation for 2D Detection}

\section{Prologue to Third Project}
\subsection{Project Publication}
\begin{itemize}
\item \textbf{Learning More with Less: Conditional PGGAN-based Data Augmentation for Brain Metastases Detection Using Highly-Rough Annotation on MR Images}. \textbf{C. Han}, K. Murao, T. Noguchi, Y. Kawata, F. Uchiyama, L. Rundo, H. Nakayama, S. Satoh, In ACM International Conference on Information and Knowledge Management (CIKM), Beijing, China, pp. 119--127, November 2019.
\end{itemize}

\subsection{Context}
Further DA applications require pathology localization for detection and advanced physician training needs atypical image generation, respectively. To meet both clinical demands, developing pathology-aware GANs (i.e., GANs conditioned on pathology position and appearance) is the best solution---the pathology-aware GANs are promising in terms of extrapolation because common and/or desired medical priors can play a key role in the conditioning~\cite{stinis2019enforcing}. However, prior to this work, researchers had focused only on improving segmentation, instead of bounding box-based detection, while the detection requires much less physicians' annotation effort~\cite{shin2018medical, jin2018ct}. Moreover, they had relied on image-to-image GANs, instead of noise-to-image GANs, which sacrifices image diversity due to an input benign image.

\subsection{Contributions}
This project's fundamental contribution is to propose a novel pathology-aware noise-to-image GAN called CPGGANs for improved 2D bounding box-based detection; it incorporates highly-rough bounding box conditions incrementally into the noise-to-image GAN (i.e., PGGANs) to place realistic/diverse brain metastases at desired positions/sizes on $256 \times 256$ MR images. By so doing, our CPGGAN-based DA boosts sensitivity 83\% to 91\% with Intersection over Union (IoU) threshold $0.25$ in tumor detection with clinically acceptable additional False Positives (FPs). Moreover, we find that GAN training on additional normal images could increase synthetic images' realism, including pathology, but decrease DA performance.

\subsection{Recent Developments}
Almost simultaneously, Kanayama \textit{et al.} also tackle bounding box-based pathology detection using the image-to-image GAN, instead of the noise-to-image GAN~\cite{kanayama2019gastric}; they translated normal endoscopic images with various image sizes ($458 \times 405$ on average) into gastric cancer ones by inputting both a benign image and a black image (i.e., pixel value: 0) with a specific lesion ROI at desired position.

\newpage

\section{Motivation}
Accurate CAD with high sensitivity can alleviate the risk of overlooking the diagnosis in a clinical environment. Specifically, CNNs have revolutionized medical imaging, such as diabetic eye disease diagnosis~\cite{gulshan2016development}, mainly thanks to large-scale annotated training data. However, obtaining such annotated medical big data is demanding; thus, better diagnosis requires intensive DA techniques, such as geometric/intensity transformations of original images~\cite{Ronneberger,Milletari}. Yet, those augmented images intrinsically have a similar distribution to the original ones, leading to limited performance improvement; in this context, GAN~\cite{Goodfellow}-based DA can boost the performance by filling the real image distribution uncovered by the original dataset, since it generates realistic but completely new samples showing good generalization ability; GANs achieved outstanding performance in computer vision, including $21\%$ performance improvement in eye-gaze estimation~\cite{Shrivastava}.

Also in medical imaging, where the primary problem lies in small and fragmented imaging datasets from various scanners~\cite{Rundo2}, GAN-based DA performs effectively: researchers improved classification by augmentation with noise-to-image GANs~\cite{frid2018gan} and segmentation with image-to-image GANs~\cite{shin2018medical,jin2018ct}. Such applications include $256 \times 256$ brain MR image generation for tumor/non-tumor classification~\cite{Han2}. Nevertheless, unlike bounding box-based object detection, simple classification cannot locate disease areas and rigorous segmentation requires physicians' expensive annotation.

\begin{figure}[t!]
  \centering
  \centerline{\includegraphics[width=\linewidth]{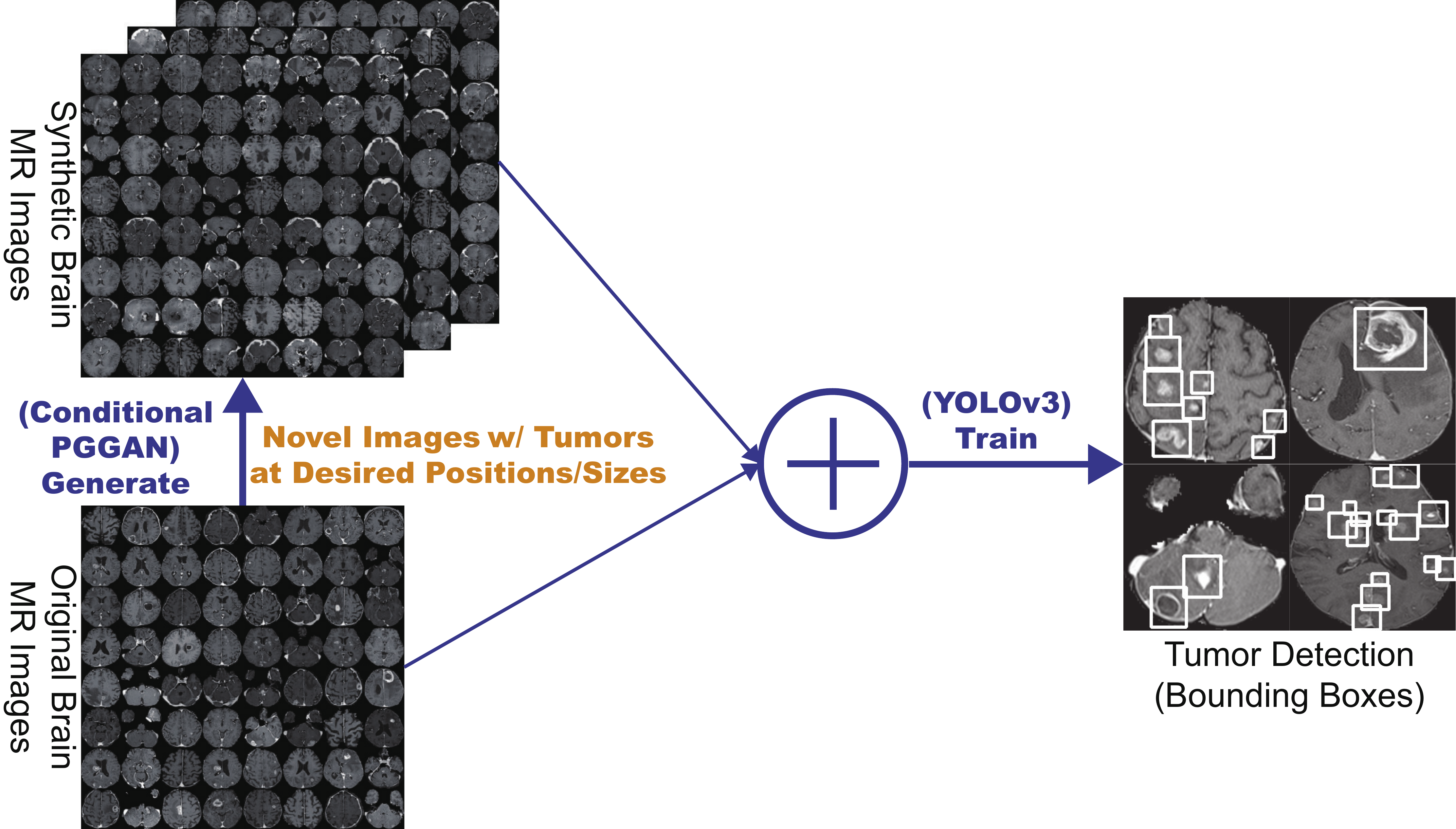}}
\caption[CPGGAN-based DA for better tumor detection.]{CPGGAN-based DA for better tumor detection: our CPGGANs generates a number of realistic/diverse brain MR images with tumors at desired positions/sizes based on bounding boxes, and the object detector uses them as additional training data.}
\label{fig6_1}
\end{figure}

So, how can we achieve high sensitivity in diagnosis using GANs with minimum annotation cost, based on highly-rough and inconsistent bounding boxes?
We aim to generate GAN-based realistic and diverse $256\times256$ brain MR images with brain metastases at desired positions/sizes for accurate CNN-based tumor detection (Fig.~\ref{fig6_1}); this is clinically valuable for better diagnosis, prognosis, and treatment, since brain metastases are the most common intra-cranial tumors, getting prevalent as oncological therapies improve cancer patients' survival~\cite{arvold2016updates}. Conventional GANs cannot generate realistic $256 \times 256$ whole brain MR images conditioned on tumor positions/sizes under limited training data/highly-rough annotation~\cite{Han2}; since noise-to-image GANs cannot directly be conditioned on an image describing desired objects, we have to use image-to-image GANs (e.g., input both conditioning image/random noise samples or the conditioning image alone with dropout noises~\cite{srivastava2014dropout} on a generator~\cite{isola2017image})---it results in unrealistic high-resolution MR images with odd artifacts due to the limited training data/rough annotation, tumor variations, and strong consistency in brain anatomy, unless we also input a benign image sacrificing image diversity.

Such a high-resolution whole image generation approach, not involving ROIs alone, however, could facilitate detection because it provides more image details and most CNN architectures adopt around $256 \times 256$ input pixels. Therefore, as a conditional noise-to-image GAN not relying on an input benign image, we propose CPGGANs, incorporating highly-rough bounding box conditions incrementally into PGGANs~\cite{Karras} to naturally place tumors of random shape at desired positions/sizes on MR images. Moreover, we evaluate the generated images' realism \textit{via} Visual Turing Test~\cite{Salimans} by three expert physicians, and visualize the data distribution \textit{via} t-SNE algorithm~\cite{Maaten}.
Using the synthetic images, our novel CPGGAN-based DA boosts $10\%$ sensitivity in diagnosis with clinically acceptable additional FPs.
Surprisingly, we confirm that further realistic tumor appearance, judged by the physicians, does not contribute to detection performance.\\

\noindent \textbf{Research Questions.} We mainly address two questions:
\begin{itemize}
\item \textbf{PGGAN Conditioning:} How can we modify PGGANs to naturally place objects of random shape, unlike rigorous segmentation, at desired positions/sizes based on highly-rough bounding box masks?
\item \textbf{Medical DA:} How can we balance the number of real and additional synthetic training data to achieve the best detection performance?\\
\end{itemize}

\noindent \textbf{Contributions.} Our main contributions are as follows:
\begin{itemize}
\item \textbf{Conditional Image Generation:} As the first bounding box-based $256 \times 256$ whole pathological image generation approach, CPGGANs can generate realistic/diverse images with objects naturally at desired positions/sizes; the generated images can play a vital role in clinical oncology applications, such as DA, data anonymization, and physician training.

\item \textbf{Misdiagnosis Prevention:} This study allows us to achieve high sensitivity in automatic CAD using small/fragmented medical imaging datasets with minimum annotation efforts based on highly-rough/inconsistent bounding boxes.

\item \textbf{Brain Metastases Detection:} This first bounding box-based brain metastases detection method successfully detects tumors with CPGGAN-based DA.
\end{itemize}

\newpage

\section{Materials and Methods}
\subsection{Brain Metastases Dataset}
As a new dataset for the first bounding box-based brain metastases detection, this project uses a dataset of T1c brain axial MR images, collected by the authors (National Center for Global Health and Medicine, Tokyo, Japan) and currently not publicly available for ethical restrictions; for robust clinical applications, it contains $180$ brain metastatic cancer cases from multiple MRI scanners---those images differ in contrast, magnetic field strength (i.e., $1.5$ T, $3.0$ T), and matrix size (i.e., $190 \times 224$, $216 \times 256$, $256 \times 256$, $460 \times 460$ pixels). We also use additional brain MR images from $193$ normal subjects only for CPGGAN training, not in tumor detection, to confirm the effect of combining the normal and pathological images for training.

\subsection{CPGGAN-based Image Generation}
\noindent \textbf{Data Preparation}
For tumor detection, our whole brain metastases dataset ($180$ patients) is divided into: (\textit{i}) a training set ($126$ patients); (\textit{ii}) a validation set ($18$ patients); (\textit{iii}) a test set ($36$ patients); only the training set is used for GAN training to be fair. Our experimental dataset consists of:
\begin{itemize}
\item Training set ($2,813$ images/$5,963$ bounding boxes);
\item Validation set ($337$ images/$616$ bounding boxes);
\item Test set ($947$ images/$3,094$ bounding boxes).
\end{itemize}

\begin{figure}[t!]
  \centering
  \centerline{\includegraphics[width=\linewidth]{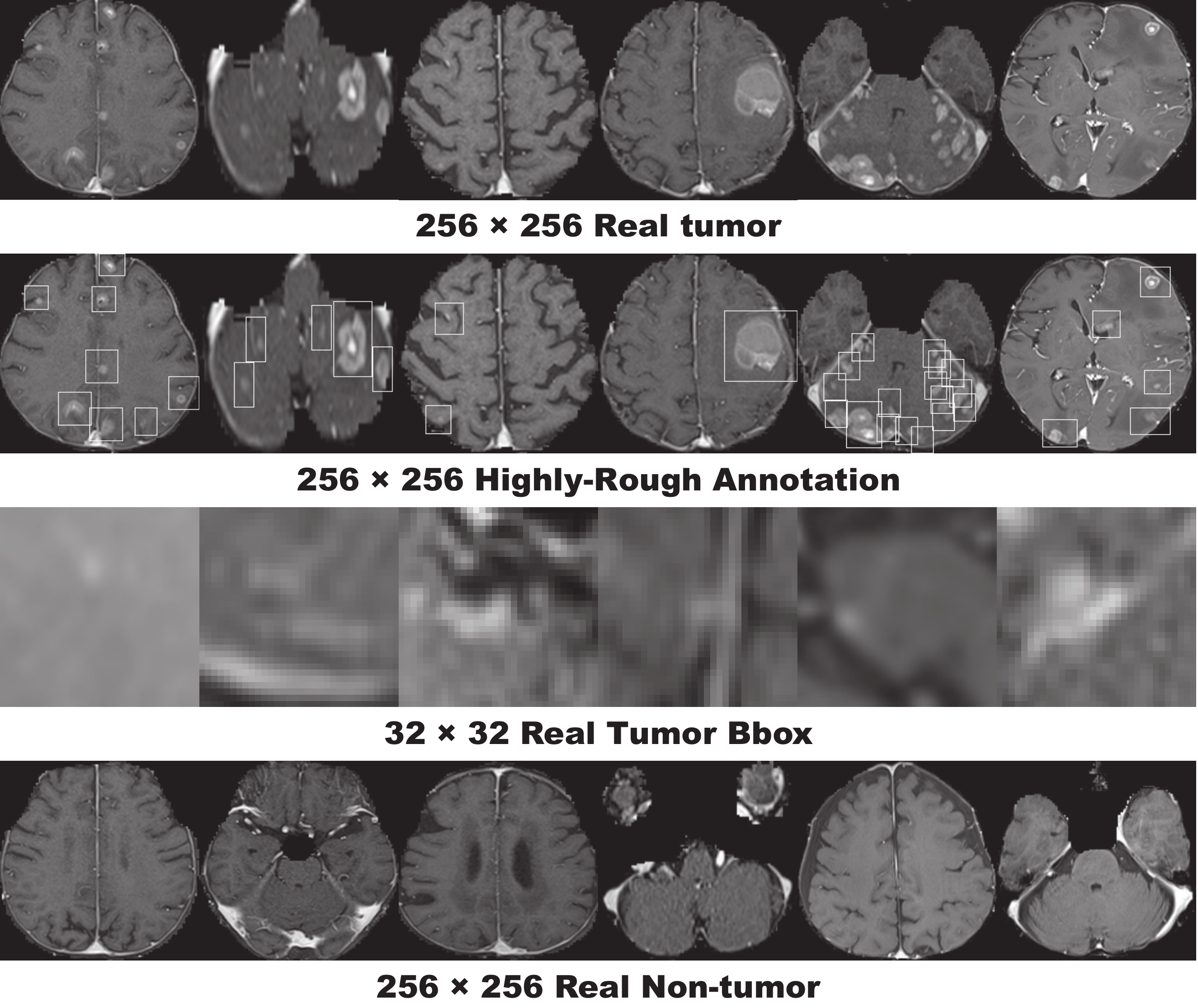}}
\caption{Example real $256 \times 256$ MR images with highly-rough annotation used for GAN training and resized $32 \times 32$ tumor bounding boxes.}
\label{fig6_2}
\end{figure}

Our training set is relatively small/fragmented for CNN-based applications, considering that the same patient's tumor slices could convey very similar information. To confirm the effect of realism and diversity---provided by combining PGGANs and bounding box conditioning---on tumor detection, we compare the following GANs: (\textit{i}) CPGGANs trained only with the brain metastases images; (\textit{ii}) CPGGANs trained also with additional $16,962$ brain images from $193$ normal subjects; (\textit{iii}) Image-to-image GAN trained only with the brain metastases images. After skull-stripping on all images with various resolution, remaining brain parts are cropped and resized to $256 \times 256$ pixels (i.e., a power of $2$ for better GAN training). As Fig.~\ref{fig6_2} shows, we lazily annotate tumors with highly-rough and inconsistent bounding boxes to minimize expert physicians' labor.\\

\noindent \textbf{CPGGANs} is a novel conditional noise-to-image training method for GANs, incorporating highly-rough bounding box conditions incrementally into PGGANs~\cite{Karras}, unlike conditional image-to-image GANs requiring rigorous segmentation masks~\cite{bailo2019red}. The original PGGANs exploits a progressively growing generator and discriminator: starting from low-resolution, newly-added layers model fine-grained details as training progresses. As Fig.~\ref{fig6_3} shows, we further condition the generator and discriminator to generate realistic and diverse $256 \times 256$ brain MR images with tumors of random shape at desired positions/sizes using only bounding boxes without an input benign image under limited training data/highly-rough annotation. Our modifications to the original PGGANs are as follows:

\begin{figure}[t!]
  \centering
  \centerline{\includegraphics[width=\linewidth]{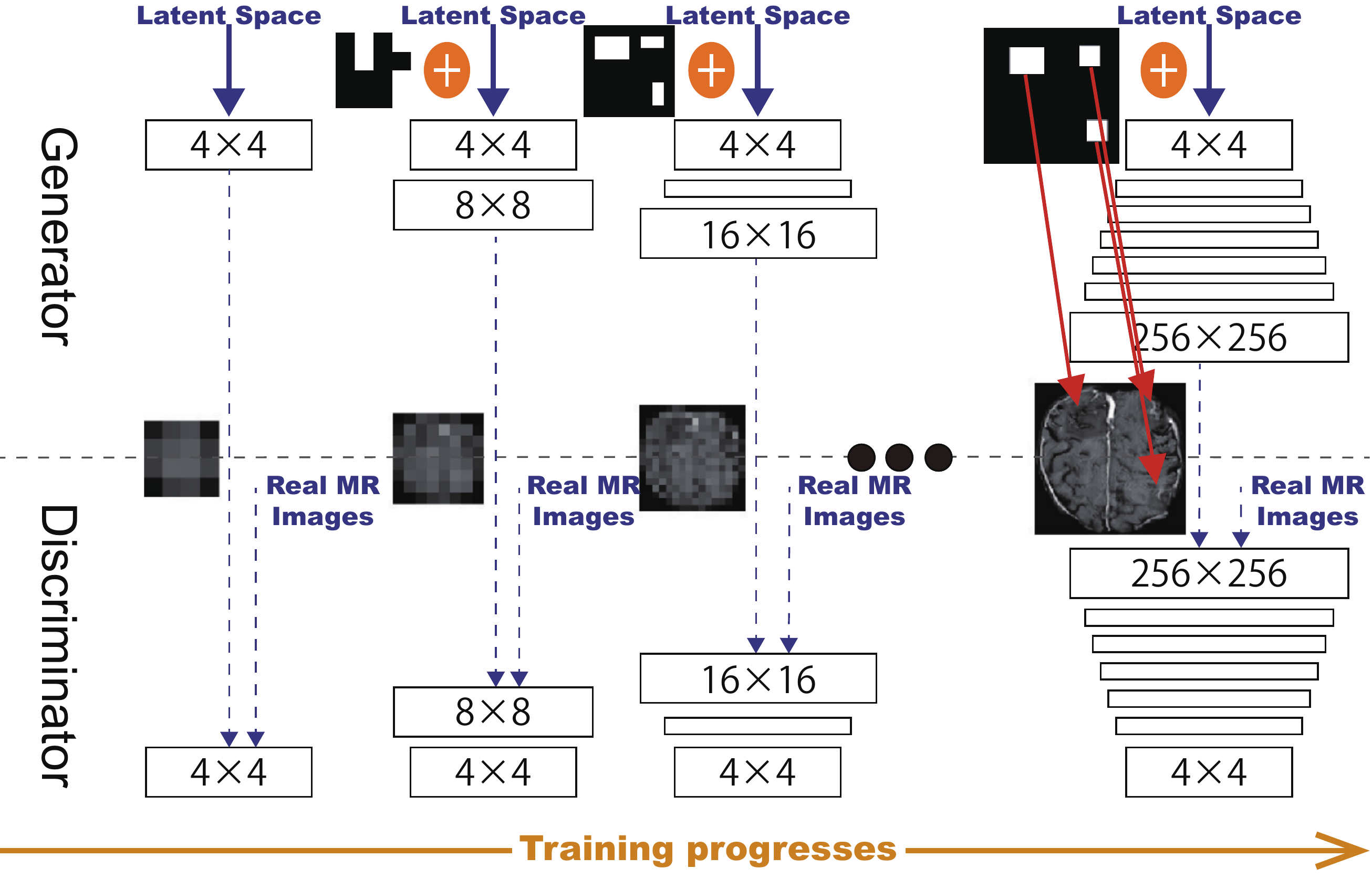}}
\caption{Proposed CPGGAN architecture for synthetic $256 \times 256$ brain MR image generation with tumors
at desired positions/sizes based on bounding boxes.}
\label{fig6_3}
\end{figure}

\begin{itemize}
\item Conditioning image: prepare a $256 \times 256$ black image (i.e., pixel value: $0$) with white bounding boxes (i.e., pixel value: $255$) describing tumor positions/sizes for attention;
\item Generator input: resize the conditioning image to the previous generator's output resolution/channel size and concatenate them (noise samples generate the first $4 \times 4$ images);
\item Discriminator input: concatenate the conditioning image with a real or synthetic image.
\end{itemize}

\noindent \textbf{CPGGAN Implementation Details}
We use the CPGGAN architecture with the WGAN-GP loss~\cite{Gulrajani}:
\begin{eqnarray}\label{eq:wgan_gp}
\underset{{\tilde{y}\sim{\mathbb{P}_g}}}{\mathbb{E}}[D(\tilde{y})]-\underset{{y\sim{\mathbb{P}_r}}}{\mathbb{E}}[D(y)] +
\lambda\underset{{{\hat{y}}\sim{\mathbb{P}_{\hat{y}}}}} {\mathbb{E}}[(\left \| \nabla_{\hat{y}}{D({\hat{y}})} \right \|_2-1)^2]
\end{eqnarray}
where the discriminator $D$ belongs to the set of $1$-Lipschitz functions, $\mathbb{P}_r$ is the data distribution by the true data sample ${y}$, and $\mathbb{P}_g$ is the model distribution by the synthetic sample ${\tilde{y}}$ generated from the conditioning image uniform noise samples in $[-1, 1]$. The last term is gradient penalty for the random sample ${\hat{y}}\sim{\mathbb{P}_{\hat{y}}}$. 

Training lasts for $3,000,000$ steps with a batch size of $4$ and $2.0 \times 10^{-4}$ learning rate for the Adam optimizer \cite{Kingma2015}. We flip the discriminator's real/synthetic labels once in three times for robustness. During testing, as tumor attention images, we use the annotation of training images with a random combination of horizontal/vertical flipping, width/height shift up to $10\%$, and zooming up to $10\%$; these CPGGAN-generated images are used as additional training images for tumor detection. \\

\noindent \textbf{Image-to-image GAN} is a conventional conditional GAN that generates brain MR images with tumors, concatenating a $256 \times 256$ conditioning image with noise samples for a generator input and concatenating the conditioning image with a real/synthetic image for a discriminator input, respectively. It uses a U-Net-like~\cite{Ronneberger} generator with $4$ convolutional/deconvolutional layers in encoders/decoders respectively with skip connections, along with a discriminator with $3$ decoders. We apply batch normalization~\cite{ioffe2015batch} to both convolution with LeakyReLU and deconvolution with ReLU. It follows the same implementation details as for the CPGGANs.

\subsection{YOLOv3-based Brain Metastases Detection}
\noindent \textbf{You Only Look Once v3 (YOLOv3)}~\cite{DBLP:journals/corr/abs-1804-02767} is a fast/accurate CNN-based object detector: unlike conventional classifier-based detectors, it divides the image into regions and predicts bounding boxes/probabilities for each region. We adopt YOLOv3 to detect brain metastases since its high efficiency can play a clinical role in real-time tumor alert; moreover, it shows very comparable results with $608 \times 608$ network resolution against other state-of-the-art detectors, such as Faster RCNN~\cite{ren2015faster}.

To confirm the effect of GAN-based DA, the following detection results are compared: (\textit{i}) $2,813$ real images without DA, (\textit{ii}), (\textit{iii}), (\textit{iv}) with  $4,000$/$8,000$/$12,000$ CPGGAN-based DA, (\textit{v}), (\textit{vi}), (\textit{vii}) with  $4,000$/$8,000$/$12,000$ CPGGAN-based DA, trained with additional normal brain images, (\textit{viii}), (\textit{ix}), (\textit{x}) with  $4,000$/$8,000$/$12,000$ image-to-image GAN-based DA. Due to the risk of overlooking the diagnosis $via$ medical imaging, higher sensitivity matters more than less FPs; thus, we aim to achieve higher sensitivity with a clinically acceptable number of FPs, adding the additional synthetic training images. Since our annotation is highly-rough, we calculate sensitivity/FPs per slice with both IoU threshold 0.5 and 0.25.\\

\noindent \textbf{YOLOv3 Implementation Details}
We use the YOLOv3 architecture with Darknet-53 as a backbone classifier and sum squared error between the predictions/ground truth as a loss:

\begin{align}
&\lambda_\text{coord} \sum_{i=0}^{S^2}\sum_{j=0}^B \mathbbm{1}_{ij}^\text{obj}\left[(x_i-\hat{x}_i)^2 + (y_i-\hat{y}_i)^2 \right] \nonumber\\&+ \lambda_\text{coord} \sum_{i=0}^{S^2}\sum_{j=0}^B \mathbbm{1}_{ij}^\text{obj}\left[\left(\sqrt{w_i}-\sqrt{\hat{w}_i}\right)^2 +\left(\sqrt{h_i}-\sqrt{\hat{h}_i}\right)^2 \right]\nonumber\\
&+ \sum_{i=0}^{S^2}\sum_{j=0}^B \mathbbm{1}_{ij}^\text{obj}(C_i - \hat{C}_i)^2 + \lambda_\text{noobj}\sum_{i=0}^{S^2}\sum_{j=0}^B \mathbbm{1}_{ij}^\text{noobj}(C_i - \hat{C}_i)^2 \nonumber\\
&+ \sum_{i=0}^{S^2} \mathbbm{1}_{i}^\text{obj}\sum_{c \in \text{classes}}(p_i(c) - \hat{p}_i(c))^2
\end{align}
where $x_i, y_i$ are the centroid location of an anchor box, $w_i, h_i$ are the width/height of the anchor, $C_i$ is the $\text{Objectness}$ (i.e., confidence score of whether an object exists), and $p_i(c)$ is the classification loss. Let $S^2$ and $B$ be the size of a feature map and  the number of anchor boxes, respectively. $\mathbbm{1}_{i}^\text{obj}$ is $1$ when an object exists in cell $i$ and otherwise $0$.

During training, we use a batch size of $64$ and $1.0 \times 10^{-3}$ learning rate for the Adam optimizer. The network resolution is set to $416 \times 416$ pixels during training and $608 \times 608$ pixels during validation/testing respectively to detect small tumors better. We recalculate the anchors at each DA setup. As classic DA, geometric/intensity transformations are also applied to both real/synthetic images during training to achieve the best performance. For testing, we pick the model with the best sensitivity on validation with detection threshold 0.1\%/IoU threshold 0.5 between $96,000$-$240,000$ steps to avoid severe FPs while achieving high sensitivity.

\subsection{Clinical Validation \textit{via} Visual Turing Test}
To quantitatively evaluate how realistic the CPGGAN-based synthetic images are, we supply, in random order, to three expert physicians a random selection of $50$ real and $50$ synthetic brain metastases images. They take four tests in ascending order: (\textit{i}), (\textit{ii}) test 1, 2: real \textit{vs} CPGGAN-generated resized $32 \times 32$ tumor bounding boxes, trained without/with additional normal brain images; (\textit{iii}), (\textit{iv}) test 3, 4: real \textit{vs} CPGGAN-generated $256 \times 256$ MR images, trained without/with additional normal brain images.

Then, the physicians constantly classify them as real/synthetic, if needed, zooming/rotating them, without previous training stages revealing which is real/synthetic.

\subsection{Visualization \textit{via} t-SNE}
To visually analyze the distribution of real/synthetic images, we use t-SNE~\cite{Maaten} on a random selection of:
\begin{itemize}
\item $500$ real tumor images;
\item $500$ CPGGAN-generated tumor images;
\item $500$ CPGGAN-generated tumor images, trained with additional normal brain images.
\end{itemize}
We normalize the input images to $[0, 1]$.


\noindent \textbf{t-SNE Implementation Details}
We use t-SNE with a perplexity of $100$ for $1,000$ iterations to get a 2D representation.

\section{Results}
This section shows how CPGGANs and image-to-image GAN generate brain MR images. The results include instances of synthetic images and their influence on tumor detection, along with CPGGAN-generated images' evaluation \textit{via} Visual Turing Test and t-SNE. 
\subsection{MR Images Generated by CPGGANs}

\begin{figure}[t!]
  \centering
  \centerline{\includegraphics[width=\linewidth]{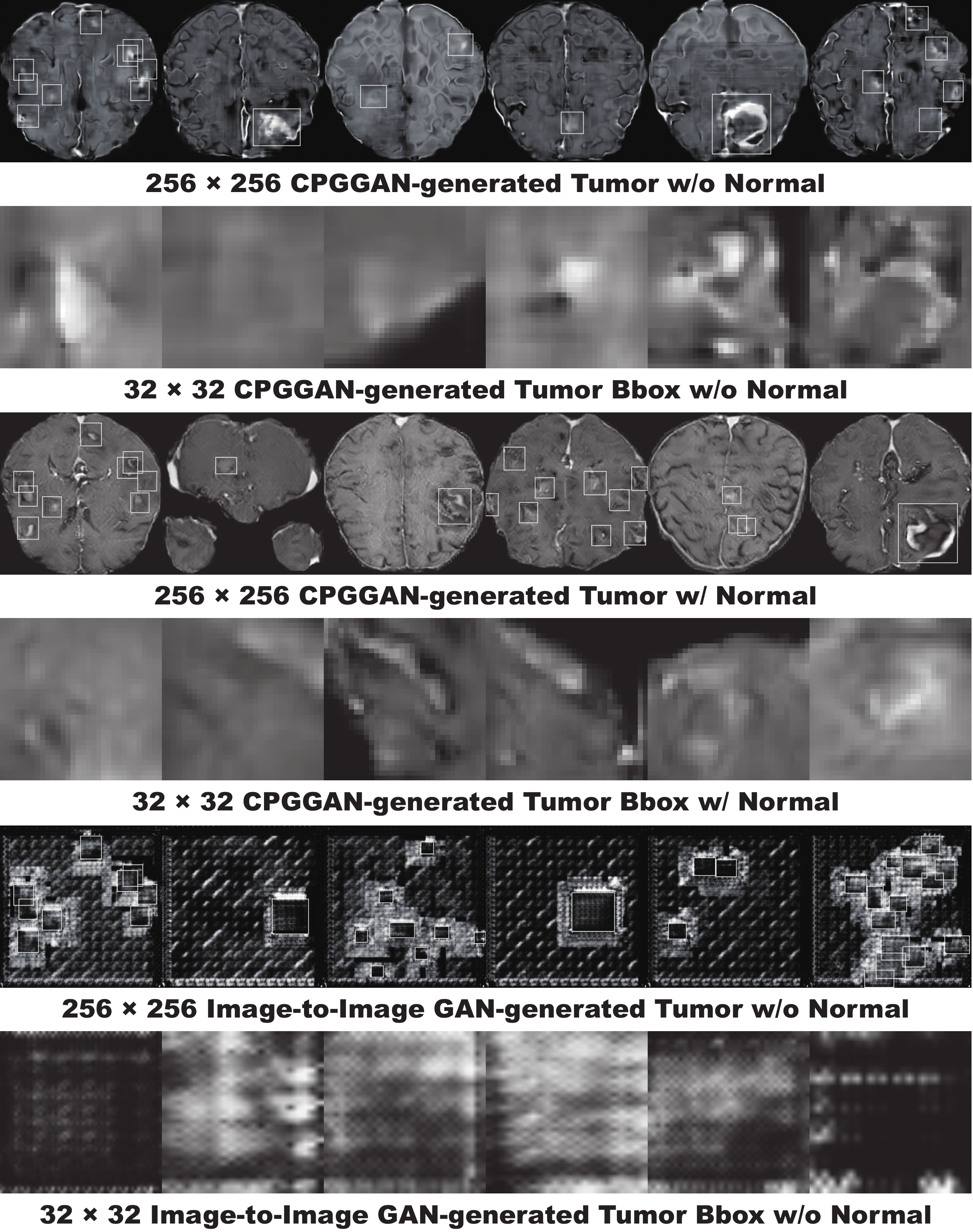}}
\caption[Example CPGGAN-generated $256 \times 256$ MR images and resized $32 \times 32$ tumor bounding boxes.]{Example synthetic $256 \times 256$ MR images and resized $32 \times 32$ tumor bounding boxes yielded by (a), (b) CPGGANs trained without/with additional normal brain images; (c) image-to-image GAN trained without normal images.}
\label{fig6_4}
\end{figure}

Fig.~\ref{fig6_4} illustrates example GAN-generated images. CPGGANs successfully captures the T1c-specific texture and tumor appearance at desired positions/sizes. Since we use highly-rough bounding boxes, the synthetic tumor shape largely varies within the boxes. When trained with additional normal brain images, it clearly maintains the realism of the original images with less odd artifacts, including tumor bounding boxes, which the additional images do not include. However, as expected, image-to-image GAN, without progressive growing, generates clearly unrealistic images without an input benign image due to the limited training data/rough annotation.

\subsection{Brain Metastases Detection Results}
Table~\ref{tab6_1} shows the tumor detection results with/without GAN-based DA. As expected, the sensitivity remarkably increases with the additional synthetic training data while FPs per slice also increase. Adding more synthetic images generally leads to a higher amount of FPs, also detecting blood vessels that are small/hyper-intense on T1c MR images, very similarly to the enhanced tumor regions (i.e., the contrast agent perfuses throughout the blood vessels). However, surprisingly, adding only $4,000$ CPGGAN-generated images achieves the best sensitivity improvement by $0.10$ with IoU threshold $0.5$ and by $0.08$ with IoU threshold $0.25$, probably due to the real/synthetic training image balance---the improved training robustness achieves sensitivity $0.91$ with moderate IoU threshold $0.25$ despite our highly-rough bounding box annotation.

\begin{table}[t!]
\caption[Bounding box-based YOLOv3 brain metastases detection results of ten DA setups.]{Bounding box-based YOLOv3 brain metastases detection results of ten DA setups (with detection threshold 0.1\%).}
\label{tab6_1}
\centering
\scalebox{0.71}{
\begin{tabular}{lrrrr}
\Hline\noalign{\smallskip}
 & \multicolumn{2}{c}{IoU $\geq$ 0.5}	& \multicolumn{2}{c}{IoU $\geq$ 0.25}\\
\bfseries  & {\bfseries Sensitivity (\%)} & \bfseries FPs per slice & {\bfseries  Sensitivity (\%)}  & \bfseries FPs per slice \\\noalign{\smallskip}\hline\noalign{\smallskip}
2,813 real images & 67 & 4.11 & 83 & 3.59\\
\noalign{\smallskip}\hline\noalign{\smallskip}
+ 4,000 CPGGAN-based DA & \textbf{77} & 7.64 & \textbf{91} & 7.18\\
+ 8,000 CPGGAN-based DA & 71 & 6.36 &87 & 5.85\\
+ 12,000 CPGGAN-based DA & 76 & 11.77 & \textbf{91} & 11.29\\
\noalign{\smallskip}\hline\noalign{\smallskip}
+ 4,000 CPGGAN-based DA (+ normal) & 69 & 7.16 & 86 & 6.60\\
+ 8,000 CPGGAN-based DA (+ normal) & 73 & 8.10 & 89 & 7.59\\
+ 12,000 CPGGAN-based DA (+ normal) & 74 & 9.42 & 89 & 8.95\\
\noalign{\smallskip}\hline\noalign{\smallskip}
+ 4,000 Image-to-Image GAN-based DA & 72 & 6.21 & 87 & 5.70\\
+ 8,000 Image-to-Image GAN-based DA & 68 & \textbf{3.50} & 84 & \textbf{2.99}\\
+ 12,000 Image-to-Image GAN-based DA & 74 & 7.20 & 89 & 6.72\\
\noalign{\smallskip}\Hline\noalign{\smallskip}
\end{tabular}
}
\end{table}

\begin{figure*}[t!]
  \centering
  \centerline{\includegraphics[width=\linewidth]{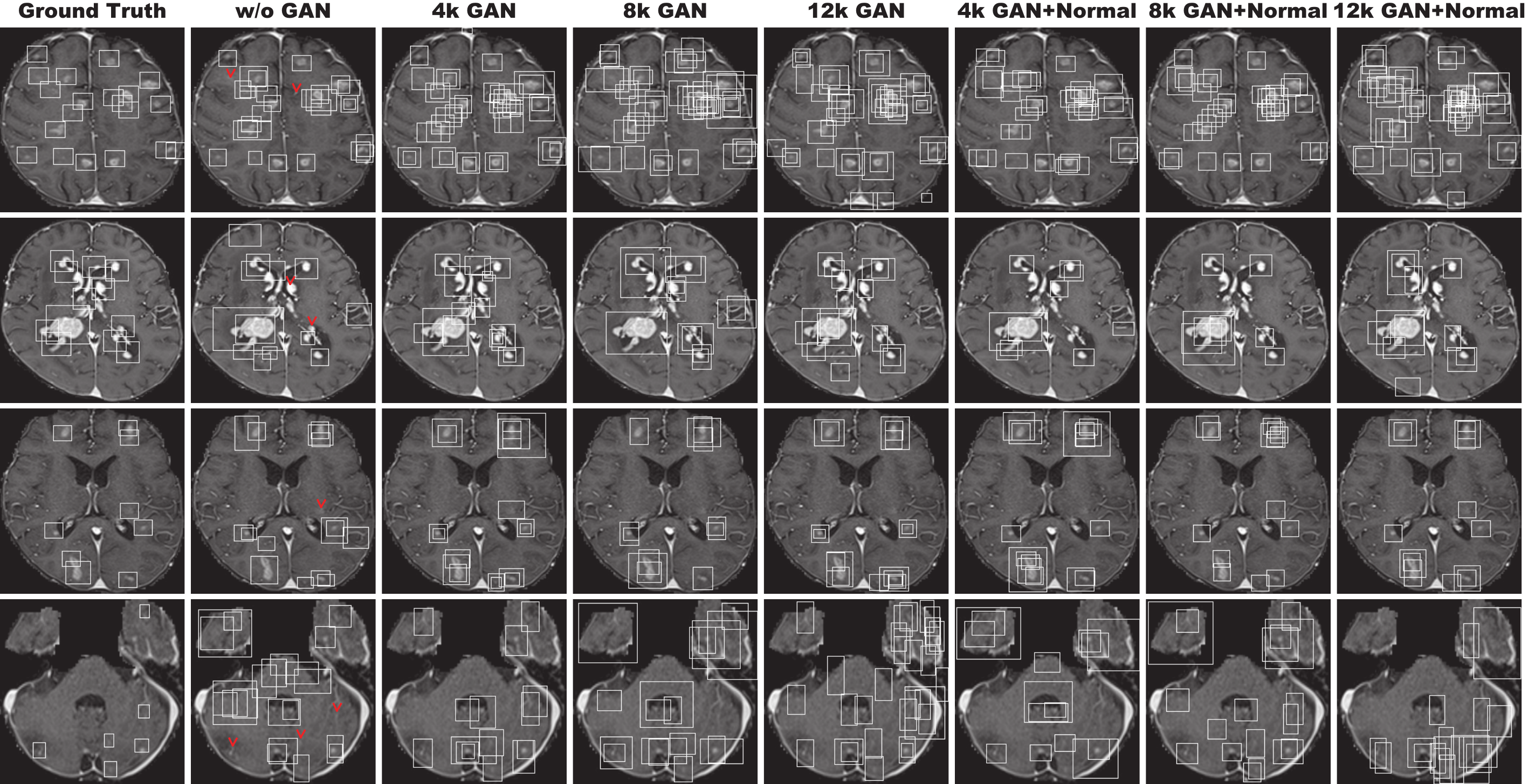}}
\caption[Example detection results of seven DA setups on four different images, compared against the ground truth.]{Example detection results of seven DA setups on four different images, compared against the ground truth: (a) ground truth; (b) without CPGGAN-based DA; (c), (d), (e) with $4\text{k}$/$8\text{k}$/$12\text{k}$ CPGGAN-based DA; (f), (g), (h) with $4\text{k}$/$8\text{k}$/$12\text{k}$ CPGGAN-based DA, trained with additional normal brain images. Red V symbols indicate the brain metastases undetected without CPGGAN-based DA, but detected with $4$k CPGGAN-based DA.}
\label{fig6_5}
\end{figure*}

Fig.~\ref{fig6_5} also visually indicates that it can alleviate the risk of overlooking the tumor diagnosis with clinically acceptable FPs; in the clinical routine, the bounding boxes, highly-overlapping around tumors, only require a physician's single check by switching on/off transparent alpha-blended annotation on MR images. It should be noted that we cannot increase FPs to achieve such high sensitivity without CPGGAN-based DA. Moreover, our results reveal that further realism---associated with the additional normal brain images during training---does not contribute to detection performance, possibly as the training focuses less on tumor generation. Image-to-image GAN-based DA just moderately facilitates detection with less additional FPs, probably because the synthetic images have a distribution far from the real ones and thus their influence on detection is limited during testing.

\begin{table*}[t!]
\caption[Visual Turing Test results by three physicians for classifying 50 real \textit{vs} 50 CPGGAN-generated images.]{Visual Turing Test results by three physicians for classifying real \textit{vs} CPGGAN-generated images: (a), (b) Test 1, 2: resized $32 \times 32$ tumor bounding boxes, trained without/with additional normal brain images; (c), (d) Test 3, 4: $256 \times 256$ MR images, trained without/with normal brain images. Accuracy denotes the physicians' successful classification ratio between the real/synthetic images.}
\label{tab6_2}
\centering
\scalebox{0.68}{
\begin{tabular}{p{0.2em}lrrrrr}
\Hline\noalign{\smallskip}
& \bfseries  & \multicolumn{1}{c}{\bfseries Accuracy (\%)} & \bfseries Real as Real (\%) & \bfseries Real as Synt (\%) & \bfseries Synt as Real (\%) & \bfseries Synt as Synt (\%) \\\noalign{\smallskip}\hline\noalign{\smallskip}
\parbox[t]{2mm}{\multirow{3}{*}{\rotatebox[origin=c]{270}{\textbf{\shortstack{\\Test 1}}}}} & Physician 1 & 88 & 80 & 20 & 4 & 96\\
& Physician 2 & 95 & 90 & 10 & 0 & 100\\
& Physician 3 & 97 & 98 & 2 & 4 & 96\\
\noalign{\smallskip}\hline\noalign{\smallskip}
\parbox[t]{2mm}{\multirow{3}{*}{\rotatebox[origin=c]{270}{\textbf{\shortstack{\\Test 2}}}}} & Physician 1 & 81 & 78 & 22 & 16 & 84\\
& Physician 2 & 83 & 86 & 14 & 20 & 80\\
& Physician 3 & 91 & 90 & 10 & 8 & 92\\
\noalign{\smallskip}\Hline\noalign{\smallskip}
\parbox[t]{2mm}{\multirow{3}{*}{\rotatebox[origin=c]{270}{\textbf{\shortstack{\\Test 3}}}}} & Physician 1 & 97 & 94 & 6 & 0 & 100\\
& Physician 2 & 96 & 92 & 8 & 0 & 100\\
& Physician 3 & 100 & 100 & 0 & 0 & 100\\
\noalign{\smallskip}\hline\noalign{\smallskip}
\parbox[t]{2mm}{\multirow{3}{*}{\rotatebox[origin=c]{270}{\textbf{\shortstack{\\Test 4}}}}} & Physician 1 & 91 & 82 & 18 & 0 & 100\\
& Physician 2 & 96 & 96 & 4 & 4 & 96\\
& Physician 3 & 100 & 100 & 0 & 0 & 100\\
\noalign{\smallskip}\Hline\noalign{\smallskip}
\end{tabular}
}
\end{table*}

\subsection{Visual Turing Test Results}

Table~\ref{tab6_2} shows the confusion matrix for the Visual Turing Test. The expert physicians easily recognize $256 \times 256$ synthetic images due to the lack of training data. However, when CPGGANs is trained with additional normal brain images, the experts classify a considerable number of synthetic tumor bounding boxes as real; it implies that the additional normal images remarkably facilitate the realism of both healthy and pathological brain parts while they do not include abnormality; thus, CPGGANs might perform as a tool to train medical students and radiology trainees when enough medical images are unavailable, such as abnormalities at rare positions/sizes. Such GAN applications are clinically prospective~\cite{finlayson2018towards}, considering the expert physicians' positive comments about the tumor realism.

\begin{figure}[t!]
  \centering
  \centerline{\includegraphics[width=\linewidth]{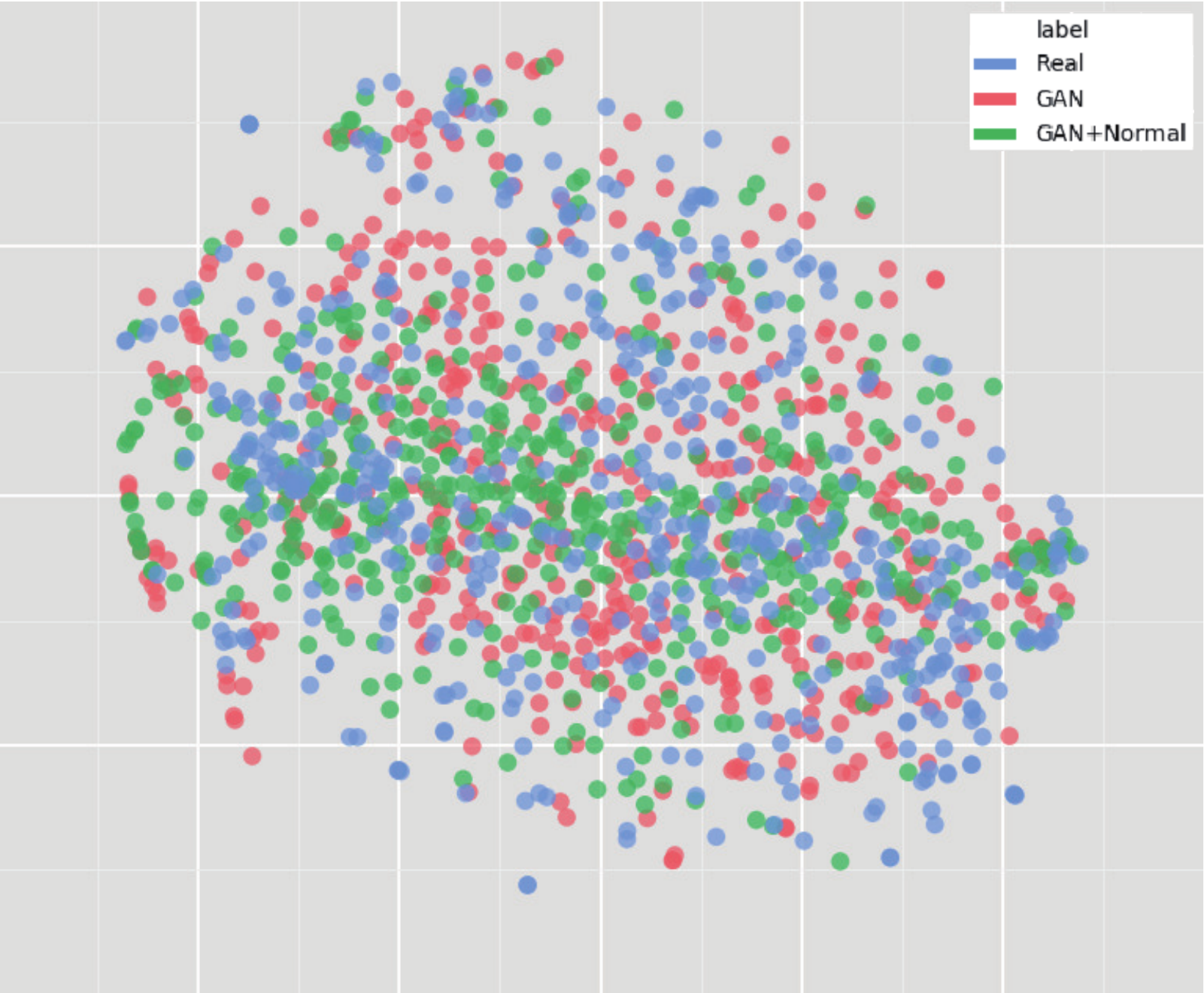}}
\caption[T-SNE plot with $500$ $32 \times 32$ resized tumor bounding box images per each category.]{T-SNE results with $500$ $32 \times 32$ resized tumor bounding box images per each category: (a) Real tumor images; (b), (c) CPGGAN-generated tumor images, trained without/with additional normal brain images.}
\label{fig6_6}
\end{figure}

\begin{figure}[t!]
  \centering
  \centerline{\includegraphics[width=\linewidth]{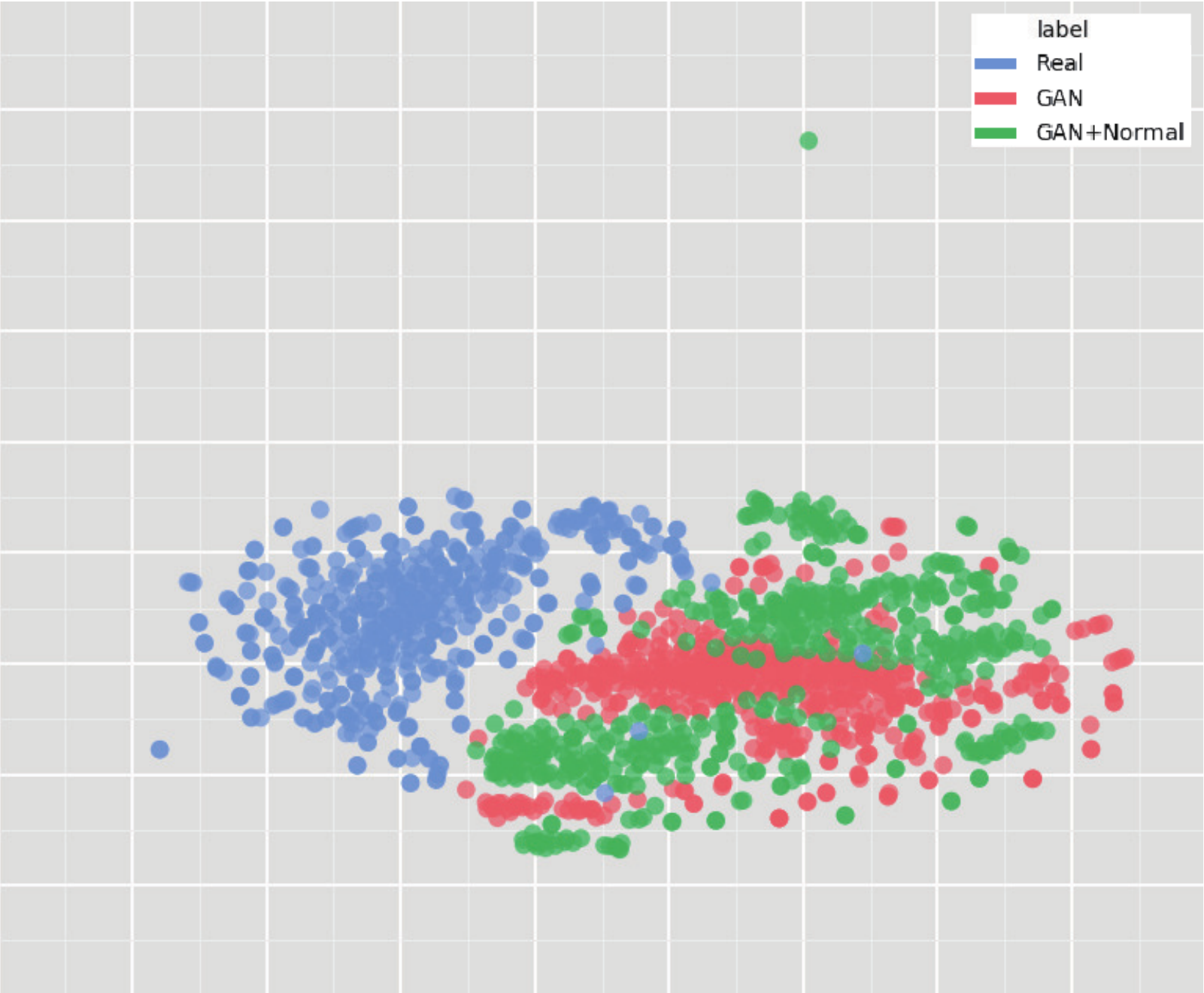}}
\caption[T-SNE plot with $500$ $256 \times 256$ images per each category.]{T-SNE results with $500$ $256 \times 256$ images per each category: (a) Real tumor images; (b), (c) CPGGAN-generated tumor images, trained without/with additional normal brain images.}
\label{fig6_7}
\end{figure}

\subsection{T-SNE Results}
As presented in Fig.~\ref{fig6_6}, synthetic tumor bounding boxes have a moderately similar distribution to real ones, but they also fill the real image distribution uncovered by the original dataset, implying their effective DA performance; especially, the CPGGAN-generated images trained without normal brain images distribute wider than the center-concentrating images trained with the normal brain images. Meanwhile, real/synthetic whole brain images clearly distribute differently, due to the real MR images' strong anatomical consistency (Fig.~\ref{fig6_7}). Considering the achieved high DA performance, the tumor (i.e., ROI) realism/diversity matter more than the whole image realism/diversity, since YOLOv3 look at an image patch instead of a whole image, similarly to most other CNN-based object detectors.

\section{Conclusion}
Without relying on an input benign image, our CPGGANs can generate realistic and diverse $256 \times 256$ MR images with brain metastases of random shape, unlike rigorous segmentation, naturally at desired positions/sizes, and achieve high sensitivity in tumor detection---even with small/fragmented training data from multiple MRI scanners and lazy annotation using highly-rough bounding boxes; in the context of intelligent data wrangling, this attributes to the CPGGANs' good generalization ability to incrementally synthesize conditional whole images with the real image distribution unfilled by the original dataset, improving the training robustness.

We confirm that the realism and diversity of the generated images, judged by three expert physicians $via$ Visual Turing Test, do not imply better detection performance; as the t-SNE results show, the CPGGAN-generated images, trained with additional non-tumor normal images, lack diversity probably because the training less focuses on tumors. Moreover, we notice that adding over-sufficient synthetic images leads to more FPs, but not always higher sensitivity, possibly due to the training data imbalance between real and synthetic images; as the t-SNE results reveal, the CPGGAN-generated tumor bonding boxes have a moderately similar---mutually complementary---distribution to the real ones; thus, GAN-overwhelming training images may decrease the necessary influence of the real samples and harm training, rather than providing robustness. Lastly, image-to-image GAN-based DA just moderately facilitates detection with less additional FPs, probably due to the lack of realism. However, further investigations are needed to maximize the effect of the CPGGAN-based medical image augmentation.


For example, we could verify the effect of further realism in return for less diversity by combining $\ell _1$ loss with the WGAN-GP loss for GAN training. We can also combine those CPGGAN-generated images, trained without/with additional brain images, similarly to ensemble learning~\cite{dietterich2002ensemble}. Lastly, we plan to define a new GAN loss function that directly optimizes the detection results, instead of realism, similarly to the three-player GAN for optimizing classification results~\cite{vandenhende2019three}.

Overall, minimizing expert physicians' annotation efforts, our novel CPGGAN-based DA approach sheds light on diagnostic and prognostic medical applications, not limited to brain metastases detection; future studies, especially on 3D bounding box detection with highly-rough annotation, are required to extend our promising results. Along with the DA, the CPGGANs has other potential clinical applications in oncology: (\textit{i}) A data anonymization tool to share patients' data outside their institution for training while preserving detection performance. Such a GAN-based application is reported in Shin~\textit{et al.}~\cite{shin2018medical}; (\textit{ii}) A physician training tool to display random synthetic medical images
with abnormalities at both common and rare positions/sizes, by training CPGGANs on highly unbalanced medical datasets (i.e., limited pathological and abundant normal samples,
respectively). It can help train medical students and radiology trainees despite infrastructural and legal constraints~\cite{finlayson2018towards}.
\chapter{\LARGE GAN-based Medical Image Augmentation for 3D Detection}

\section{Prologue to Fourth Project}
\subsection{Project Publication}
\begin{itemize}
\item \textbf{Synthesizing Diverse Lung Nodules Wherever Massively: 3D Multi-Conditional GAN-based CT Image Augmentation for Object Detection}. \textbf{C. Han}, Y. Kitamura, A. Kudo, A. Ichinose, L. Rundo, Y. Furukawa, K. Umemoto, H. Nakayama, Y. Li, In International Conference on 3D Vision (3DV), Qu\'ebec City, Canada, pp. 729--737, September 2019.
\end{itemize}

\subsection{Context}
Prior to this work, no researchers had tackled 3D GANs for general bounding box-based detection whereas 3D Medical Image Analysis can improve diagnosis by capturing anatomical and functional information. Jin \textit{et al.} had used an image-to-image GAN to generate $64 \times 64 \times 64$ CT images of lung nodules including the surrounding tissues by inputting a VOI centered at a lung nodule, but with a central sphere region erased~\cite{jin2018ct}; however, they had targeted annotation-expensive segmentation, instead of the detection, also translating both nodules/surroundings \textit{via} expensive computation. Without conditioning a noise-to-image GAN with nodule position, Gao \textit{et al.} had generated $40 \times  40 \times 18$ 3D nodule subvolumes only applicable to their subvolume-based detector using binary classification~\cite{gao2019augmenting}. Unfortunately, no research had focused on multiple GAN conditions for more versatile 3D GANs while lesions vary in position, size, and attenuation.

\subsection{Contributions}
This project's primary contribution is to propose a novel 3D pathology-aware multi-conditional GAN called 3D MCGAN for improved 3D bounding box-based detection in general; it translates noise boxes into realistic/diverse $32 \times 32 \times 32$ lung nodules placed naturally at desired position, size, and attenuation on CT scans---inputting the noise box with the surrounding tissues has the effect of combining the noise-to-image and image-to-image GANs. The $32 \times 32 \times 32$ nodule-only generation, not translating the $64 \times 64 \times 64$ surroundings, can decrease computational cost. By so doing, our 3D MCGAN-based DA boosts sensitivity in nodule detection under any nodule size/attenuation at fixed FP rates. Moreover, we find that GAN training with $\ell _1$ loss could increase synthetic images' realism, but decrease DA performance. Using proper augmentation ratio (i.e., $1 : 1$) could improve the DA performance. Considering the outstanding realism confirmed by physicians, it could perform as a physician training tool to display realistic medical images with desired abnormalities (i.e., position, size, and attenuation).

\subsection{Recent Developments}
According to their arXiv paper, Xu \textit{et al.} have generated realistic/diverse $64 \times 64 \times 64$ CT images of lung nodules combining the image-to-image GAN with gene expression profiles~\cite{xu2019correlation}.

\newpage

\section{Motivation}
Accurate CAD, thanks to recent CNNs, can alleviate the risk of overlooking the diagnosis in a clinical environment.
Such great success of CNNs, including diabetic eye disease diagnosis~\cite{gulshan2016development}, primarily derives from large-scale annotated training data to sufficiently cover the real data distribution. However, obtaining and annotating such diverse pathological images are laborious tasks; thus, the massive generation of proper synthetic training images matters for reliable diagnosis. Researchers usually use classical DA techniques, such as geometric/intensity transformations~\cite{Ronneberger,Milletari}. However, those one-to-one translated images have intrinsically similar appearance and cannot sufficiently cover the real image distribution, causing limited performance improvement; in this regard, thanks to their good generalization ability, GANs~\cite{Goodfellow} can generate realistic but completely new samples using many-to-many mappings for further performance improvement; GANs showed excellent DA performance in computer vision, including $21\%$ performance improvement in eye-gaze estimation~\cite{Shrivastava}.

This GAN-based DA trend especially applies to medical imaging, where the biggest problem lies in small and fragmented datasets from various scanners. For performance boost in various 2D medical imaging tasks, some researchers used noise-to-image GANs for classification~\cite{frid2018gan, han2019combining, Han2}; others used image-to-image GANs for object detection~\cite{han2019learning} and segmentation~\cite{bailo2019red}. However, although 3D imaging is spreading in radiology (e.g., CT and MRI), such 3D medical GAN-based DA approaches are limited, and mostly focus on segmentation~\cite{shin2018medical,jin2018ct}---3D medical image generation is more challenging  than 2D one due to expensive computational cost and strong anatomical consistency. Accordingly, no 3D conditional GAN-based DA approach exists for general bounding box-based 3D object detection, while it can locate disease areas with physicians' minimum annotation cost, unlike rigorous 3D segmentation. Moreover, since lesions vary in position/size/attenuation, further GAN-based DA performance requires multiple conditions.

\newpage

\begin{figure}[t!]
  \centering
  \centerline{\includegraphics[width=1\linewidth]{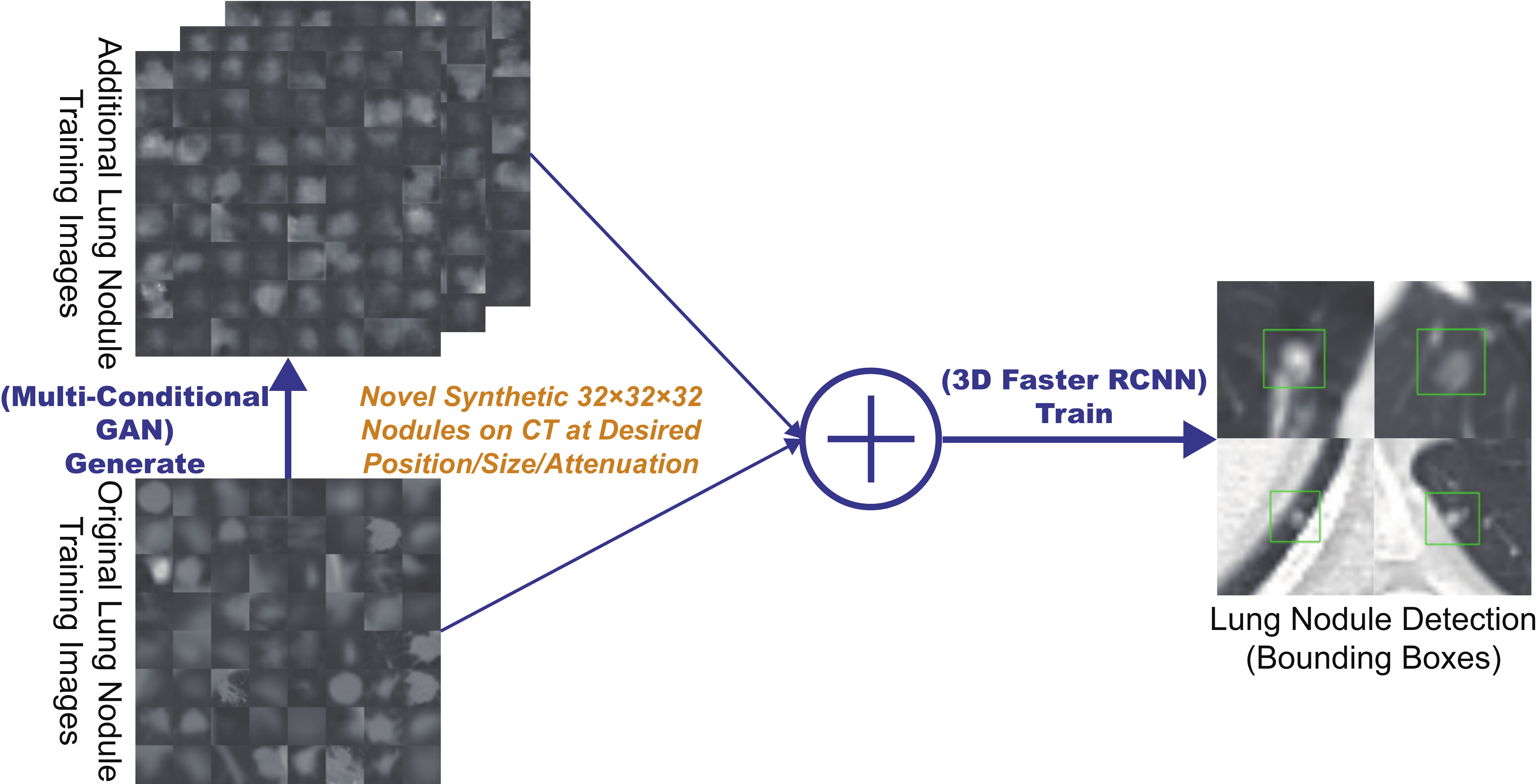}}
\caption[3D MCGAN-based DA for better nodule detection.]{3D MCGAN-based DA for better nodule detection: Our MCGAN generates realistic and diverse nodules naturally on lung CT scans at desired position, size, and attenuation based on bounding boxes, and the CNN-based object detector uses them as additional training data.}
\label{fig7_1}
\end{figure}

So, how can GAN generate realistic/diverse 3D nodules placed naturally on lung CT with multiple conditions to boost sensitivity in any 3D object detector? For accurate 3D CNN-based nodule detection (Fig.~\ref{fig7_1}), we propose 3D MCGAN to generate $32 \times 32 \times 32$ nodules---such nodule detection is clinically valuable for the early diagnosis/treatment of lung cancer, the deadliest cancer~\cite{siegel2019cancer}. Since nodules vary in position/size/attenuation, to improve CNN's robustness, we adopt two discriminators with different loss functions for conditioning: the context discriminator learns to classify real \textit{vs} synthetic nodule/surrounding pairs with noise box-centered surroundings; the nodule discriminator attempts to classify real \textit{vs} synthetic nodules with size and attenuation conditions. We also evaluate the synthetic images' realism \textit{via} Visual Turing Test~\cite{Salimans} by two expert physicians, and visualize the data distribution \textit{via} t-SNE~\cite{Maaten}. The 3D MCGAN-generated additional training images can achieve higher sensitivity under any nodule size/attenuation at fixed FP rates. Lastly, this study suggests training GANs without $\ell _1$ loss and using proper augmentation ratio (i.e., $1:1$) for better medical GAN-based DA performance.\\


\noindent \textbf{Research Questions.} We mainly address two questions:
\begin{itemize}
\item \textbf{3D Multiple GAN Conditioning:} How can we condition 3D GANs to naturally place objects of random shape, unlike rigorous segmentation, at desired position/size/attenuation based on bounding box masks?
\item \textbf{Synthetic Images for DA:} How can we set the number of real/synthetic training data and GAN loss functions to achieve the best detection performance?
\end{itemize}

\noindent \textbf{Contributions.} Our main contributions are as follows:
\begin{itemize}
\item \textbf{3D Multi-conditional Image Generation:} This first multi-conditional pathological image generation approach shows that 3D MCGAN can generate realistic and diverse nodules placed on lung CT at desired position/size/attenuation, which even expert physicians cannot distinguish from real ones.

\item \textbf{Misdiagnosis Prevention:} This first GAN-based DA method available for any 3D object detector allows to boost sensitivity at fixed FP rates in CAD with limited medical images/annotation.

\item \textbf{Medical GAN-based DA:} This study implies that training GANs without $\ell _1$ loss and using proper augmentation ratio (i.e., $1:1$) may boost CNN-based detection performance with higher sensitivity and less FPs in medical imaging.
\end{itemize}

\section{Materials and Methods}
\subsection{3D MCGAN-based Image Generation}
\noindent \textbf{Data Preparation}
This study exploits the Lung Image Database Consortium image collection (LIDC) dataset~\cite{armato2011lung} containing $1,018$ chest CT scans with lung nodules. Since the American College of Radiology recommends lung nodule evaluation using thin-slice CT scans~\cite{setio2017validation}, we only use scans with the slice thickness $\leq 3$ mm and $0.5$ mm $\leq$ in-plane pixel spacing $\leq 0.9$ mm. Then, we interpolate the slice thickness to $1.0$ mm and exclude scans with slice number $> 400$.

To explicitly provide MCGAN with meaningful nodule appearance information and thus boost DA performance, the authors further annotate those nodules by size and attenuation for GAN training with multiple conditions: small (slice thickness $\leq 10$ mm); medium ($10$ mm $\leq$ slice thickness $\leq 20$ mm); large (slice thickness $>$ $20$ mm); solid; part-solid; Ground-Glass Nodule (GGN). Afterwards, the remaining dataset ($745$ scans) is divided into: (\textit{i}) a training set ($632$ scans/$3,727$ nodules); (\textit{ii}) a validation set ($37$ scans/$143$ nodules); (\textit{iii}) a test set ($76$ scans/$265$ nodules); only the training set is used for MCGAN training to be methodologically sound. The training set contains more average nodules since we exclude patients with too many nodules for the validation/test sets; we arrange a clinical environment-like situation, where we could find more healthy patients than highly diseased ones to conduct anomaly detection.

\begin{figure}[t!]
  \centering
  \centerline{\includegraphics[width=1\linewidth]{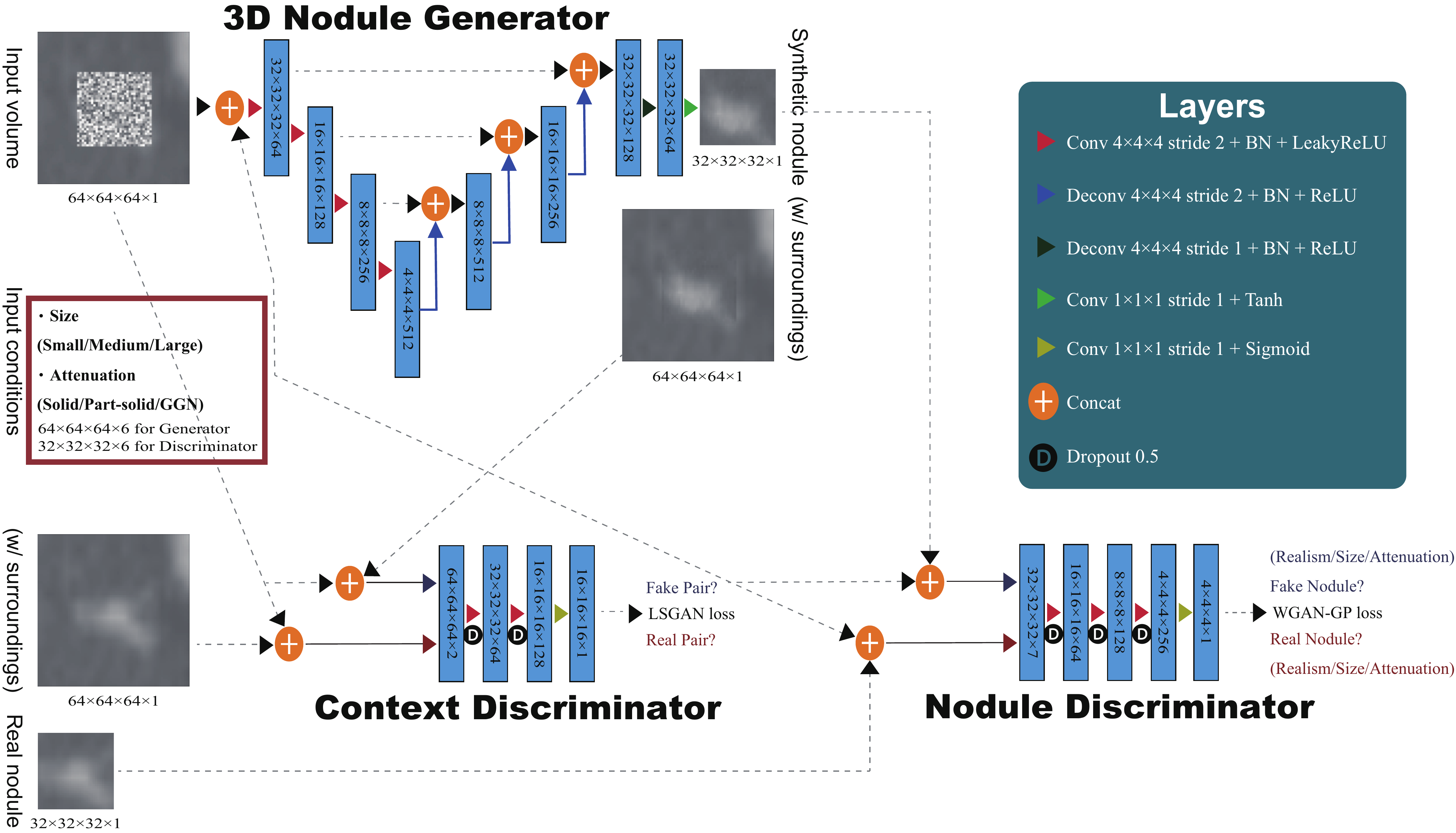}}
\caption[Proposed 3D MCGAN architecture for realistic/diverse $32 \times 32 \times 32$ lung CT scan of nodule generation.]{Proposed 3D MCGAN architecture for realistic/diverse $32 \times 32 \times 32$ lung CT scan of nodule generation: the context discriminator learns to classify real \textit{vs} synthetic nodule/surrounding pairs while the nodule discriminator learns to classify real \textit{vs} synthetic nodules.}
\label{fig7_2}
\end{figure}

\noindent \textbf{3D MCGAN} is a novel GAN training method for DA, generating realistic but new nodules at desired position/size/attenuation, naturally blending with surrounding tissues (Fig.~\ref{fig7_2}). We crop/resize various nodules to $32 \times 32 \times 32$ voxels and replace them with noise boxes from a uniform distribution between $[-0.5,0.5]$, while maintaining their $64 \times 64 \times 64$ surroundings as VOIs---using those noise boxes, instead of boxes filled with the same voxel values, improves the training robustness; then, we concatenate the VOIs with $6$ size/attenuation conditions tiled to $64 \times 64 \times 64$ voxels (e.g., if the size is small, each voxel of the small condition is filled with $1$, while the medium/large condition voxels are filled with $0$ to consider the effect of scaling factor). So, our generator uses the $64 \times 64 \times 64 \times 7$ inputs to generate desired nodules in the noise box regions. The 3D U-Net~\cite{cciccek20163d}-like generator adopts $4$ convolutional layers in encoders and $4$ deconvolutional layers in decoders respectively with skip connections to effectively capture both nodule/context information.

We adopt two \textit{Pix2Pix} GAN~\cite{isola2017image}-like discriminators with different loss functions: the context discriminator learns to classify real \textit{vs} synthetic nodule/surrounding pairs with noise box-centered surroundings using Least Squares loss (LSGANs)~\cite{mao2017least}; the nodule discriminator attempts to classify real \textit{vs} synthetic nodules with size/attenuation conditions using WGAN-GP~\cite{Gulrajani}. The LSGANs in the context discriminator forces the model to learn surrounding tissue background by reacting more sensitively to every pixel in images than regular GANs. The WGAN-GP in the nodule discriminator allows the model to generate realistic/diverse nodules without focusing too much on details. Empirically, we confirm that such multiple discriminators with the mutually complementary loss functions, along with size/attenuation conditioning, help generate realistic/diverse nodules naturally placed at desired positions on CT scans; similar results are also reported by this work~\cite{ouyang2018pedestrian} for 2D pedestrian detection without label conditioning. We apply dropout to inject randomness and balance the generator/discriminators. Batch normalization is applied to both convolution (using LeakyReLU) and deconvolution (using ReLU).


Most GAN-based DA approaches use reconstruction $\ell _1$ loss~\cite{gao2019augmenting} to generate realistic images, even modifying it for further realism~\cite{jin2018ct}. However, no one has ever validated whether it really helps DA---it assures synthetic images resembling the original ones, sacrificing diversity; thus, to confirm its influence during classifier training, we compare our MCGAN objective without/with it, respectively:
\begin{eqnarray}
	G^*&=&\arg\min_{G}\max_{D1, D2}\mathcal{L}_{\textbf{LSGANs}}(G,D1)\nonumber \\&+&\mathcal{L}_{\textbf{WGAN-GP}}(G,D2),\\
		G^*&=&\arg\min_{G}\max_{D1, D2}\mathcal{L}_{\textbf{LSGANs}}(G,D1)\nonumber \\&+&\mathcal{L}_{\textbf{WGAN-GP}}(G,D2) + 100 \mathcal{L}_{\ell_1}(G).
\end{eqnarray}
We set 100 as a weight for the $\ell _1$ loss, since empirically it works well for reducing visual artifacts introduced by the GAN loss and most GAN works adopt the weight~\cite{isola2017image, ouyang2018pedestrian}.

\noindent \textbf{3D MCGAN Implementation Details}
Training lasts for $6,000,000$ steps with a batch size of $16$ and $2.0 \times 10^{-4}$ learning rate for the Adam optimizer. We use horizontal/vertical flipping as DA and flip real/synthetic labels once in three times for robustness. During testing, we augment nodules with the same size/attenuation conditions by applying a random combination to real nodules of width/height/depth shift up to $10\%$ and zooming up to $10\%$ for better DA. As post-processing, we blend bounding boxes' $3$ nearest surfaces from all the boundaries by averaging the values of $6$ nearest voxels/itself for $5$ iterations. We resample the resulting nodules to their original resolution and map back onto the original CT scans to prepare additional training data.

\subsection{3D Faster RCNN-based Lung Nodule Detection}
\noindent \textbf{3D Faster RCNN} is a 3D version of Faster RCNN~\cite{ren2015faster} using multi-task loss with a $27$-layer Region Proposal Network of 3D convolutional/batch normalization/ReLU layers. To confirm the effect of MCGAN-based DA, we compare the following detection results trained on (\textit{i}) $632$ real images without GAN-based DA, (\textit{ii}), (\textit{iii}), (\textit{iv}) with $1\times$/$2\times$/$3\times$ MCGAN-based DA (i.e., $632$/$1,264$/$1,896$ additional synthetic training images) , (\textit{v}), (\textit{vi}), (\textit{vii}) with $1\times$/$2\times$/$3\times$ MCGAN-based DA trained with $\ell _1$ loss. During training, we shuffle the real/synthetic image order. We evaluate the detection performance as follows: (\textit{i}) Free Receiver Operation Characteristic (FROC) analysis, sensitivity as a function of FPs per scan; (\textit{ii}) Competition Performance
Metric (CPM) score~\cite{Niemeijer}, average sensitivity at seven pre-defined FP rates: 1/8, 1/4, 1/2, 1, 2, 4, and 8 FPs per scan---this quantifies if a CAD system can identify a significant percentage of nodules with both very few FPs and moderate FPs.



\noindent \textbf{3D Faster RCNN Implementation Details}
During training, we use a batch size of $2$ and $1.0 \times 10^{-3}$ learning rate ($1.0 \times 10^{-4}$ after $20,000$ steps) for the SGD optimizer with momentum. The input volume size to the network is set to $160 \times 176 \times 224$ voxels. As classical DA, a random combination of width/height/depth shift up to $15\%$ and zooming up to $15\%$ are also applied to both real/synthetic images to achieve the best performance. For testing, we pick the model with the highest sensitivity on validation between $30,000$-$40,000$ steps under IoU threshold $0.25$/detection threshold $0.5$ to avoid severe FPs.

\subsection{Clinical Validation \textit{via} Visual Turing Test}
To quantitatively evaluate the realism of MCGAN-generated images, we supply, in random order, to two expert physicians a random selection of $50$ real and $50$ synthetic lung nodule images with all of 2D axial/coronal/sagittal views at the center. They take four classification tests in ascending order: Test1, 2: real \textit{vs} MCGAN-generated $32 \times 32 \times 32$ nodules, trained without/with $\ell _1$ loss; Test3, 4: real \textit{vs} MCGAN-generated $64 \times 64 \times 64$ nodules with surroundings without/with $\ell _1$ loss.

\subsection{Visualization \textit{via} t-SNE}
To visually analyze the distribution of real/synthetic images, we use t-SNE~\cite{Maaten} on a random selection of $500$ real, $500$ synthetic, and $500$ $\ell _1$ loss-added synthetic nodule images, with a perplexity of $100$ for $1,000$ iterations to get a 2D representation. We normalize the input images to $[0, 1]$.

\begin{figure}[t!]
  \centering
  \centerline{\includegraphics[width=1\columnwidth]{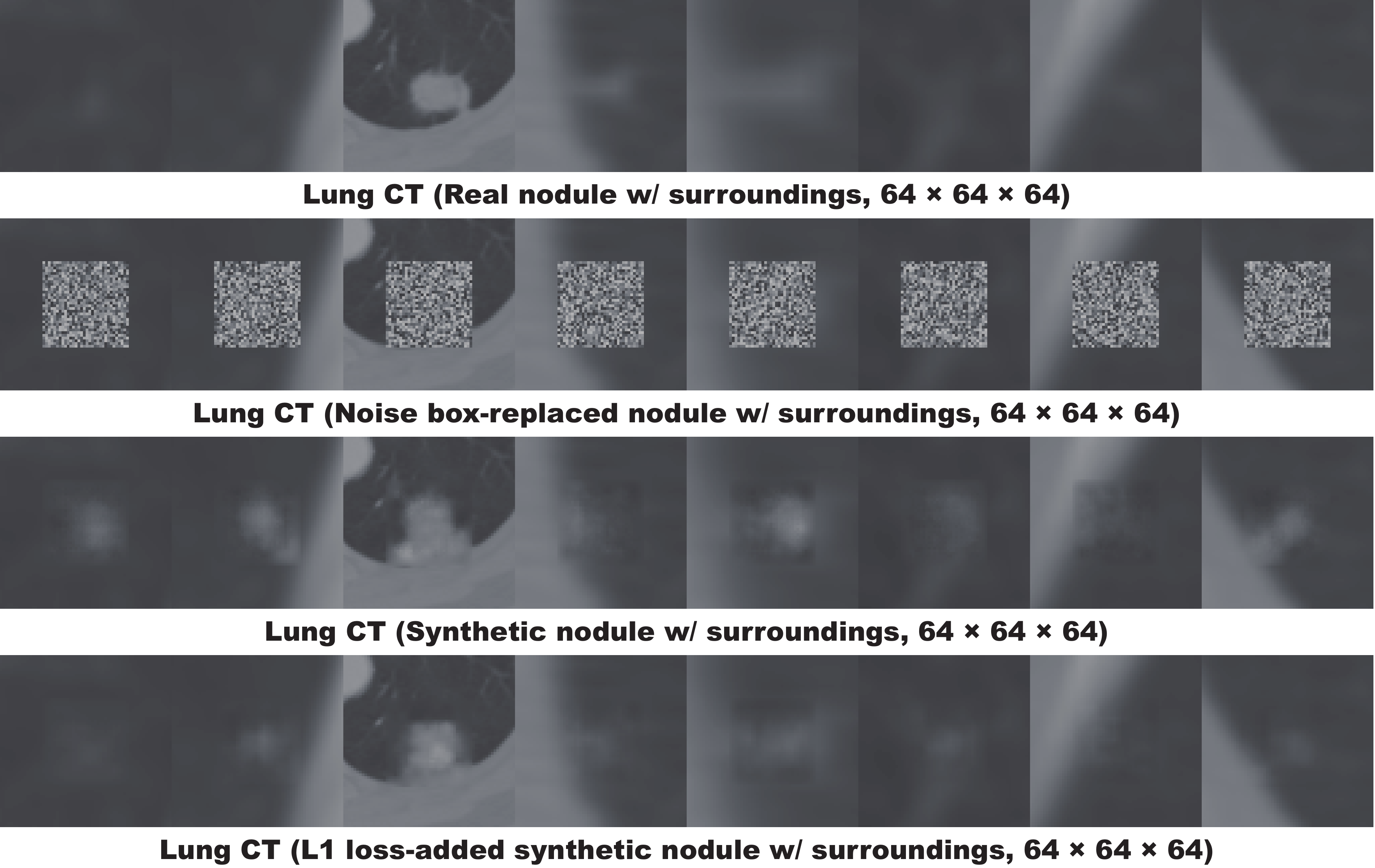}}
\caption[2D axial view of example real/synthetic $64 \times 64 \times 64$ CT scans of nodules with surrounding tissues.]{2D axial view of example real/synthetic $64 \times 64 \times 64$ CT scans of lung nodules with surrounding tissues; 3D MCGANs generate only $32 \times 32 \times 32$ nodules.}
\label{fig7_3}
\end{figure}

\section{Results}
\subsection{Lung Nodules Generated by 3D MCGAN}

We generate realistic nodules in noise box regions at various position/size/attenuation, naturally blending with surrounding tissues including vessels, soft tissues, and thoracic walls (Fig.~\ref{fig7_3}). Especially, when trained without $\ell _1$ loss, those synthetic nodules look clearly more different from the original real ones, including slight shading difference.

\begin{table}[t!]
\caption[3D Faster RCNN nodule detection results (CPM) of seven DA setups.]{3D Faster RCNN nodule detection results (CPM) of seven DA setups (IoU $\geq$ $0.25$). Both results without/with $\ell _1$ loss at different augmentation ratio are compared. CPM is average sensitivity at $1/8, 1/4, 1/2, 1, 2, 4,$ and $8$ FPs per scan.}
\label{tab7_1}
\centering
\scalebox{0.76}{
\begin{tabular}{lr|rrr|rrr}
\Hline\noalign{\smallskip}
\multicolumn{2}{c}{} & \multicolumn{3}{c}{CPM by Size (\%)}& \multicolumn{3}{c}{CPM by Attenuation (\%)}\\
& {\bfseries CPM} (\%) & {\bfseries Small} & {\bfseries Medium} & {\bfseries Large} & {\bfseries Solid} & {\bfseries Part-solid} & {\bfseries GGN} \\\noalign{\smallskip}\hline\noalign{\smallskip}
 632 real images & 51.8 & 44.7 & 61.8 & 62.4  & 65.5  & 46.4  & 24.2 \\
 + $1\times$ 3D MCGAN-based DA & \textbf{55.0} & \textbf{45.2} & \textbf{68.3} & \textbf{66.2} & \textbf{69.9} & 52.1 & 24.4\\
 + $2\times$ 3D MCGAN-based DA & 52.7 & 44.7 & 67.4 & 42.9 & 65.5 & 40.7 & \textbf{28.9} \\
 + $3\times$ 3D MCGAN-based DA & 51.2 & 41.1 & 64.4 & 66.2 &	61.6 &	\textbf{57.9} & 27.7 \\
 + $1\times$ 3D MCGAN-based DA w/ $\ell _1$ & 50.8 & 43.0 & 63.3 & 55.6 & 62.6 & 47.1 & 27.1 \\
 + $2\times$ 3D MCGAN-based DA w/ $\ell _1$ & 50.9 & 40.6 & 64.4 & 65.4 & 64.9 & 43.6 & 23.3 \\
 + $3\times$ 3D MCGAN-based DA w/ $\ell _1$ & 47.9 & 38.9 & 59.4 & 61.7 & 59.6 & 50.7 & 22.6 \\
\noalign{\smallskip}\Hline\noalign{\smallskip}
\end{tabular}
}
\end{table}

\begin{figure}[t!]
  \centering
  \centerline{\includegraphics[width=1\linewidth]{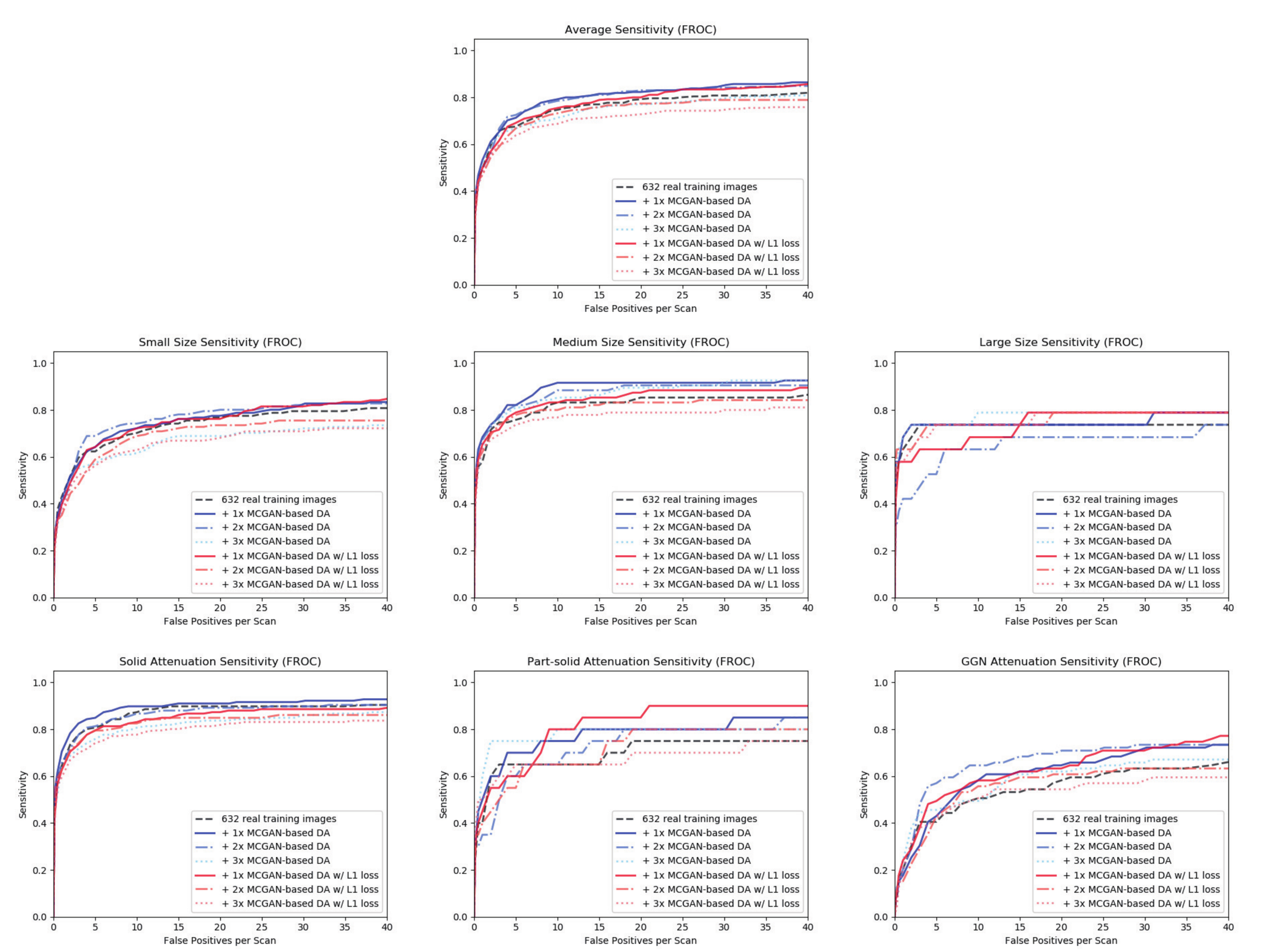}}
\caption{FROC curves of seven DA setups by average/size/attenuation.}
\label{fig7_4}
\end{figure}

\begin{figure}[t!]
  \centering
  \centerline{\includegraphics[width=1\linewidth]{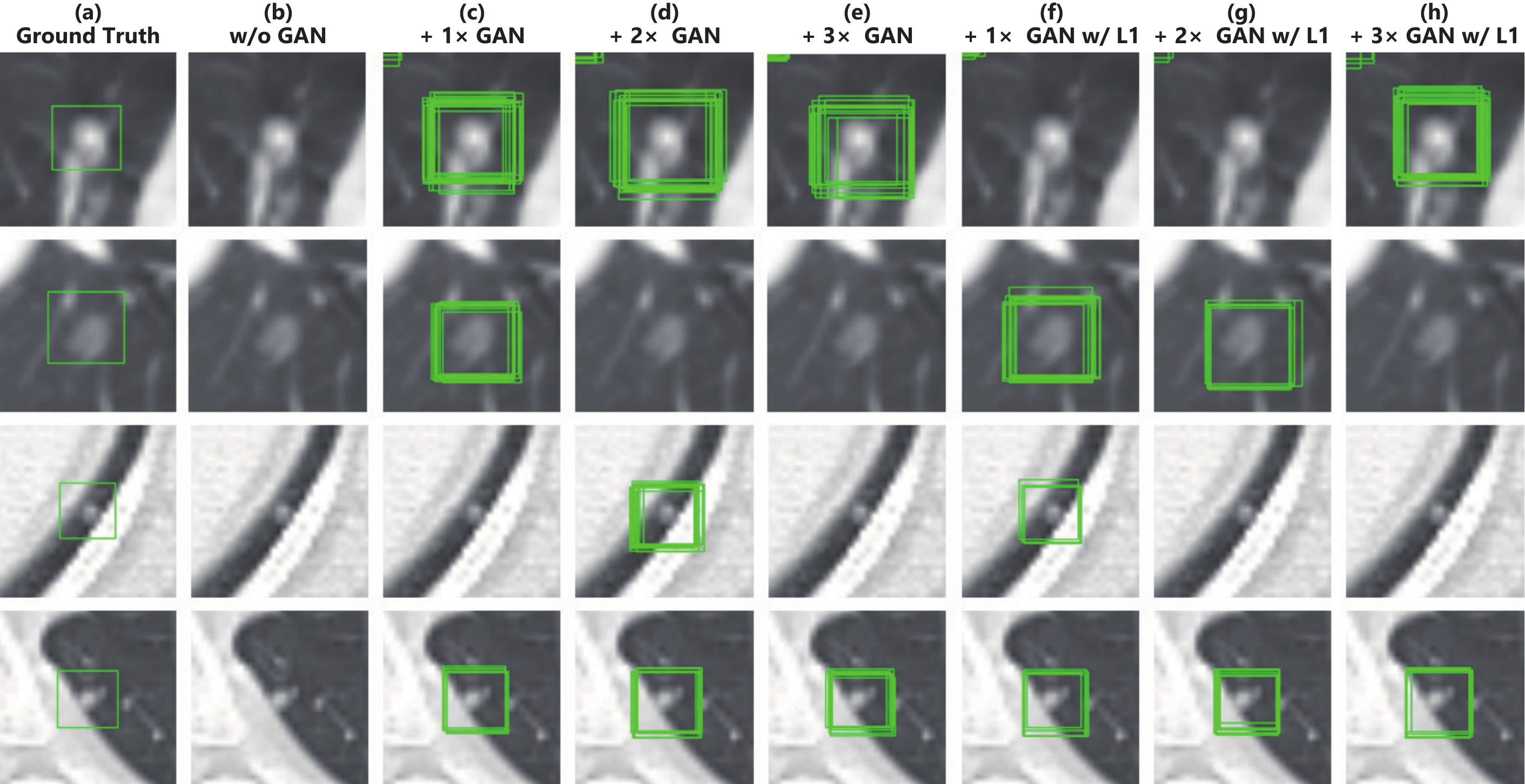}}
\caption[Example detection results of seven DA setups on four different images, compared against the ground truth.]{Example detection results of seven DA setups on four different images, compared against the ground truth (detection threshold $0.5$): (a) ground truth; (b) without GAN-based DA; (c), (d), (e) with $1\times$/$2\times$/$3\times$ 3D MCGAN-based DA; (f), (g), (h) with $1\times$/$2\times$/$3\times$ $\ell _1$ loss-added 3D MCGAN-based DA.}
\label{fig7_5}
\end{figure}

\newpage

\subsection{Lung Nodule Detection Results}
Table~\ref{tab7_1} and Fig.~\ref{fig7_4} show that it is easier to detect nodules with larger size/lower attenuation due to their clear appearance. 3D MCGAN-based DA with less augmentation ratio consistently increases sensitivity at fixed FP rates---especially, training with $1\times$ MCGAN-based DA without $\ell _1$ loss outperforms training only with real images under any size/attenuation in terms of CPM, achieving average CPM improvement by 0.032. It especially boosts nodule detection performance with larger size and lower attenuation. Fig.~\ref{fig7_5} visually reveals its ability to alleviate the risk of overlooking the nodule diagnosis with clinically acceptable FPs (i.e., the highly-overlapping bounding boxes around nodules only require a physician's single check by switching on/off transparent alpha-blended annotation on CT scans). Surprisingly, adding more synthetic images tends to decrease sensitivity, due to the real/synthetic training image balance. Moreover, further nodule realism introduced by $\ell _1$ loss rather decreases sensitivity as $\ell _1$ loss sacrifices diversity in return for the realism.

\newpage

\subsection{Visual Turing Test Results}
As Table~\ref{tab7_2} shows, expert physicians fail to classify real \textit{vs} MCGAN-generated nodules without surrounding tissues---even regarding the synthetic nodules trained without $\ell _1$ loss more realistic than the real ones. Contrarily, they relatively recognize the synthetic nodules with surroundings due to slight shading difference between the nodules/surroundings, especially when trained without the reconstruction $\ell _1$ loss. Considering the synthetic images' realism, CPGGANs might perform as a tool to train medical students and radiology trainees when enough medical images are unavailable, such as abnormalities at rare position/size/attenuation. Such GAN applications are clinically promising~\cite{finlayson2018towards}.

\begin{table}[t!]
\caption[Visual Turing Test results by two physicians for classifying $50$ real \textit{vs} $50$ 3D MCGAN-generated images.]{Visual Turing Test results by two physicians for classifying $50$ real \textit{vs} $50$ 3D MCGAN-generated images: Test1, 2: $32 \times 32 \times 32$ lung nodules, trained without/with $\ell _1$ loss; Test3, 4: $64 \times 64 \times 64$ nodules with surrounding tissues, trained without/with $\ell _1$ loss. Accuracy denotes the physicians' successful classification ratio between the real/synthetic images.}
\label{tab7_2}
\centering
\scalebox{0.63}{
\begingroup
\renewcommand{\arraystretch}{1.2}
\begin{tabular}{p{1.5em}lrrrrr}
\Hline\noalign{\smallskip}
& \bfseries  & \multicolumn{1}{c}{\bfseries Accuracy (\%)} \ \ & \bfseries Real as Real (\%)\ \ \  & \bfseries Real as Synt (\%)\ \ \  & \bfseries Synt as Real (\%)\ \ \  & \bfseries Synt as Synt (\%) \\\noalign{\smallskip}\hline\noalign{\smallskip}
\parbox[t]{2mm}{\multirow{2}{*}{\rotatebox[origin=c]{270}{\textbf{\shortstack{\\Test 1}}}}} & Physician 1 & 43 \ \ & 38 \ \ & 62 \ \ & 52 \ \ & 48\\
& Physician 2 & 43\ \ \  & 26\ \ \  & 74\ \ \  & 40\ \ \  & 60\\
\noalign{\smallskip}\hline\noalign{\smallskip}
\parbox[t]{2mm}{\multirow{2}{*}{\rotatebox[origin=c]{270}{\textbf{\shortstack{\\Test 2}}}}} & Physician 1 & 57 \ \ & 44 \ \ & 56 \ \ & 30 \ \ & 70\\
& Physician 2 & 53 \ \ & 22 \ \ & 78 \ \ & 16 \ \ & 84\\
\noalign{\smallskip}\hline\noalign{\smallskip}
\parbox[t]{2mm}{\multirow{2}{*}{\rotatebox[origin=c]{270}{\textbf{\shortstack{\\Test 3}}}}} & Physician 1 & 62 \ \ & 50 \ \ & 50 \ \ & 26 \ \ & 74\\
& Physician 2 & 79 \ \ & 64 \ \ & 36 \ \ & 6 \ \ & 94\\
\noalign{\smallskip}\hline\noalign{\smallskip}
\parbox[t]{2mm}{\multirow{2}{*}{\rotatebox[origin=c]{270}{\textbf{\shortstack{\\Test 4}}}}} & Physician 1 & 58 \ \ & 42 \ \ & 58 \ \ & 26 \ \ & 74\\
& Physician 2 & 66 \ \ & 72 \ \ & 28 \ \ & 40 \ \ & 60\\
\noalign{\smallskip}\Hline\noalign{\smallskip}
\end{tabular}
\endgroup
}
\end{table}

\subsection{T-SNE Results}
\begin{figure}[t!]
  \centering
  \centerline{\includegraphics[width=1\linewidth]{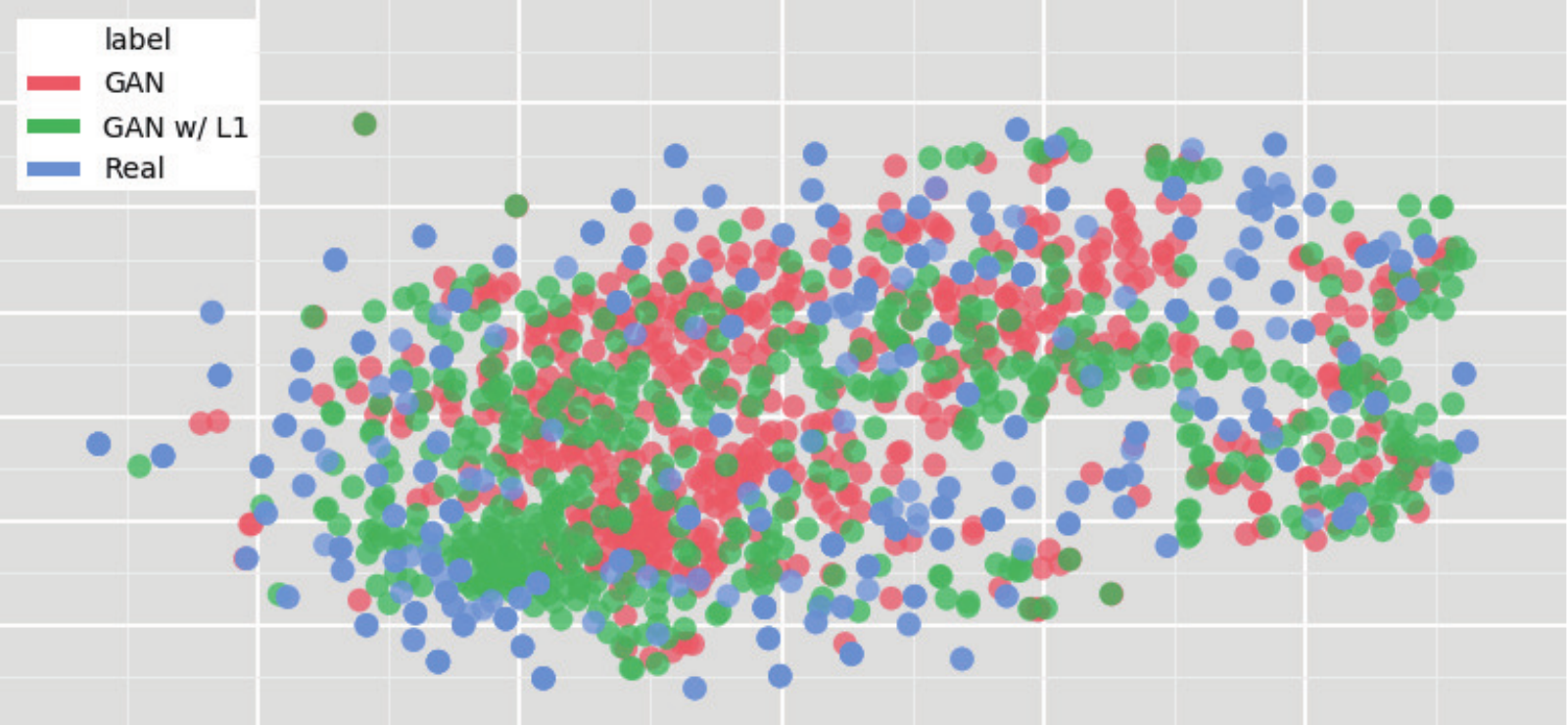}}
\caption[T-SNE plot with $500$ $32 \times 32 \times 32$ nodule images per each category.]{T-SNE plot with $500$ $32 \times 32 \times 32$ nodule images per each category: (a), (b) 3D MCGAN-generated nodules, trained without/with $\ell _1$ loss; (c) real nodules.}
\label{fig7_6}
\end{figure}

Implying their  effective  DA  performance, synthetic nodules have a similar distribution to real ones, but concentrated in left inner areas with less real ones especially when trained without $\ell_1$ loss (Fig.~\ref{fig7_6})--using only GAN loss during training can avoid overwhelming influence from the real image samples, resulting in a moderately similar distribution; thus, those synthetic images can partially fill the real image distribution uncovered by the original dataset.

\section{Conclusion}
Our bounding box-based 3D MCGAN can generate diverse CT-realistic nodules at desired position/size/attenuation, naturally blending with surrounding tissues---those synthetic training data boost sensitivity under any size/attenuation at fixed FP rates in 3D CNN-based nodule detection. This attributes to the MCGAN's good generalization ability coming from multiple discriminators with mutually complementary loss functions, along with informative size/attenuation conditioning; they allow to cover the real image distribution unfilled by the original dataset, improving the training robustness.

Surprisingly, we find that adding over-sufficient synthetic images produces worse results due to the real/synthetic image balance; as t-SNE results show, the synthetic images only partially cover the real image distribution, and thus GAN-overwhelming training images rather harm training. Moreover, we notice that GAN training without $\ell _1$ loss obtains better DA performance thanks to increased diversity providing robustness; also, expert physicians confirm their sufficient realism without $\ell _1$ loss.

Overall, our 3D MCGAN could help minimize expert physicians' time-consuming annotation tasks and overcome the general medical data paucity, not limited to lung CT nodules. As future work, we will investigate the MCGAN-based DA results without size/attenuation conditioning to confirm their influence on DA performance. Moreover, we will compare our DA results with other non-GAN-based recent DA approaches, such as Mixup~\cite{zhang2017mixup} and Cutout~\cite{devries2017improved}. For further performance boost, we plan to directly optimize the detection results for MCGANs, instead of realism, similarly to the three-player GAN for classification~\cite{vandenhende2019three}. Lastly, we will investigate how our MCGAN can perform as a physician training tool to display random realistic medical images
with desired abnormalities (i.e., position/size/attenuation conditions) to help train medical students and radiology trainees despite infrastructural and legal constraints~\cite{finlayson2018towards}.
\chapter{\LARGE Discussions on Developing Clinically\\Relevant AI-Powered Diagnosis Systems}

\section{Prologue to First Project}
\subsection{Project Publication}
\begin{itemize}
\item \textbf{Bridging the gap between AI and healthcare sides: towards developing clinically relevant AI-powered diagnosis systems}. \textbf{C. Han}, L. Rundo, K. Murao, T. Nemoto, H. Nakayama, In \textit{IFIP International Conference on Artificial Intelligence Applications and Innovations (AIAI)}, pp. 320--333, June 2020.
\end{itemize}

\newpage

\section{Feedback from Physicians}
\subsection{Methods for Questionnaire Evaluation}
To confirm the clinical relevance for diagnosis of our proposed pathology-aware GAN methods for DA and physician training respectively, we conduct a questionnaire survey for $9$ Japanese physicians who interpret MR and CT images in daily practice. The experimental settings are the following:
\begin{itemize}
\item \textbf{Subjects:} $3$ physicians (i.e., a radiologist, a psychiatrist, and a physiatrist) committed to (at least one of) our pathology-aware GAN projects and $6$ project non-related radiologists without much AI background.
\item \textbf{Experiments:} Physicians are asked to answer the following questionnaire within 2 weeks from December 6th, 2019 after reading 10 summary slides written in Japanese\footnote{{\scriptsize Available \textit{via} Dropbox: \url{https://www.dropbox.com/sh/bacowc3ilz1p1r3/AABNS9SyjArHq8BntgaODLb2a?dl=0}}} about general Medical Image Analysis and our pathology-aware GAN projects along with example synthesized images. We conduct both qualitative (i.e., free comments) and quantitative (i.e., five-point Likert scale~\cite{allen2007likert}) evaluation: Likert scale 1 $=$ very negative, 2 $=$ negative, 3 $=$ neutral, 4 $=$ positive, 5 $=$ very positive.

\item \textbf{Question 1:} Are you keen to exploit medical AI in general when it achieves accurate and reliable performance in the near future? (five-point Likert scale) Please tell us your expectations, wishes, and worries (free comments).
\item \textbf{Question 2:} What do you think about using GAN-generated images for DA? (five-point Likert scale) Please tell us your expectations, wishes, and worries (free comments).
\item \textbf{Question 3:} What do you think about using GAN-generated images for physician training? (five-point Likert scale) Please tell us your expectations, wishes, and worries (free comments).
\item \textbf{Question 4:} Any comments or suggestions about our projects towards developing clinically relevant AI-powered systems based on your daily diagnosis experience?
\end{itemize}

\subsection{Results}
We show the questions and Japanese physicians' corresponding answers.

\noindent \textbf{Question 1:} Are you keen to exploit medical AI in general when it achieves accurate and reliable performance in the near future?
\begin{itemize}
\item \textbf{Likert scale} Project-related physicians: 5 5 5 (average: 5)\\Project non-related radiologists: 5 5 3 4 5 5 (average: 4.5)
\item \textbf{Free comments} (one comment for each physician)
\item As radiologists, we need AI-based diagnosis during image interpretation as soon as possible.
\item It is common to conduct further medical examinations when identifying disease is difficult from CT/MR images; thus, if AI-based diagnosis outperforms that of physicians, such clinical decision support systems could prevent unnecessary examinations. Moreover, recently lung cancer misdiagnosis occurred in Japan, but AI technologies may prevent such death caused by misdiagnosis.
\item The lack of diagnosticians is very evident in Healthcare, so AI has great potential to support us. It may be already applicable without severe problems for typical disease cases.
\item I am looking forward to its practical applications, especially at low or zero price.
\item I would like to use AI-based diagnosis as a kind of data, but it is yet uncertain how much I trust AI.
\item I am wondering whether such systems will become popular due to practical problems such as introduction cost.
\item The definition of \textit{accurate and reliable} is unclear. Since a physician's annotation is always subjective, we cannot claim that AI-based diagnosis is really correct even if AI diagnoses similarly to the specific physician. Because I do not believe other physicians' diagnosis, but my own eyes, I would use AI just to identify abnormal candidates.
\end{itemize}

As expected, the project-related physicians are AI-enthusiastic while the project non-related radiologists are also generally very positive about the medical AI. Many of them appeal the necessity of AI-based diagnosis for more reliable diagnosis because of the lack of physicians. Meanwhile, other physicians worry about its cost and reliability. We may be able to persuade them by showing expected profitability (e.g., currently CT scanners have an earning rate 16\% and CT scans require 2-20 minutes for interpretation in Japan); similarly, we can explain how experts annotate medical images and AI diagnoses disease based on them (e.g., multiple physicians, not a single one, can annotate the images \textit{via} discussion).\\

\newpage

\noindent \textbf{Question 2:} What do you think about using GAN-generated images for DA?
\begin{itemize}
\item \textbf{Likert scale} Project-related physicians: 5 5 4 (average: 4.7)\\Project non-related radiologists: 4 5 4 4 4 4 (average: 4.2)
\item \textbf{Free comments} (one comment for each physician)
\item Achieved accuracy improvement shows its superiority in identifying diverse disease.
\item It would be effective, especially as rare disease training data.
\item I am looking forward to the future with advanced GAN technology.
\item It significantly improves detection sensitivity; but I am also curious about its influence on other metrics, such as specificity.
\item If Deep Learning could be more effective, we should introduce it; but anonymization would be important for privacy preservation.
\item Achieved accuracy improvement shows its superiority in identifying diverse disease.
\item It would be effective to train AI on data-limited disease, but which means that AI is inferior to humans.
\item It would be helpful if such DA improves accuracy and reliability. Since I am not familiar with AI and a generator/classifier's failure judgment mechanisms, I am uncertain whether it will really increase reliability though.
\end{itemize}

As expected, the project-related physicians are very positive about the GAN-based DA while the project non-related radiologists are also positive. Many of them are satisfied with its achieved accuracy/sensitivity improvement when available annotated images are limited. However, similarly to their opinions on general Medical Image Analysis, some physicians question its reliability.

\newpage

\noindent \textbf{Question 3:} What do you think about using GAN-generated images for physician training?
\begin{itemize}
\item \textbf{Likert scale} Project-related physicians: 3 4 3 (average: 3.3)\\Project non-related radiologists: 3 5 2 3 2 3 (average: 3)
\item \textbf{Free comments} (one comment for each physician)
\item In future medical care, physicians should actively introduce and learn new technology; in this sense, GAN technology should be actively used for physician training in rare diseases.
\item It could be useful for medical student training, which aims for 85\% accuracy by covering typical cases. But expert physician training aims for over 85\% accuracy by comparing typical/atypical cases and acquiring new understanding---real atypical images are essential.
\item In physician training, we use radiological images after definite diagnosis, such as pathological examination---but, we actually lack rare disease cases. Since the GAN-generated images' realism fluctuates based on image augmentation schemes and available training images, further realistic image generation of the rare cases would help the physician training.
\item It depends on how to construct the system.
\item Which specific usage is assumed for such physician training?
\item I cannot state an opinion before actually using the system, but I strongly recognize the importance of looking at real images.
\item I do not 	exactly understand in which situation such physician training is used, but eventually training with realistic images would be also helpful. However, if real images are available, using them would be better.
\end{itemize}

We generally receive neutral feedback because we do not provide a concrete physician training tool, but instead general pathology-aware generation ideas with example synthesized images---thus, some physicians are positive, and some are not. A physician provides a key idea about a pathology-coverage rate for medical student/expert physician training, respectively; for extensive physician training with GAN-generated atypical images, along with pathology-aware GAN-based extrapolation, further GAN-based extrapolation would be valuable.\\

\noindent \textbf{Question 4:} Any comments or suggestions about our pathology-aware GAN projects towards developing clinically relevant AI-powered systems based on your daily diagnosis experience?
\begin{itemize}
\item This approach will change the way physicians work. I have high expectations for AI-based diagnosis, so I hope it to overcome the legal barrier.
\item For now, please show small abnormal findings, such as nodules and ground glass opacities---it would halve radiologists' efforts. Then, we could develop accurate diagnosis step by step.
\item Showing abnormal findings with their shapes/sizes/disease names would increase diagnosis accuracy. But I also would like to know how diagnosticians' roles change after all.
\item I hope that this approach will lead to physicians' work reduction in the future.
\item Please develop reliable AI systems by increasing accuracy with the GAN-based image augmentation.
\item GANs can generate typical images, but not atypical images; this would be the next challenge.
\item AI can alert physicians to detect typical cases, and thus decrease interpretation time; however, it may lead to the diagnosticians' easy diagnosis without much consideration. Especially in Japan, we currently often conduct unnecessary diagnostic tests, so the diagnosticians should be more responsible of their own duties after introducing AI.
\end{itemize}

Most physicians look excited about our pathology-aware GAN-based image augmentation projects and express their clinically relevant requests. The next steps lie in performing further GAN-based extrapolation, developing clinician-friendly systems with new practice guidelines, and overcoming legal/financial constraints.

\section{AI and Healthcare Workshop}
\subsection{Methods for Workshop Evaluation}
Convolutional Neural Networks (CNNs) have achieved accurate and reliable Computer-Aided Diagnosis (CAD), occasionally outperforming expert physicians~\cite{hwang2018development,wu2019TMI,mckinney2020}. However, such research results cannot be easily applied to a clinical environment: AI and Healthcare sides have a huge gap around technology, funding, and people, such as clinical significance/interpretation, data acquisition, commercial purpose, and anxiety about AI. Aiming to identify/bridge the gap between AI and Healthcare sides in Japan towards develop medical AI fitting into a clinical environment in five years, we hold a workshop for $7$ Japanese professionals with various AI and/or Healthcare background. The experimental settings are the following:

\begin{table*}[t!]
\caption[Workshop program to (\textit{i}) know the overview of Medical Image Analysis and (\textit{ii}) find the intrinsic gap and solutions between AI researchers and Healthcare workers.]{Workshop program to \textit{i}) know the overview of Medical Image Analysis and \textit{ii}) find the intrinsic gap and its solutions between AI researchers and Healthcare workers. \textbf{*} indicates activities given by a facilitator (i.e., the first author), such as lectures.}
\label{tab8_1}
\centering
\scalebox{0.97}{
\begin{tabular}{ll}
\Hline\noalign{\smallskip}
{\bfseries Time} (mins) & \bfseries Activity \\\noalign{\smallskip}\hline\noalign{\smallskip}
& \bfseries Introduction \\
10 & 1. Explanation of the workshop's purpose and flow\textbf{*} \\
10 & 2. Self-introduction and explanation of motivation for participation \\
5 & 3. Grouping into two groups based on background\textbf{*}\\
\noalign{\smallskip}\hline\noalign{\smallskip}
& \bfseries Learning: Knowing Medical Image Analysis \\
15 & 1. TED speech video watching: \textit{Artificial Intelligence Can Change}\\ & \textit{the future of Medical Diagnosis}\textbf{*}\\
35 & 2. Lecture: Overview of Medical Image Analysis including\\ & state-of-the-art research, well-known challenges/solutions, and our\\ & pathology-aware GAN projects summary\textbf{*}\\ &  (its video in Japanese: https://youtu.be/rTQLknPvnqs)\\
10 & 3. Sharing expectations, wishes, and worries about Medical Image\\ & Analysis (its video in Japanese: https://youtu.be/ILPEGga-hkY)\\
10 & Intermission\\
\noalign{\smallskip}\hline\noalign{\smallskip}
& \bfseries Thinking: Finding How to Develop Robust Medical AI\\
25 & 1. Identifying the intrinsic gap between AI/Healthcare sides \\ & after sharing their common and different thinking/working styles\\
60 & 2. Finding how to develop gap-bridging medical AI fitting into \\ &  a clinical environment in five years\\
10 & Intermission\\
\noalign{\smallskip}\hline\noalign{\smallskip}
& \bfseries Summary \\
25 & 1. Presentation\\
10 & 2. Sharing workshop impressions and ideas to apply obtained \\ & knowledge (its video in Japanese: https://youtu.be/F31tPR3m8hs)\\
5 & 3. Answering a questionnaire about satisfaction/further comments\\
5 & 4. Closing remarks\textbf{*}\\
\noalign{\smallskip}\Hline\noalign{\smallskip}
\end{tabular}
}
\end{table*}

\begin{itemize}
\item \textbf{Subjects:} $2$ Medical Imaging experts (i.e., a Medical Imaging researcher and a medical AI startup entrepreneur), $2$ physicians (i.e., a radiologist and a psychiatrist), and $3$ generalists between Healthcare and Informatics (i.e., a nurse and researcher in medical information standardization, a general practitioner and researcher in medical communication, and a medical technology manufacturer's owner and researcher in health disparities)

\item \textbf{Experiments:} As its program shows (Table~\ref{tab8_1}), during the workshop, we conduct 2 activities: (\textit{Learning}) Know the overview of Medical Image Analysis, including state-of-the-art research, well-known challenges/solutions, and the summary of our pathology-aware GAN projects; (\textit{Thinking}) Find the intrinsic gap and its solutions between AI researchers and Healthcare workers after sharing their common and different thinking/working styles. Supported by GCL program, this workshop was held on March 17th, 2019 at Nakayama Future Factory, Open Studio, The University of Tokyo, Tokyo, Japan.
\end{itemize}

\subsection{Results}
We show the summary of clinically-relevant findings from this Japanese workshop.
\subsubsection{Gap Between AI and Healthcare Sides}
\noindent \textbf{Gap 1:} AI, including Deep Learning, does not provide clear decision criteria, does it make physicians reluctant to use it in a clinical environment, especially for diagnosis?

\begin{itemize}
\item \textbf{Healthcare side}: We rather expect applications other than diagnosis. If we use AI for diagnosis, instead of replacing physicians, we suppose a \textit{reliable second opinion}, such as alert to avoid misdiagnosis, based on various clinical data not limited to images---every single diagnostician is anxious about their diagnosis. AI only provides minimum explanation, such as a heatmap showing attention, which makes persuading not only the physicians but also patients difficult; so, the physicians' intervention is essential for intuitive explanation. Methodological safety and feeling safe are different. In this sense, pursuing explainable AI generally decreases AI's diagnostic accuracy~\cite{adadi2018peeking}, so physicians should still serve as mediators by engaging in high-level conversation or interaction with patients. Moreover, according to the medical law in most countries including Japan, only doctors can make the final decision. The first autonomous AI-based diagnosis without a physician was cleared by the Food and Drug Administration in the US in 2018~\cite{abramoff2018pivotal}, but such a case is exceptional.

\item \textbf{AI side}: Compared with other systems or physicians, Deep Learning's explanation is not particularly poor, so we require too severe standards for AI; the word \textit{AI} is excessively promoting anxiety and perfection. If we could thoroughly verify the reliability of its diagnosis against physicians by exploring uncertainty measures~\cite{nair2020exploring}, such intuitive explanation would be optional.
\end{itemize}

\noindent \textbf{Gap 2:} Are there any benefits to actually introducing medical AI?

\begin{itemize}
\item \textbf{Healthcare side}: After all, even if AI can achieve high accuracy and convenient operation, hospitals would not introduce it without any commercial benefits. Moreover, small clinics, where physicians are desperately needed, often do not have CT or MRI scanners~\cite{jankharia2008commentary}.

\item \textbf{AI side}: The commercial deployment of medical AI is strongly tied to diagnostic accuracy~\cite{vollmer2018machine}; so, if it can achieve significantly outstanding accuracy at various tasks in the near future, patients would not visit hospitals/clinics without AI. Accordingly, introducing medical AI would become profitable in five years.
\end{itemize}

\noindent \textbf{Gap 3:} Is medical AI's diagnostic accuracy reliable?

\begin{itemize}
\item \textbf{Healthcare side}: To evaluate AI's diagnostic performance, we should consider many metrics, such as sensitivity and specificity. Moreover, its generalization ability for medical data highly relies on inter-scanner/inter-individual variability~\cite{oConnor2017healthy}. How can we evaluate whether it is suitable as a clinically applicable system?

\item \textbf{AI side}: Generally, alleviating the risk of overlooking the diagnosis is the most important, so sensitivity matters more than specificity unless their balance is highly disturbed. Recently, such research on medical AI that is robust to different datasets is active~\cite{Rundo2}.
\end{itemize}

\subsubsection{How to Develop Medical AI Fitting into a Clinical Environment in Five Years}

\noindent \textbf{Why:} Clinical significance/interpretation

\begin{itemize}
\item \textbf{Challenges}: We need to clarify which clinical situations actually require AI introduction. Moreover, AI's early diagnosis might not be always beneficial for patients.

\item \textbf{Solutions}: Due to nearly endless disease types and frequent misdiagnosis coming from physicians' fatigue, we should use it as alert to avoid misdiagnosis~\cite{vandenberghe2017relevance} (e.g., reliable second opinion), instead of replacing physicians. It should help prevent oversight in diagnostic tests not only with CT and MRI, but also with blood data, chest X-ray, and mammography before taking CT and MRI~\cite{li2019medical}. It could be also applied to segmentation for radiation therapy~\cite{agn2016generative}, neurosurgery navigation~\cite{abi2018machine}, and pressure ulcers' echo evaluation. Along with improving the diagnosis, it would also make the physicians' workflow easier, such as by denoising~\cite{yang2018low}. Patients should decide whether they accept AI-based diagnosis under informed consent.

\end{itemize}

\noindent \textbf{How:} Data acquisition

\begin{itemize}
\item \textbf{Challenges}: Ethical screening in Japan is exceptionally strict, so acquiring and sharing large-scale medical data/annotation are challenging---it also applies to Europe due to General Data Protection Regulation~\cite{rumbold2017effect}. Considering the speed of technological advances in AI, adopting it for medical devices is difficult in Japan, unlike in medical AI-ready countries, such as the US, where the ethical screening is relatively loose in return for the responsibility of monitoring system stability. Moreover, whenever diagnostic criteria changes, we need further reviews and software modifications; for example, the Tumor-lymph Node-Metastasis (TNM) classification~\cite{sobin2011tnm} criteria changed for oropharyngeal cancer in 2018 and for lung cancer in 2017, respectively. Diagnostic equipment/target changes also require large-scale data/annotation acquisition again.

\item \textbf{Solutions}: 
For Japan to keep pace, the ethical screening should be adequate to the other leading countries. Currently, overseas research and clinical trials are proceeding much faster, so it seems better to collaborate with overseas companies than to do it in Japan alone. Moreover, complete medical checkup, which is extremely costly, is unique in East Asia, so Japan could be superior in individuals' multiple medical data---Japan is the only country, where most workers 40 or older are required to have medical checkups once a year independent of their health conditions by the Industrial Safety and Health Act~\cite{nawata2016evaluation}. To handle changes in diagnostic criteria/equipment and overcome dataset/task dependency, it is necessary to establish a common database creation workflow~\cite{mansour2019visual} by regularly entering electronic medical records into the database. For reducing data acquisition/annotation cost, AI techniques, such as GAN-based DA~\cite{han2019synthesizing} and domain adaptation~\cite{ghafoorian2017transfer}, would be effective.
\end{itemize}

\noindent \textbf{How:} Commercial deployment

\begin{itemize}
\item \textbf{Challenges}: Hospitals currently do not have commercial benefits to actually introduce medical AI.

\item \textbf{Solutions}: For example, it would be possible to build AI-powered hospitals~\cite{chen2019feasibility} operated with less staff. Medical manufacturers could also standardize data format~\cite{laplante2016hearing}, such as for X-ray, and provide some AI services. Many IT giants like Google are now working on medical AI to collect massive biomedical data~\cite{morley2019google}, so they could help rural areas and developing countries, where physician shortage is severe~\cite{jankharia2008commentary}, at relatively low cost.
\end{itemize}

\noindent \textbf{How:} Safety and feeling safe

\begin{itemize}
\item \textbf{Challenges}: Considering multiple metrics, such as sensitivity and specificity~\cite{rossini2016diagnostic}, and dataset/task dependency~\cite{huang2018medical}, accuracy could be unreliable, so ensuring safety is challenging. Moreover, reassuring physicians and patients is important to actually use AI in a clinical environment~\cite{krittanawong2018rise}.

\item \textbf{Solutions}: We should integrate various clinical data, such as blood test biomarkers and multiomics, with images~\cite{li2019medical}. Moreover, developing bias-robust technology is important since confounding factors are inevitable~\cite{li2018fully}. To prevent oversight, prioritizing sensitivity over specificity is essential while maintaining a balance~\cite{jain2017addressing}. We should also devise education for medical AI users, such as result interpretation, to reassure patients~\cite{wartman2019reimagining}.
\end{itemize}
\chapter{\LARGE Conclusion}

\section{Final Remarks}
Inspired by their excellent ability to generate realistic and diverse images, we propose to use noise-to-image GANs for (\textit{i}) Medical DA and (\textit{ii}) physician training~\cite{Han1}. Through information conversion, such applications can relieve the lack of pathological data and their annotation; this is uniquely and intrinsically important in Medical Image Analysis, as CNN generalization becomes unstable on unseen data due to large inter-subject, inter-pathology, and cross-modality variability~\cite{Rundo2, rundo2020cnn, pooch2019can}. Towards clinically relevant implementation for the DA and physician training, we find effective loss functions and training schemes for each of them~\cite{Han2,han2019combining}---the diversity matters more for the DA to sufficiently fill the real image distribution whereas the realism matters more for the physician training not to confuse medical students and radiology trainees.

Specifically, our results imply that GAN training without $\ell _1$ loss, using proper augmentation ratio (i.e., $1 : 1$), and further refining synthetic images' texture/shape could improve the DA performance, whereas discarding weird-looking synthetic images to humans is unnecessary; for example, adding over-sufficient GAN-generated training images leads to more FPs in detection, but not always higher sensitivity, due to the real/synthetic training data balance (both of their distributions are biased, but differently). Regarding the physician training, GAN training with $\ell _1$ loss, GAN training on additional normal images, and post-processing, such as by Poisson image editing~\cite{perez2003poisson}, could improve the synthetic images' realism; for instance, the GAN training on normal images along with pathological ones, remarkably facilitates the realism of both healthy and pathological parts while they do not include abnormality.

Because such excellent realism and diversity can be achieved by GAN-based interpolation and extrapolation, we propose novel 2D/3D pathology-aware GANs for bounding box-based pathology detection~\cite{han2019learning,han2019synthesizing}: (\textit{Interpolation}) The GAN-based medical image augmentation is reliable because medical modalities (e.g., X-ray, CT, MRI) can display the human body's strong anatomical consistency at fixed position while clearly reflecting inter-subject variability~\cite{hsieh2009computed,brown2014magnetic}---this is different from natural images, where various objects can appear at any position; (\textit{Extrapolation}) The pathology-aware GANs are promising because common and/or desired medical priors can play a key role in the conditioning---theoretically, infinite conditioning instances, external to the training data, exist and enforcing such constraints have an extrapolation effect \textit{via} model reduction~\cite{stinis2019enforcing}.

After conducting a questionnaire survey about our GAN projects for 9 physicians and holding a workshop about how to develop medical AI fitting into a clinical environment for 7 professionals with various AI and/or Healthcare background, we confirm our pathology-aware GANs' clinical relevance for diagnosis: (\textit{DA}) They could be integrated into a clinical decision support system; since CT has a much higher earning rate and longer interpretation time than MRI (16\% to 3\% and 2-20 minutes to 1 minute in Japan), alerting abnormal findings on CT, such as nodules/ground glass opacities, would halve radiologists' efforts and increase hospitals' financial outcomes; (\textit{Physician training}) They could perform as a non-expert physician training tool; when the normal training images are sufficiently available, we can stably generate typical pathological images useful for medical student training, thanks to the excellent interpolation; but it is still challenging to generate atypical images needed for expert physician training. Whereas our pathology-aware bounding box conditioning largely improves extrapolation ability, better DA and physician training would require further GAN-based extrapolation.

\section{Future Work}
We believe that the next steps towards GAN-based extrapolation and thus atypical pathological image generation lie in (\textit{i}) generation by parts with coordinate conditions~\cite{lin2019coco}, (\textit{ii}) generation with both image and gene expression conditions~\cite{xu2019correlation}, and (\textit{iii}) transfer learning among different body parts and disease types~\cite{chen2019med3d}. Due to biological constraints, human interaction is restricted to part of the surrounding environment. Accordingly, we must reason spatial relationships across the surrounding parts to piece them together. Similarly, since machine performance also depends on computational constraints, it is plausible for a generator to generate partial images using the corresponding spatial coordinate conditions---meanwhile, a discriminator attempts to judge realism across the assembled patches by global coherence, local appearance, and edge-crossing continuity. This approach allowed COnditional COordinate GAN (COCO-GAN) to generate state-of-the-art realistic and seamless full images~\cite{lin2019coco}. Since human anatomy has a much stronger local consistency than various object relationships in natural images, reasoning the body's spatial relationships, like the COCO-GAN, would perform effective extrapolation both for medical DA and physician training.

We can also condition the GANs both on the image features and gene expression profiles to non-invasively identify molecular properties
of disease. By so doing, Xu \textit{et al.} succeeded to produce $60 \times 60 \times 60$ realistic synthetic CT images of lung nodules~\cite{xu2019correlation}. If the gene expression data are available, such condition fusing could be helpful for the medical DA and physician training.

Such information conversion, not limited to the GAN conditioning, should locate in the core of future Medical Image Analysis to overcome the data paucity. Whereas the transfer learning from large-scale natural image/video datasets for CNNs is already common in Medical Image Analysis, such pre-trained models cannot extract general human anatomical features. Accordingly, pre-training on large-scale 3D medical volumes for CNNs, such as CT and MRI, significantly outperformed the pre-training on natural videos or training from scratch for classification and segmentation~\cite{chen2019med3d} both by accuracy and training convergence speed. Similarly, transfer learning from mammography for the CNNs also significantly improved mass detection on digital breast tomosynthesis slices~\cite{samala2016mass}. Such transfer learning across different body parts and disease types for the GANs would also largely improve their extrapolation ability.
\begin{singlespace}
\bibliography{main}
\bibliographystyle{ieeetr}
\end{singlespace}

\appendix
\chapter{\LARGE Scientific Production}
\section{Related Publications/Presentations}

\textbf{Journal papers}
\begin{itemize}
\item \textbf{C. Han}, L. Rundo, R. Araki, Y. Nagano, Y. Furukawa, G. Mauri, H. Nakayama, H. Hayashi, Combining Noise-to-Image and Image-to-Image GANs: Brain MR Image Augmentation for Tumor Detection, \textit{IEEE Access}, October 2019 (\textbf{Project 2}).
\item \textbf{C. Han}, K. Murao, S. Satoh, H. Nakayama, Learning More with Less: GAN-based Medical Image Augmentation, \textit{Medical Imaging Technology}, Japanese Society of Medical Imaging Technology, June 2019 (\textbf{Tutorial Paper}).
\end{itemize}
\textbf{Book chapter}
\begin{itemize}
\item \textbf{C. Han}, L. Rundo, R. Araki, Y. Furukawa, G. Mauri, H. Nakayama, H. Hayashi, Infinite Brain MR Images: PGGAN-based Data Augmentation for Tumor Detection, In A. Esposito, M. Faundez-Zanuy, F. C. Morabito, E. Pasero (eds.) \textit{Neural Approaches to Dynamics of Signal Exchanges}, Springer, September 2019 (\textbf{Project 2}).
\end{itemize}

\newpage

\textbf{Conference proceedings}
\begin{itemize}
\item \textbf{C. Han}, L. Rundo, K. Murao, T. Nemoto, H. Nakayama, Bridging the gap between AI and healthcare sides: towards developing clinically relevant AI-powered diagnosis systems, In \textit{IFIP International Conference on Artificial Intelligence Applications and Innovations (AIAI)}, pp. 320--333, June 2020\\ (\textbf{Discussion Paper}).
\item \textbf{C. Han}, K. Murao, T. Noguchi, Y. Kawata, F. Uchiyama, L. Rundo, H. Nakayama, S. Satoh, Learning More with Less: Conditional PGGAN-based Data Augmentation for Brain Metastases Detection Using Highly-Rough Annotation on MR Images, In \textit{ACM International Conference on Information and Knowledge Management (CIKM)}, Beijing, China, November 2019 (\textbf{Project 3}).
\item \textbf{C. Han}, Y. Kitamura, A. Kudo, A. Ichinose, L. Rundo, Y. Furukawa, K. Umemoto, H. Nakayama, Y. Li, Synthesizing Diverse Lung Nodules Wherever Massively: 3D Multi-Conditional GAN-based CT Image Augmentation for Object Detection, In \textit{International Conference on 3D Vision (3DV)}, Qu\'ebec City, Canada, September 2019 (\textbf{Project 4}).
\item \textbf{C. Han}, H. Hayashi, L. Rundo, R. Araki, Y. Furukawa, W. Shimoda, S. Muramatsu, G. Mauri, H. Nakayama, GAN-based Synthetic Brain MR Image Generation, In \textit{IEEE International Symposium on Biomedical Imaging (ISBI)}, Washington, D.C., The United States, April 2018 (\textbf{Project 1}).
\end{itemize}

\section{Other Publications/Presentations}
\textbf{Journal paper}
\begin{itemize}
\item  L. Rundo*, \textbf{C. Han}*, Y. Nagano, J. Zhang, R. Hataya, C. Militello, A. Tangherloni, M. S. Nobile, C. Ferretti, D. Besozzi, M. C. Gilardi, S. Vitabile, G. Mauri, H. Nakayama, P. Cazzaniga, USE-Net: incorporating Squeeze-and-Excitation blocks into U-Net for prostate zonal segmentation of multi-institutional MRI datasets, \textit{Neurocomputing}, July 2019 (* denotes co-first authors).
\end{itemize}

\textbf{Book chapters}
\begin{itemize}
\item L. Rundo, \textbf{C. Han}, J. Zhang, R. Hataya, Y. Nagano, C. Militello, C. Ferretti, M. S. Nobile, A. Tangherloni, M. C. Gilardi, S. Vitabile, H. Nakayama, G. Mauri, CNN-based Prostate Zonal Segmentation on T2-weighted MR Images: A Cross-dataset Study, A. Esposito, M. Faundez-Zanuy, F. C. Morabito, E. Pasero (eds.) \textit{Neural Approaches to Dynamics of Signal Exchanges}, Springer, September 2019. 
\item \textbf{C. Han}, K. Tsuge, H. Iba, Application of Learning Classifier Systems to Gene Expression Analysis in Synthetic Biology, In S. Patnaik, X. Yang, and K. Nakamatsu (eds.) \textit{Nature Inspired Computing and Optimization: Theory and Applications}, Springer, March 2017.
\end{itemize}

\textbf{Conference proceedings}
\begin{itemize}
\item K. Murao, Y. Ninomiya, \textbf{C. Han}, K. Aida, S. Satoh, Cloud platform for deep learning-based CAD via collaboration between Japanese medical societies and institutes of informatics, In \textit{SPIE Medical Imaging} (oral presentation), Houston, The United States, February 2020.
\item \textbf{C. Han}, L. Rundo, K. Murao, Z. Á. Milacski, K. Umemoto, H. Nakayama, S. Satoh, GAN-based Multiple Adjacent Brain MRI Slice Reconstruction for Unsupervised Alzheimer's Disease Diagnosis, In \textit{Computational Intelligence methods for Bioinformatics and Biostatistics (CIBB)}, Bergamo, Italy, September 2019.
\item \textbf{C. Han}, K. Tsuge, H. Iba, Optimization of Artificial Operon Construction by Consultation Algorithms Utilizing LCS, In \textit{IEEE Congress on Evolutionary Computation (CEC)}, Vancouver, Canada, July 2016.
\end{itemize}

\end{document}